\documentclass[prd,preprintnumbers,floatfix,
nofootinbib,superscriptaddress]{revtex4}

\usepackage{float}
\usepackage{nicefrac}
\usepackage{mathtools}
\usepackage{amsfonts} 
\usepackage{amssymb} 
\usepackage{amsmath} 
\usepackage{graphicx} 
\usepackage{subfigure} 
\usepackage{array} 
\usepackage{dcolumn} 
\usepackage{bm} 
\usepackage{latexsym} 
\usepackage{longtable} 
\usepackage{hyperref} 
\usepackage{verbatim}
\usepackage{epsfig}
\usepackage{slashed}
\usepackage{color}

\usepackage{glossaries-extra}
\setabbreviationstyle[acronym]{long-short}
\glssetcategoryattribute{acronym}{nohyperfirst}{true}
\renewcommand{\vec}[1]{\mathbf{#1}}
\usepackage{braket}

\newcommand{\Eqref}[1]{(\ref{#1})}

\newcommand{\cB}[0]{\mathcal B}
\newcommand{\cD}[0]{\mathcal D}
\newcommand{\cI}[0]{\mathcal I}
\newcommand{\cK}[0]{\mathcal K}
\newcommand{\cL}[0]{\mathcal L}
\newcommand{\cM}[0]{\mathcal M}
\newcommand{\cO}[0]{\mathcal O}

\newcommand{\cR}[0]{\mathcal R}

\newcommand{\cT}[0]{\mathcal T}
\newcommand{\cY}[0]{\mathcal Y}

\newcommand{\wt}[0]{\widetilde}
\newcommand{\mc}[0]{\mathcal}
\newcommand{\df}[0]{\mathrm{df}}
\newcommand{\on}[0]{\mathrm{on}}

\newcommand{\uu}[0]{{(u,u)}}
\newcommand{\oneu}[0]{{(u)}}
\newcommand{\ones}[0]{{(s)}}
\newcommand{\onets}[0]{{(\tilde s)}}

\newcommand{\Kdf}[0]{{\cK_{\df,3}}}
\newcommand{\Kdfuu}[0]{{\widetilde\cK_{\df,3}^{\uu}}}
\newcommand{\Kdfuun}[0]{{\cK_{\df,3}^{\prime\uu}}}
\newcommand{\Kdfuunn}[0]{{\wt\cK_{\df,3}^{\prime\uu}}}
\newcommand{\KdfuuHS}[0]{{\cK_{\df,3}^{\uu}}}
\newcommand{\PV}[0]{{\mathrm{PV}}}

\newcommand{\Kalluu}[0]{{\widetilde\cK_{\df,23,L}^\uu}}
\newcommand{\Z}{\mathbb{Z}}

\newcommand{\eps}{\epsilon}
\newcommand{\iy}{\infty}

\newcommand{\K}{\mc{K}}
\newcommand{\Kdft}{\wt{\mc{K}}_{\text{df},3}}
\newcommand{\I}{\mc{I}}
\newcommand{\UV}{\mathrm{UV}}

\makeatletter
\newcommand{\wtt}[1]{{%
  \mathpalette\double@widetilde{#1}
}}
\newcommand{\double@widetilde}[2]{%
  \sbox\z@{$\m@th#1\widetilde{#2\, }$}%
  \ht\z@=.9\ht\z@
  \widetilde{\box\z@}%
}
\makeatother

\newcommand{\LtoK}[0]{Hansen:2014eka}
\newcommand{\KtoM}[0]{Hansen:2015zga}
\newcommand{\KSS}[0]{Kim:2005gf}
\newcommand{\SPT}[0]{Sharpe:2017jej}
\newcommand{\HSBS}[0]{Hansen:2016ync}
\newcommand{\HSPT}[0]{Hansen:2015zta}
\newcommand{\HSTH}[0]{Hansen:2016fzj}
\newcommand{\BHSQC}[0]{Briceno:2017tce}
\newcommand{\BHSnum}[0]{Briceno:2018mlh}
\newcommand{\BHSK}[0]{Briceno:2018aml}
\newcommand{\HSQCa}[0]{Hansen:2014eka}

\newcommand{\dwave}[0]{Blanton:2019igq}
\newcommand{\largera}[0]{Romero-Lopez:2019qrt}
\newcommand{\HSrev}[0]{Hansen:2019nir}

\newcommand{\HHanal}[0]{Blanton:2019vdk}
\newcommand{\isospin}[0]{Hansen:2020zhy}

\newcommand{\Akakia}[0]{Hammer:2017uqm}
\newcommand{\Akakib}[0]{Hammer:2017kms}

\newcommand{\MDpi}[0]{Mai:2018djl}
\newcommand{\HH}[0]{Horz:2019rrn}

\newcommand{\MD}[0]{Mai:2017bge}
\newcommand{\MDHH}[0]{Mai:2019fba}
\newcommand{\Akakinum}[0]{Doring:2018xxx}
\newcommand{\isobar}[0]{Jackura:2018xnx}
\newcommand{\Maiisobar}[0]{Mai:2017vot}
\newacronym{CMF}{CMF}{center-of-momentum frame}

\newcommand{\steve}[0]{\bf \color{magenta}}

\begin{document}


\title{Alternative derivation of the relativistic three-particle quantization condition}

\author{Tyler D. Blanton}
\email[e-mail: ]{blanton1@uw.edu}
\affiliation{Physics Department, University of Washington, Seattle, WA 98195-1560, USA}

\author{Stephen R. Sharpe}
\email[e-mail: ]{srsharpe@uw.edu}
\affiliation{Physics Department, University of Washington, Seattle, WA 98195-1560, USA}


\date{\today}

\begin{abstract}
We present a simplified derivation of the relativistic three-particle quantization
condition for identical, spinless particles 
described by a generic relativistic field theory
satisfying a $\mathbb Z_2$ symmetry.
The simplification is afforded by using
a three-particle quasilocal K matrix that is not fully symmetrized,
$\Kdfuu$, and makes extensive use of time-ordered perturbation theory (TOPT).
We obtain a new form of the quantization condition.
This new form can then be related algebraically to
the standard quantization condition,
which depends on a fully
symmetric three-particle K matrix, $\Kdf$.
The new derivation is fully explicit, allowing, for example, a closed-form expression 
for $\Kdf$ to be given in terms of TOPT amplitudes.
The new form of the quantization condition is similar in structure to that obtained in
the ``finite-volume unitarity'' approach, and in a companion paper we make
this connection concrete.
Our simplified approach
should also allow a more straightforward generalization of the quantization condition to 
nondegenerate particles, and perhaps also to more than three particles.
 \end{abstract}


\nopagebreak

\maketitle


\section{Introduction\label{sec:intro}}

One of the present frontiers of lattice QCD (LQCD) is the study of systems containing three or more particles.
The aims include the determination of the three-nucleon interaction and the study of resonances decaying
to three or more particles.
Advances have been made both in the ability to calculate multiple finite-volume energy levels
using numerical simulations,
and in the theoretical formalism needed to interpret the results.
Examples of the successful combination of these methods are in Refs.~\cite{%
Beane:2007es,Detmold:2008fn,Detmold:2011kw,Detmold:2013wda,Detmold:2013gua,Mai:2018djl,\HH,\HHanal,Mai:2019fba,Culver:2019vvu,Beane:2020ycc}.\footnote{%
For related applications in lattice simulations of $\phi^4$ theory see Ref.~\cite{Romero-Lopez:2018rcb}.}

A key output of the theoretical formalism is a quantization condition, an equation whose solutions
give the finite-volume three-particle energy levels in terms of infinite-volume scattering quantities.
The latter quantities are then related to infinite-volume scattering amplitudes in a second step 
that involves solving integral equations.
Our aim in this work is to provide a simplified method for deriving the quantization condition
in a generic relativistic effective field theory (RFT).
Our hope is that our new method will simplify the generalization of the quantization condition
to systems not heretofore studied, for example
to three nondegenerate particles and to more than three particles,
as well as allow the unification of the different approaches used to develop the three-particle
formalism (to be described below).

The two-particle quantization condition has been known for decades and
is now a standard tool in LQCD~\cite{Luscher:1986n2,Luscher:1991n1,Rummukainen:1995vs,Kim:2005gf, He:2005ey, Lage:2009,Fu:2011,Hansen:2012tf, Briceno:2012yi,Gockeler:2012yj,Briceno:2014oea}.
(See Ref.~\cite{Briceno:2017max} for a review.) 
The three-particle formalism has been developed more recently, 
using three main approaches:\footnote{%
See also the foundational work of Ref.~\cite{Polejaeva:2012ut},
the threshold expansions of Refs.~\cite{Tan:2007bg,Beane:2007qr,Detmold:2008gh,\HSTH,Detmold:2014fpa},
and the alternative approaches in 
Refs.~\cite{Briceno:2012rv,Guo:2017ism,Klos:2018sen,Guo:2018ibd}.
For recent reviews, see Refs.~\cite{\HSrev,Rusetsky:2019gyk}.
}
\begin{enumerate}
\item 
The RFT approach, which is the most general and also
the most complicated. This formalism was derived in Refs.~\cite{\LtoK,\KtoM}
for the case of identical scalar particles with a $\Z_2$, G-parity-like, symmetry.
We refer to these papers in the following  as HS1 and HS2, respectively.
The formalism has been 
subsequently generalized to allow $2\leftrightarrow 3$ transitions~\cite{\BHSQC},
K matrix poles~\cite{\BHSK,\largera},
and nonidentical but degenerate particles~\cite{\isospin}.
The numerical implementation of the formalism has been studied in
Refs.~\cite{\BHSnum,\dwave,\largera},
and  recently applied to extract the $3\pi^+$ interaction~\cite{\HHanal}
using results for the $3\pi^+$ spectrum from Ref.~\cite{\HH}.
\item 
The nonrelativistic effective field theory (NREFT) approach of Refs.~\cite{\Akakia,\Akakib,\Akakinum,Pang:2019dfe}.
Here the derivation is much simpler, but to date the formalism has been developed only for
two-particle interactions restricted to the s-wave. 
\item 
The ``finite-volume unitarity'' (FVU) approach of Refs.~\cite{\MD,\MDpi,Culver:2019vvu,\MDHH}.
This is a relativistic formalism that is based on general forms for the three-particle scattering amplitude
developed in Refs.~\cite{\Maiisobar,\isobar}
that incorporate s-channel unitarity.
As for the NREFT approach, it has to date been developed only
for s-wave two-particle interactions. It has been applied to the $3\pi^+$ spectrum from LQCD in
Refs.~\cite{Mai:2019fba,Culver:2019vvu}.
\end{enumerate}
Our aim, as already noted, is to simplify the RFT derivation given in HS1 and HS2. 
As a side benefit, our result provides a further check on the original formalism,
adding to the checks provided in Refs.~\cite{\HSTH,\HSBS,\HSPT,\SPT}.

The derivation of HS1 and HS2 uses TOPT to identify the loop integrals that lead to power-law
finite-volume effects, but the main analysis is based on a skeleton expansion using Feynman diagrams.
The strategy is, crudely speaking, to convert loop sums into integrals plus a volume-dependent remainder
at every opportunity. This leads to complicated intermediate expressions
involving kernels that are not symmetric under interchanges of external momenta, and are
given by implicit, constructive definitions. Considerable effort is then required to rewrite the final
quantization condition in terms of a symmetric three-particle K matrix (called $\Kdf$).
Our alternative approach uses TOPT throughout,\footnote{%
Aside from an initial use of Feynman diagrams to analyze self-energy diagrams, as
described in Appendix~\ref{app:TOPT}.
We note that extension of the derivation of HS1 to theories without the $\Z_2$ symmetry,
given in Ref.~\cite{\BHSQC}, makes more extensive use of TOPT, and we use several results concerning TOPT 
from that work.}
converts fewer sums into integrals, and does not
aim for full symmetrization in the initial quantization condition. This allows for a simpler derivation
that is completely explicit. A second step then leads to the HS1 form of
the quantization condition in terms of $\Kdf$.
We stress that, despite the differences, many technical steps in our approach are based closely
on the developments and technical results of HS1 and HS2.

We close the introduction with a summary of our approach, which also serves to describe
the organization of the paper.
The new derivation of the quantization condition is presented in Sec.~\ref{sec:derivation}, 
and is broken into several steps.
We begin in Sec.~\ref{sec:TOPT} with a recap of the essentials of TOPT, 
and then, in Sec.~\ref{sec:skeleton},
explain how the  three-particle correlation function can be written in terms of 
two- and three-particle irreducible TOPT amplitudes.
This is the analog in TOPT of the Feynman-diagram-based skeleton expansion used in HS1.
The advantage of the TOPT approach is that the result, given in Eq.~\Eqref{eq:CL_TOPT},
is a simple geometric series that is straightforward to derive.
The next step, described in Sec.~\ref{sec:onshell}, is to rewrite the expression for the correlator
in terms of on-shell quantities. 
This is achieved by the results in Eqs.~\Eqref{eq:Gsplit} and \Eqref{eq:Fsplit}, which 
introduce two finite-volume quantities $\wt G$ and $\wt F$ 
that are closely related to the $G$ and $F$ appearing in HS1.
Using these results, the form for the correlator can be reorganized into a geometric series
involving on-shell, infinite-volume kernels, Eq.~(\ref{eq:CL_final}).
As described in Sec.~\ref{sec:QC}, this immediately leads to our new form of the quantization condition,
Eq.~\Eqref{eq:QC_Ft}. While simple in form,
this has the disadvantage of involving an asymmetric three-particle K matrix ($\Kdfuu$).
The next sections of this work show how the quantization condition
can be rewritten in terms of a symmetrized three-particle K matrix using relatively simple algebraic steps,
with the final result, presented in Eq.~\Eqref{eq:QCnn}, 
having exactly the same form as that of HS1.
A spin-off from this analysis is that we obtain an explicit expression for the symmetric
three-particle K matrix of HS1, $\Kdf$, in terms of TOPT amplitudes connected by a sequence
of integral operators.

Although these final symmetrization steps are straightforward, we take a somewhat indirect
path to obtain them. This involves first, in Sec.~\ref{sec:M3L}, using the TOPT methodology
to determine an expression for an asymmetric finite-volume three-particle scattering amplitude, $\wt \cM_{3,L}^\uu$,
in terms of $\Kdfuu$, Eq.~(\ref{eq:M3Lb});
second, in Sec.~\ref{sec:M3LuuHS}, comparing that to the result for the similar amplitude
$\cM_{3,L}^\uu$ introduced in HS2;
third, in Sec.~\ref{sec:KtoK},
 asymmetrizing the HS1 result so that it is written in terms of an asymmetric amplitude
$\KdfuuHS$, and using this to show that the HS1 quantization condition can be recast in
exactly the same form as our new version;
and, finally, in Sec.~\ref{sec:symmQC}, reversing the algebraic steps to show that our quantization
condition can be rewritten in symmetric HS1 form.
We close the paper with a summary and outlook.

We include several appendices collecting technical results.
Appendix~\ref{app:TOPT} concerns TOPT.
Appendix~\ref{app:K2} describes the relation of three- and two-particle finite-volume amplitudes,
and gives details on the pole prescription that we use.
Appendix~\ref{app:algebra} explains a set of complicated though straightforward matrix manipulations
that are needed in the main text.
Appendix~\ref{app:asym} derives the asymmetrization identities needed in the main text.
Appendix~\ref{app:inteqs} derives the relation between $\Kdfuu$ and $\cM_3$.
Appendix~\ref{app:KtoK} gives details of the steps needed to relate $\Kdfuu$ and
$\KdfuuHS$.
We also include Appendix~\ref{app:opt1} in which we briefly describe a variant of our approach
that leads to a slightly different form of the quantization condition, and which may be advantageous
when considering generalizations.

In a companion paper~\cite{BS2}, we show that our new form of the quantization condition
can be written in terms of the R matrix of Refs.~\cite{\Maiisobar,\isobar}.
The provides a generalization of the FVU quantization condition to all two-particle partial waves.

\section{Derivation of the new form of the quantization condition}

\label{sec:derivation}



We work in a generic relativistic effective field theory (RFT) of identical scalar particles with physical mass $m$
in 3+1-dimensional Minkowski spacetime. 
We assume the Lagrangian has a G-parity-like $\Z_2$ symmetry so that only even-legged vertices are allowed.
This is exactly the same setup as in HS1. 
Our aim is to derive an expression---the quantization condition---that 
determines the energy levels when this theory is considered in a finite spatial box.
Following HS1, we choose a cubic volume of side length $L$, with periodic boundary conditions.

The tool we use is the finite-volume (FV) correlation function 
for fixed total four-momentum $P^\mu=(E,\vec P)$:
\begin{align}
	C_{3,L}(E,\vec{P}) \equiv \int_L d^4x \; e^{i(Ex^0 - \vec{P}\cdot\vec{x})} \braket{0|\text{T}\sigma(x)\sigma^\dagger(0)|0} \,,
\end{align}
where $\vec{P}=\frac{2\pi}{L}\vec{n}_P$ with $\vec{n}_P\in\Z^3$, the integral is over $x^0\in\mathbb{R}$ and $\vec{x}\in[0,L)^3$,
$\ket{0}$ is the true vacuum state of the interacting theory, and $\sigma(x)$ is an 
interpolating field coupling to three-particle states.
Throughout this paper, we assume the kinematic constraint $m<E^*<5m$, where $E^*=\sqrt{E^2-\vec{P}^2}$ is the total energy in the overall center-of-mass (CM) frame.
This constraint ensures that only three-particle states may go entirely on shell, a restriction that is crucial to our derivation (as well as that of HS1).

For fixed $\vec P$, the correlator $C_{3,L}(E,\vec P)$ will have poles in $E$ at the energies of the 
FV states that have the quantum numbers of $\sigma^\dagger$, namely all states with odd particle number.
By deriving an expression for the pole positions, we will obtain the desired FV quantization condition.

The derivation in HS1 involves summing over all Feynman diagrams contributing to $C_{3,L}(E,\vec{P})$.
The expressions for each diagram differ from those in infinite volume only  by the replacement of spatial
loop integrals with sums over discrete spatial momenta, $\vec{k}\in \frac{2\pi}{L}\Z^3$.
One of the key initial steps in the derivation is to note that some spatial-momentum sums have negligible 
dependence on $L$ and can therefore be replaced with (infinite-volume) integrals.
More precisely, if a summand/integrand $f(\vec{k})$ is smooth over the integration domain,
with the derivatives scaling as appropriate powers of $m$,
then, from the Poisson summation formula, the sum-integral difference is exponentially suppressed in $mL$:%
\footnote{Or, more precisely, falls faster than any power of $mL$.}
\begin{align}
	\frac{1}{L^3}\sum_{\vec{k}}^\UV f(\vec{k}) = \int^\UV\frac{d^3k}{(2\pi)^3} f(\vec{k}) + \mc{O}(e^{-mL}) \label{eq:Poisson} \,.
\end{align}
Here we have included an ultraviolet (UV) cutoff, the nature of which is unimportant as the sum-integral difference is an
infrared effect.
We assume throughout that $mL$ is large enough that $\mc{O}(e^{-mL})$ terms can be safely neglected.
An immediate consequence of this assumption is that the difference between a FV quantity and its infinite-volume analog is only non-negligible if the quantity contains a singularity in the spatial-momentum summand.

The next key step in HS1 is to utilize certain results from TOPT as a tool for identifying which Feynman diagrams can possess singularities and which cannot, as this greatly restricts the classes of Feynman diagrams that need to be considered to capture all significant FV contributions.
However, this is the full extent to which TOPT is applied in HS1; actual time-ordered diagrams are never used.

\subsection{TOPT basics}
\label{sec:TOPT}

We begin the derivation proper by a brief recapitulation of the essential features of TOPT.
A good source for the derivation of these results is Ref.~\cite{Sterman:1994ce};
further discussion in a context closely  related to that considered here is given in Appendix B of Ref.~\cite{\BHSQC}.
The main subtlety in applying TOPT concerns the use of renormalized propagators,
and we discuss this technical point in Appendix~\ref{app:TOPT}.

In TOPT, one rewrites each Feynman diagram contributing to $C_{3,L}$ as a sum of all time orderings of the vertices, with each ordering corresponding to a unique time-ordered diagram.
Each propagator in the diagram is associated with an on-shell four-momentum
$p^\mu= (\omega_p, \vec p)$, with
positive energy
$\omega_p\equiv\sqrt{\vec p^2+m^2}$,
and gives rise to a factor of $1/(2\omega_p)$.
Spatial momentum is conserved at vertices, but energy is not.
Each ``cut'' between consecutive vertices gives a kinematic factor 
\begin{align}
	iK_t \equiv \frac{i}{E_t-\sum\limits_{p_\on\in \mc{P}_t} \omega_{p} + i\epsilon}  \,,
	\label{eq:cut}
\end{align}
where $\mc{P}_t$ is the set of spatial momenta passing through the cut at time $t$, and
\begin{align}
	E_t \equiv
		\begin{cases}
			+E & \text{if }~  t_{\sigma} > t > t_{\sigma^\dagger} \\
			-E & \text{if }~  t_{\sigma^\dagger} > t > t_\sigma  \\
			0 & \text{otherwise},
		\end{cases}
\end{align}
with $t_{\sigma^\dagger}$ and $t_{\sigma}$ denoting the times at which $\sigma^\dagger$ and $\sigma$ occur in the diagram, respectively.\footnote{%
The precise values of the times are irrelevant to the value of the diagram; they are being used here only to label
the ordering of vertices.}
The factor of $i\epsilon$ has no impact in finite volume, and can be set to zero in that case.
When evaluating vertices with derivatives (which are present with arbitrary order in the generic RFT),
the corresponding momenta are placed on shell.
Symmetry factors are included as for Feynman diagrams.
Finally, all spatial momenta are summed/integrated with the standard measure.

\begin{figure}[tbh]
\begin{center}
\vspace{-10pt}
\includegraphics[width=\textwidth]{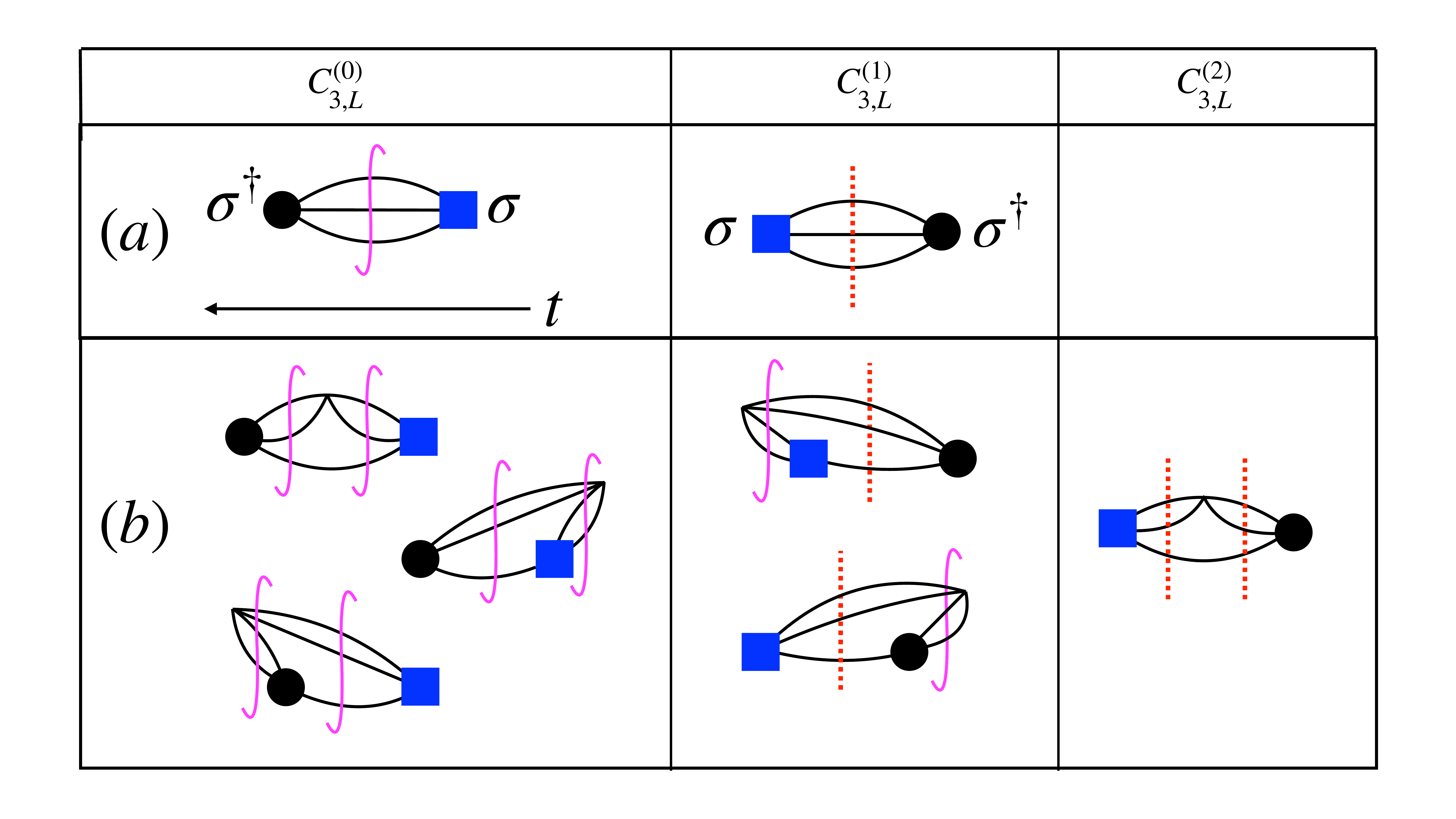}
\caption{Examples of time orderings in diagrams contributing to $C_{3,L}$. Time flows from right to left,
with the black circle (blue square) representing $\sigma^\dagger$ ($\sigma$).
Relevant cuts are shown by vertical (red)
dashed lines, while irrelevant cuts are shown by solid (magenta) integral signs.
The factors associated with these cuts are described in the text.
Vertical columns divide contributions according to the number of relevant cuts. 
Horizontal rows contain the time orderings of
 (a) the leading-order Feynman diagram and (b) a Feynman 
 diagram with a single four-point vertex. \label{fig:topt1}}
\end{center}
\end{figure}

As discussed in Appendix~\ref{app:TOPT}, by using an appropriate on-shell renormalization scheme
and restricting to our kinematic regime in which only three particles can go on shell simultaneously, 
all self-energy diagrams can be absorbed into changes in the vertices, and it is the physical mass that
enters into the factors of $\omega_p$.

Given these rules, the only singularities in diagrams are due to the kinematic factors $K_t$; by assumption,
the vertices are polynomials in momenta and thus nonsingular.
Furthermore, within our range of $E^*$, the only singularities occur if $E_t=+E$
and $|\mc{P}_t|=3$, i.e., if the cut at time $t$ contains three lines
and comes after $t_{\sigma^\dagger}$ and before $t_{\sigma}$.
We shall refer to such cuts as ``relevant three-particle cuts" or simply ``relevant cuts."\footnote{%
A potential second type of relevant cut, arising from a single dressed propagator, is discussed in
Appendix~\ref{app:TOPT}, and shown to be absent.}
The remaining cuts, which we refer to as  ``irrelevant,"  occur if either (i) $E_t=+E$ with $|\mc{P}_t|\ne3$,
or (ii) $E_t\in\{-E,0\}$.
Since $|\mc{P}_t|$ is necessarily odd for all cuts through $C_{3,L}$, case (i)
corresponds to having either a single particle in the cut, leading to a singularity at $E^*=m$, 
or at least five particles, for which singularities occur for $E^*\ge 5 m$. 
Both of these possibilities lie outside our kinematic range.
For case (ii), there are no singularities because the denominator of $K_t$ is negative definite.
Examples of relevant and irrelevant cuts are shown for simple diagrams in Fig.~\ref{fig:topt1}.

Given this classification, irrelevant cuts yield a smooth $K_t$, and
we can use Eq.~\Eqref{eq:Poisson} to take the infinite-volume limits of all 
spatial-momentum sums involving the cut, i.e. to replace the sums with integrals.
This was the key result from TOPT that HS1 used to identify which Feynman diagrams contained singularities,
but here we continue working with time-ordered diagrams.

Later in this work (in Sec.~\ref{sec:M3L}) 
we will consider scattering amplitudes, i.e. amputated correlation functions, both
in finite and infinite volume.
The corresponding Feynman diagrams can also be broken up into time-ordered components using TOPT.
The rules are the same as above, except that the operators creating incoming (destroying outgoing) 
particles are always placed at the earliest (latest) time. In addition, there are no $1/(2\omega)$ factors for
the external propagators; the first cut occurs after the first vertex, and the last cut before the final vertex.
In general such an amplitude is off shell. The on-shell amplitude, whose absolute square is related to
the scattering cross section, is obtained by choosing the initial and final spatial momenta such that
the initial- and final-state energies both sum to $E$. In this on-shell limit, the result for the amplitude is identical to that
given by the expression obtained using Feynman diagrams, and, in particular, is Lorentz invariant.

\subsection{Expansion of $C_{3,L}(E,\vec{P})$ in relevant cuts}
\label{sec:skeleton}

Our strategy is to organize the (renormalized) time-ordered diagrams that contribute to $C_{3,L}(E,\vec{P})$ by the number of relevant three-particle cuts they contain,
\begin{align}
	C_{3,L}(E,\vec{P}) &= \sum_{n=0}^\iy C_{3,L}^{(n)}(E,\vec{P}) \,,
\end{align}
where $C_{3,L}^{(n)}(E,\vec{P})$ is the sum of all diagrams containing exactly $n$ relevant cuts.
Examples of this organization are shown in Fig.~\ref{fig:topt1}.

The $n=0$ term $C_{3,L}^{(0)}(E,\vec{P})$ denotes the sum of all diagrams with no such cuts.
Since all cuts in each diagram give smooth $iK_t$ contributions, we can use Eq.~\Eqref{eq:Poisson} to replace all discrete momentum sums with integrals and take the infinite-volume limit:
\begin{align}
	C_{3,L}^{(0)}(E,\vec{P})=C_{3,\iy}^{(0)}(E,\vec{P})+\mc{O}(e^{-mL}) \,.
\end{align}
From now on we will no longer track terms that are exponentially suppressed in $mL$.

\begin{figure}[tbh]
\begin{center}
\vspace{-10pt}
\includegraphics[width=\textwidth]{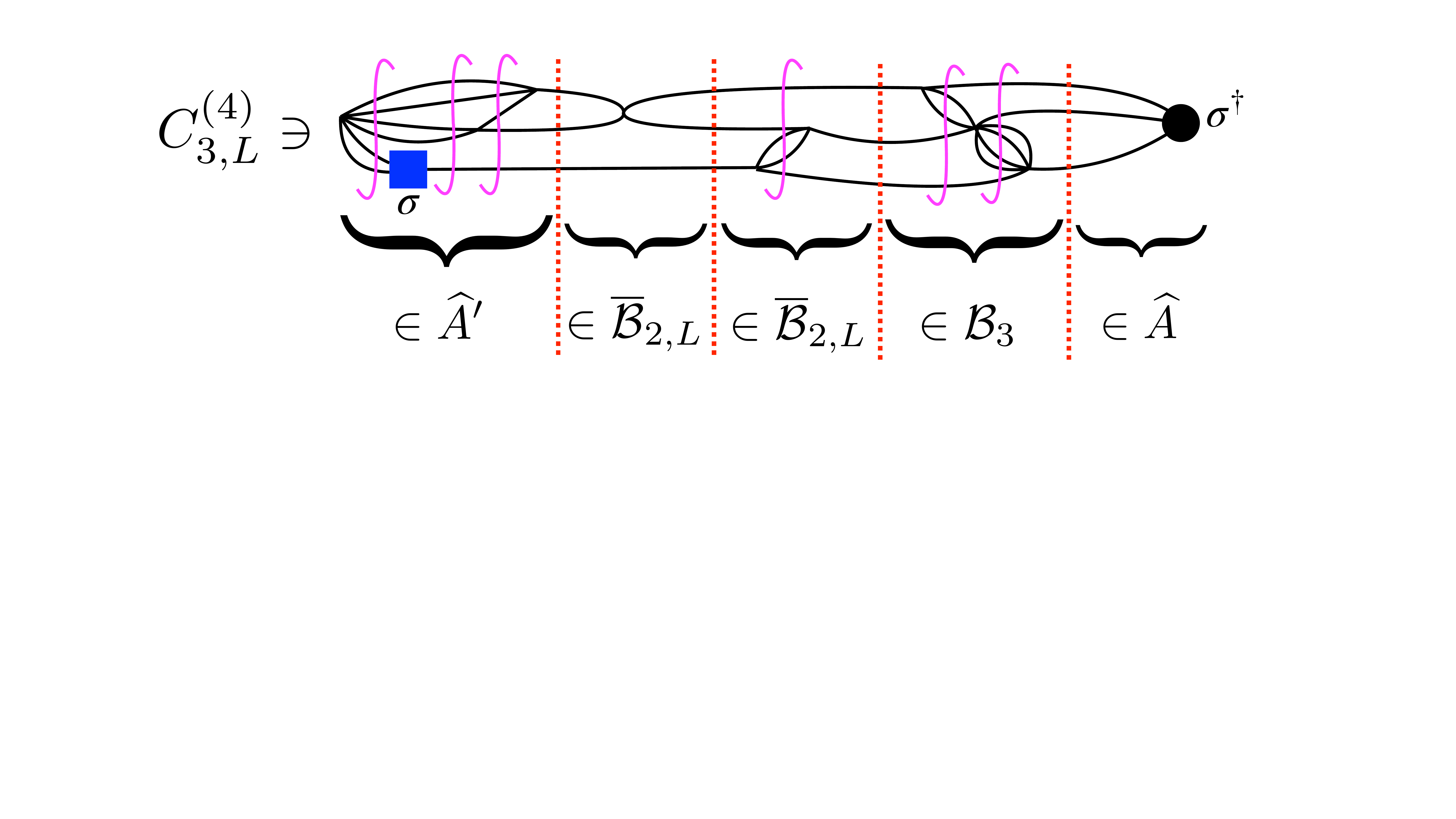}
\vspace{-2.3truein}
\caption{Example of a contribution to $C_{3,L}^{(4)}$. Notation is as in Fig.~\ref{fig:topt1}.
The names for the different types of segment are indicated by the underbraces.
\label{fig:CL_example}
}
\end{center}
\end{figure}

For diagrams with at least one relevant cut, the expressions factorize into a form with a right-hand ``endcap",
followed by some (possibly zero) number of $3\to3$ segments, followed by a left-hand endcap.
These pieces are separated by relevant cuts.
This factorization can be seen in the examples 
in the $C^{(1)}_{3,L}$ and $C^{(2)}_{3,L}$ columns of Fig.~\ref{fig:topt1}.
A more extensive example for $C_{3,L}^{(4)}$ is shown in Fig.~\ref{fig:CL_example}, which shows that
there are two types of nontrivial $3\to3$ segments: one in which two particles interact with the third particle spectating,
and the other in which all three particles interact.
We label these $\overline{\cB}_{2,L}$ and $\cB_3$ respectively, while the left (right) endcaps are denoted
$\widehat{A}'$ ($\widehat{A}$).\footnote{%
The ``hat" on these quantities is used to distinguish them from similar, but different, endcaps
denoted $A'$ and $A$ in HS1.  Similarly, we label the intermediate segments with a calligraphic $\cB$, 
in order to distinguish them from the Bethe-Salpeter kernels $B_2$ and $B_3$ used in HS1.}
Before presenting the general expression for $C_{3,L}$, we first give the definitions of these four segments.

\subsubsection{Contributing segments}

\begin{figure}[tbh]
\begin{center}
\vspace{-10pt}
\includegraphics[width=\textwidth]{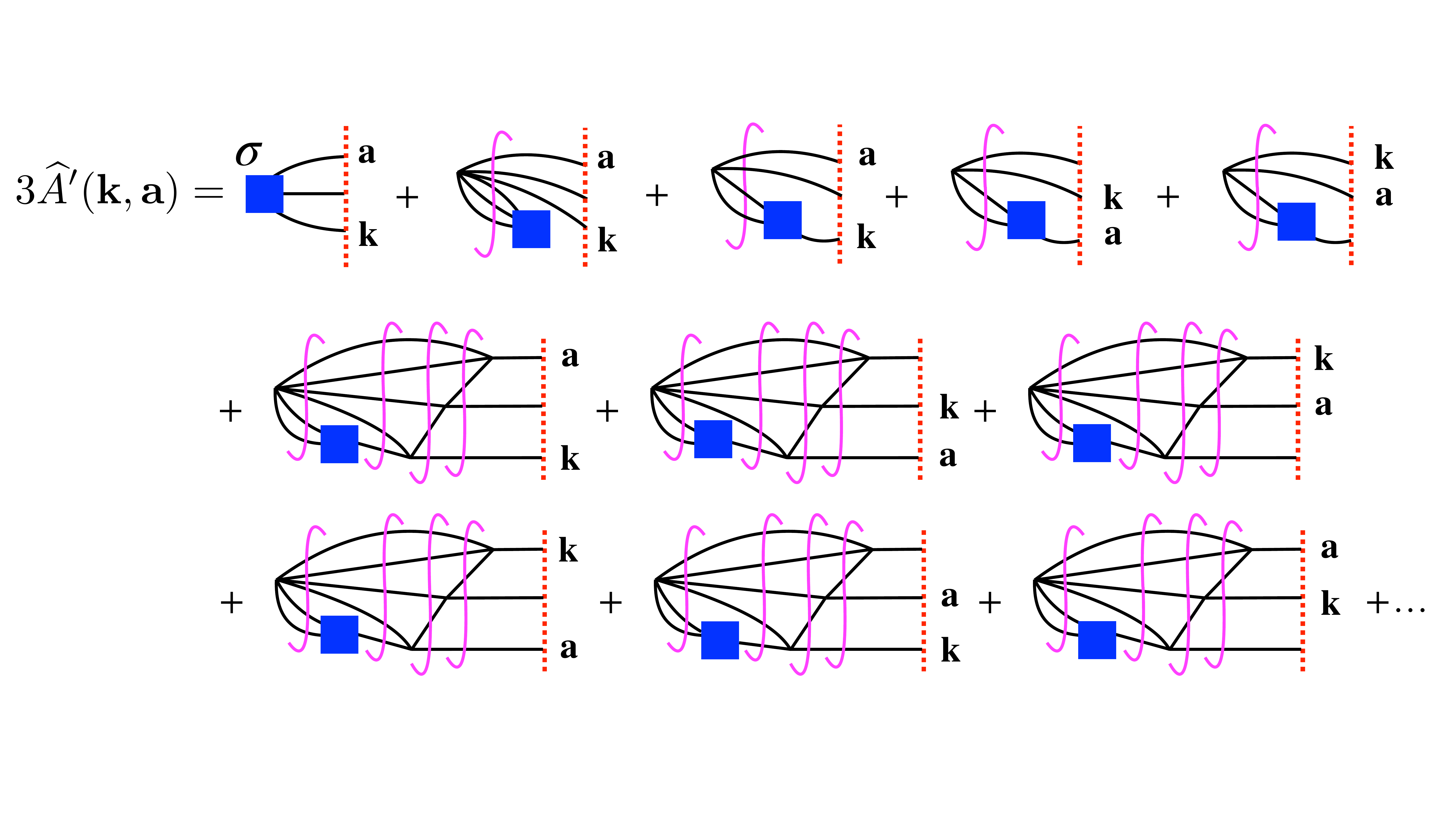}
\vspace{-0.7truein}
\caption{Examples of contributions to the left endcap $\widehat A'(\vec k,\vec a)$,
with notation as in Fig.~\ref{fig:topt1}.
When evaluating these diagrams, the external lines (those that end at the relevant cut) are amputated,
with the factors of $1/(2\omega)$ dropped, and the energy denominator is also not included.
The factor of $3$ on the left-hand side is discussed in the text. 
The vertices in the diagram can represent interactions with or without derivatives, but in all cases are fully
symmetric.
The number of relabelings of a
diagram depends on its intrinsic symmetry under interchange of the external particles. 
For the first two diagrams, which are symmetric under external particle interchange, there is only one labeling.
For the next diagram, which is symmetric under interchange of the upper two particles, there are three labelings,
as shown.
The final diagram, shown on the second and third lines,  is completely asymmetric, and all six relabelings must be included.
\label{fig:endcap}
}
\end{center}
\end{figure}

We begin with the left endcap $\widehat{A}'$. It  is given by the sum of
all time-ordered diagrams that contain $\sigma$, are three-particle irreducible in the s-channel (3PIs),\footnote{%
Our definition of 3PIs includes, in principle, the possibility of a cut through a single propagator that is carrying
the full momentum $\vec P$.
However, as explained in Appendix~\ref{app:TOPT}, in our kinematic regime all such single
propagators can be collapsed into vertices, so the issue does not arise.}
and begin with an amputated three-particle cut.
The amputation removes the factors of $1/(2\omega)$ from each line, as well as the energy denominator of
the relevant cut. These factors will be added back when we join the segments.
For fixed $E$ and $\vec P$, $\widehat{A}'$ depends on the spatial momenta of two of the particles in the
relevant cut, usually denoted $\vec k$ and $\vec a$. The momentum of the third, usually denoted
$\vec{b}_{ka}$, or simply $\vec b$ for short,
is given by $\vec{b}_{ka} = \vec{P}-\vec{k}-\vec{a}$.
We define $\widehat{A}'(\vec k, \vec a)$ to include all distinct attachments of momentum labels to the
external lines.
This is exactly what would result were we to define this as an amplitude with three single-particle creation operators
and included all Wick contractions.
This implies that it is fully symmetric\footnote{%
A reader making a detailed comparison with HS1 will observe that the endcaps used initially in that work are
asymmetric, and considerable effort is needed at a later stage to symmetrize them. One of the advantages
of the TOPT approach is that we do not need to use asymmetric quantities at this stage, with the exception
of $\overline{\cB}_{2,L}$.}
\begin{equation}
\widehat{A}'(\vec k, \vec a) = \widehat{A}'(\vec a,\vec b) =\widehat{A}'(\vec b, \vec k) 
= \widehat{A}'(\vec a,\vec k) = \widehat{A}'(\vec b,\vec a) =\widehat{A}'(\vec k, \vec b) \,.
\label{eq:fullysymmetric}
\end{equation}
It turns out to be convenient to multiply the sum of all diagrams by a factor of $1/3$, 
as this will cancel a labeling degeneracy that we describe below.
This factor can be intuitively understood as placing this inherently symmetric object on the same
footing as the inherently asymmetric segment $\overline{\cB}_{2,L}$ to be defined below.
Examples of diagrams contributing to this endcap are shown in Fig.~\ref{fig:endcap}.

The right endcap $\widehat{A}$ is defined as the sum of all amputated 
3PIs TOPT diagrams containing a $\sigma^\dagger$ and ending with an amputated three-particle cut, multiplied by $1/3$.
Diagrammatically, it is simply
the horizontal ``reflection'' of $\widehat{A}'$. It is fully symmetric.

The fully connected $3\to3$ segment $i \cB_3(\vec k',\vec a';\vec k, \vec a)$ 
(with fixed $E$ and $\vec P$ implicit) 
is defined  as the sum of all amputated, connected, 3PIs TOPT
diagrams beginning and ending at a relevant cut.\footnote{%
The factor of $i$ is included to match the standard definition of a scattering amplitude, and follows the
notation of HS1.}
All momentum assignments are included, and it is multiplied
by $1/3$ for each cut, i.e.~by $1/9$ in total. It is fully symmetric separately for both initial and final momenta,
\begin{equation}
\cB_3(\vec k',\vec a';\vec k, \vec a) = \cB_3(\vec a',\vec k';\vec k, \vec a) =
\cB_3(\vec k',\vec a';\vec a, \vec b) = \cdots\,.
\label{eq:C3symmetry}
\end{equation}

\begin{figure}[tb]
\begin{center}
\vspace{-10pt}
\includegraphics[width=\textwidth]{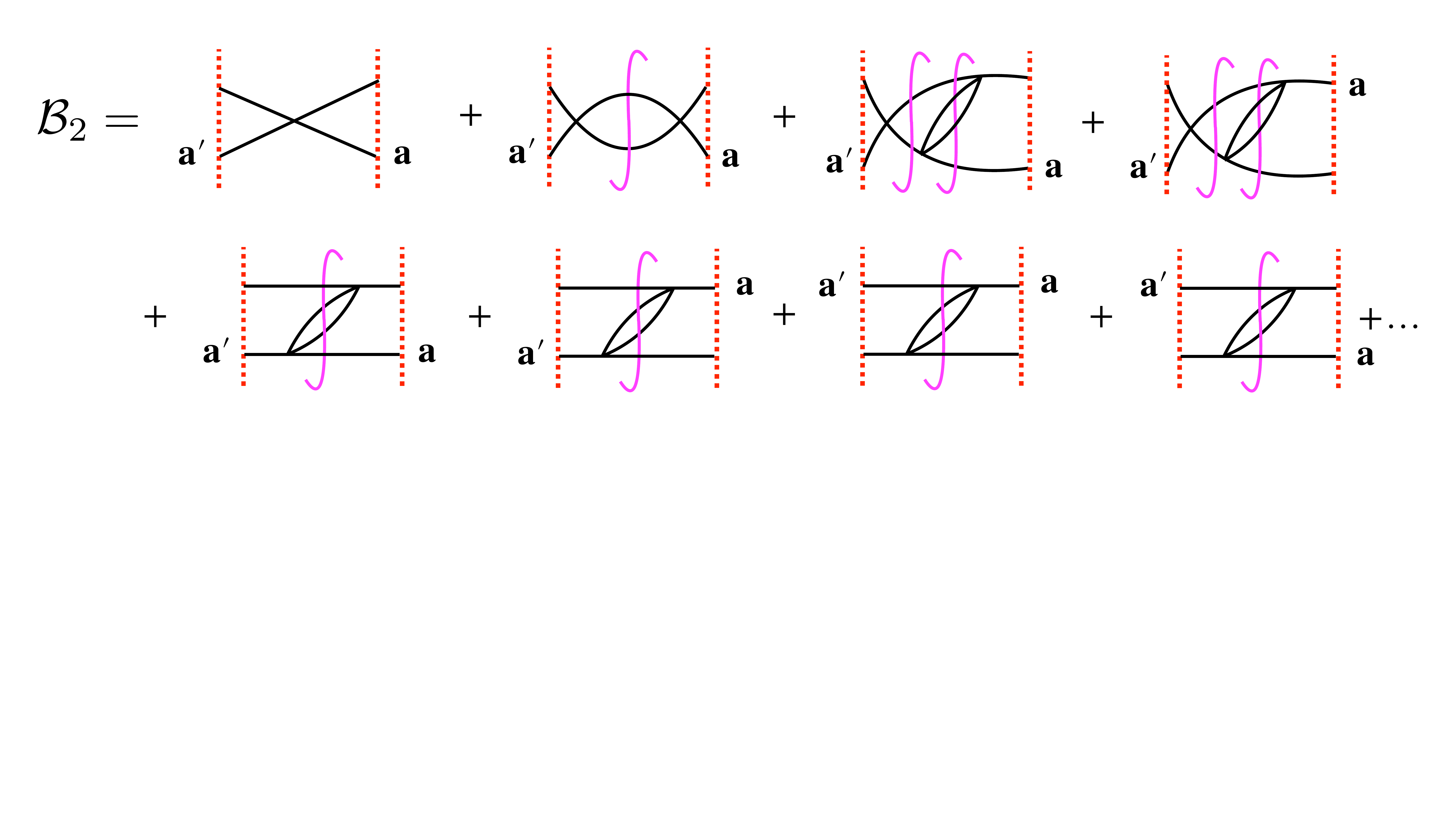}
\vspace{-2.1truein}
\caption{Examples of contributions to $i \cB_2(E_2,\vec P_2;\vec a';\vec a)$,
with notation as in Figs.~\ref{fig:topt1} and \ref{fig:endcap}.
(In this and following figures we do not keep track of factors of $i$.)
The number of relabelings of a
diagram depends on its intrinsic symmetry under interchange of the external particles. 
For the first two diagrams, which are symmetric under external particle interchange, there is only one labeling.
The next diagram is symmetric on the left, but not on the right, and so there are two relabelings, as shown.
The final diagram is asymmetric under interchange of both initial and final particles, and thus there are
four relabelings, as shown.
\label{fig:C2}
}
\end{center}
\end{figure}

The final segment is $i\overline{\cB}_{2,L}$, 
in which only two of the particles are connected---the ``interacting pair''---while the third spectates.
This segment is intrinsically asymmetric, and we choose the spectator momentum to be
$\vec k=\vec k'$,
\begin{equation}
i\overline{\cB}_{2,L}(E,\vec P;\vec{k}',\vec{a}';\vec k,\vec a) \equiv \delta_{k'k} 2\omega_k L^3 
i\cB_2(E_{2,k},\vec P_{2,k};\vec a'; \vec a)\,,\qquad
E_{2,k}\equiv E-\omega_k\,,\ \ \vec P_{2,k}\equiv \vec P-\vec k\,,
\label{eq:C2bar}
\end{equation}
where $\delta_{k' k} = \delta^3_{\vec k',\vec k}$ is the three-dimensional Kronecker delta.
Here $\cB_2$ is the sum over all amputated TOPT diagrams describing connected $2\to2$ scattering
that are 2PI in the s channel, which we denote as 2PIs diagrams. 
Note that the four-momentum flowing through $\cB_2$ is 
$(E_{2,k},\vec P_{2,k})$, since the spectator four-momentum $k^\mu=(\omega_k,\vec k)$ is subtracted from the total.
We also include an $L$ in the subscript to emphasize the 
presence of an explicit factor of $L^3$.
Examples of diagrams contributing to $\cB_2$ are shown in Fig.~\ref{fig:C2}.
It is symmetric under separate interchange of the momenta within initial and final pairs, i.e.
\begin{equation}
\cB_2(E_2,\vec P_2;\vec a';\vec a) = \cB_2(E_2,\vec P_2;\vec b';\vec a) =
\cB_2(E_2,\vec P_2;\vec a';\vec b) = \cB_2(E_2,\vec P_2;\vec b';\vec b)\,,
\qquad
\vec b'= \vec P_2-\vec a'\,,\ \ \vec b= \vec P_2-\vec a\,.
\end{equation}
It is defined without any overall factors (unlike $\cB_3$).

The factor of $2\omega_k L^3$ in the numerator of Eq.~(\ref{eq:C2bar}) is 
needed because the cuts on both sides of $\cB_2$ include a spectator propagator,
so one must be canceled. This is explained in greater detail below.

For the quantities $\widehat{A}'$, $\widehat{A}$, $\cB_3$, and ${\cB}_2$, we can
proceed as for $C_{3,L}^{(0)}$ and take the infinite-volume limit, since they contain no relevant cuts in
our kinematic region.
The lack of such cuts is by construction for the 3PIs quantities $\widehat{A}'$, $\widehat{A}$, and $\cB_3$.
For $\cB_2$, the lack of relevant three-particle cuts in $\overline{\cB}_{2,L}$ implies  that $\cB_2$ can have no
on-shell two particle cuts, since its CM energy $E_2^*$ cannot exceed $4m$.
Because of the absence of relevant cuts,
$\widehat{A}'$, $\widehat{A}$, $\cB_3$, and ${\cB}_2$ are all real in our kinematic range.
As discussed in Appendix~\ref{app:TOPT}, we implicitly use a diagram by diagram regularization and renormalization
scheme, so that all quantities are UV finite.

\subsubsection{Evaluating $C_{3,L}(E,\vec{P})$}

With the 3PIs segments in hand, we can now proceed toward a general expression for the correlator.
To write this compactly, we introduce a matrix notation, in which the indices are the two summed
finite-volume momenta at each cut, $\{ka\}\equiv \{\vec k, \vec a\}$.
Thus $\widehat{A}'$ becomes a row vector, $\cB_3$ and $\overline{\cB}_{2,L}$ become matrices,
and $\widehat{A}$ becomes a column vector:
\begin{gather}
\begin{aligned}
\widehat{A}'_{k'a'} &\equiv \widehat{A}'(\vec k',\vec a') \,, \quad
&
\left[\cB_3\right]_{k'a'; ka}&\equiv \cB_3(\vec k',\vec a';\vec k,\vec a)\,,
\\
\widehat{A}_{ka} &\equiv \widehat{A}(\vec k,\vec a)
\,, \quad
&
\left[\overline{\cB}_{2,L}\right]_{k'a'; ka} &\equiv \overline{\cB}_{2,L}(E,\vec P; \vec k',\vec a';\vec k,\vec a)
\,.
\label{eq:C2L}
\end{aligned}
\end{gather}
The sum of contributions containing exactly one relevant cut can then be written
\begin{align}
C_{3,L}^{(1)}(E,\vec P) 
&= \sum_{\vec{k}',\vec{a}'}^\UV \sum_{\vec{k},\vec{a}}^\UV 
\widehat{A}'_{k'a'} 3 i [D_F]_{k'a';ka} \widehat{A}_{ka}\,,
\\
&= \widehat{A}' \, 3 i D_F \widehat{A}\,, \label{eq:CL1}
\end{align}
where in the second line we have left matrix indices implicit.
The matrix associated with the relevant cut is 
\begin{align}
[D_F]_{k'a';ka} &\equiv \frac1{2!} \delta_{k'k}\delta_{a'a} D_{ka}\,,
\\
D_{ka} &\equiv \frac{1}{2\omega_k L^3} \frac{1}{2\omega_{b}(E-\omega_k-\omega_a-\omega_{b})} 
	\frac{1}{2\omega_a L^3}
	= D_{ak} 
	\,.
\end{align}
The reason for the subscript $F$ will become clear below.

The various terms in Eq.~(\ref{eq:CL1}) are chosen so that the correct TOPT expression 
for $C_{3,L}^{(1)}$ is obtained when sewing $\widehat{A}'$ and $\widehat{A}$ together.
In particular, $D_{ka}$ plays three roles. 
First, it puts back in the propagator factors that have been removed 
when amputating $\widehat{A}'$ and $\widehat{A}$.
Second, it contains the two factors of $1/L^3$ that are the standard
measure associated with the sums over the loop momenta $\vec k$ and $\vec a$.
Third, it includes the energy denominator from the kinematic factor $K_t$. 

The numerical factors of $3$ [multiplying $D_F$ in Eq.~(\ref{eq:CL1})]
and $1/2!$ (in the definition of $D_F$),
are needed to account for symmetry factors and labeling degeneracy.
Recalling that $\widehat{A}'$ and $\widehat{A}$ each come with a factor of $1/3$ included by hand,
the product of these factors is $N_S=(1/3)\times (3/2!) \times (1/3) = 1/3!$.
If one is considering contributions to $\widehat{A}'$ and $\widehat{A}$ that both
correspond to completely asymmetric underlying diagrams with $3!$ possible labelings
(as in the example shown in the last two lines of Fig.~\ref{fig:endcap}),
then the $(3!)^2$ terms in the product overcount the number of diagrams contributing to $C_{3,L}^{(1)}$ by
a factor of $3!$. This is canceled by $N_S$.
If instead the contributions to both $\widehat{A}'$ and $\widehat{A}$ are completely symmetric
(as in the first two examples in Fig.~\ref{fig:endcap}), then there is only one diagram that appears in the
product, but it needs a symmetry factor of $1/3!$ that is provided $N_S$.
Cases of intermediate symmetry work similarly.

For reasons that will become clear shortly, it is useful to rewrite Eq.~(\ref{eq:CL1}) as
\begin{align}
C_{3,L}^{(1)}
&= \widehat{A}' \,  i (D_F + D_G) \widehat{A}\,, 
\label{eq:CL1FG}
\\
[D_G]_{k'a';ka}  &\equiv \delta_{k'a}\delta_{ka'} D_{ka}\,.
\end{align}
The new matrix $D_G$ differs from $D_F$ in the manner in which the momenta are contracted, 
and also because it lacks the factor of $1/2!$. When adjacent to a symmetric quantity (on either side) one can replace
$D_G$ with $2 D_F$, because the corresponding momentum indices $k$ and $a$ can be freely interchanged,
and because $D_{ka}=D_{ak}$.
Given this substitution, the equivalence of Eqs.~(\ref{eq:CL1}) and (\ref{eq:CL1FG}) is immediate.

We now move on to the contributions with two relevant cuts.
These can involve either a $\cB_3$ or a $\overline{\cB}_2$ between the endcaps.
For the former, using exactly the same arguments as just described, one finds that
\begin{equation}
C_{3,L}^{(2)} \ni 
\widehat{A}' \,  3 i D_F  i\cB_3 \,3 iD_F \widehat{A}
=
\widehat{A}' \,  i (D_F + D_G) i\cB_3 i(D_F+D_G) \widehat{A}\,. 
\label{eq:CL2C3}
\end{equation}
For the latter, it turns out that the same form holds
\begin{equation}
C_{3,L}^{(2)} \ni 
\widehat{A}' \,  3 i D_F  i\overline{\cB}_{2,L} 3 iD_F \widehat{A}
=
\widehat{A}' \,  i (D_F + D_G) i\overline{\cB}_{2,L} i(D_F+D_G) \widehat{A}\,,
\label{eq:CL2C2}
\end{equation}
where the second result follows from the first because both $D_F$'s are adjacent to a symmetric endcap.
To understand why Eq.~(\ref{eq:CL2C2}) gives the correct contribution to $C_{3,L}^{(2)}$,
consider the diagram shown in the final column of Fig.~\ref{fig:topt1}.
The $2\omega_k L^3$ in $\overline{\cB}_{2,L}$, Eq.~(\ref{eq:C2bar}), cancels that in one of the $D_F$'s, so
that there is only one such factor in the overall diagram, as appropriate for the spectator line.
Using the first form in Eq.~(\ref{eq:CL2C2}),
the numerical factors combine to give $N'_S=(1/3) \times (3/2!) \times (3/2!) \times (1/3)= 1/(2!)^2$,
which is the correct symmetry factor for the diagram as a whole.
For completely asymmetric contributions to $\widehat{A}'$, $\overline{\cB}_{2,L}$, and $\widehat{A}$,
$N'_S$ serves to cancel the labeling degeneracy.
For cases with intermediate symmetry, $N'_S$ provides a mix of the needed symmetry factors and cancellation
of labeling degeneracies.
The total result with two relevant cuts can thus be written
\begin{equation}
C_{3,L}^{(2)} =
\widehat{A}' \,  i (D_F + D_G) i(\overline{\cB}_{2,L}+\cB_3) i(D_F+D_G) \widehat{A}\,.
\end{equation}

\begin{figure}[tb]
\begin{center}
\vspace{-10pt}
\includegraphics[width=\textwidth]{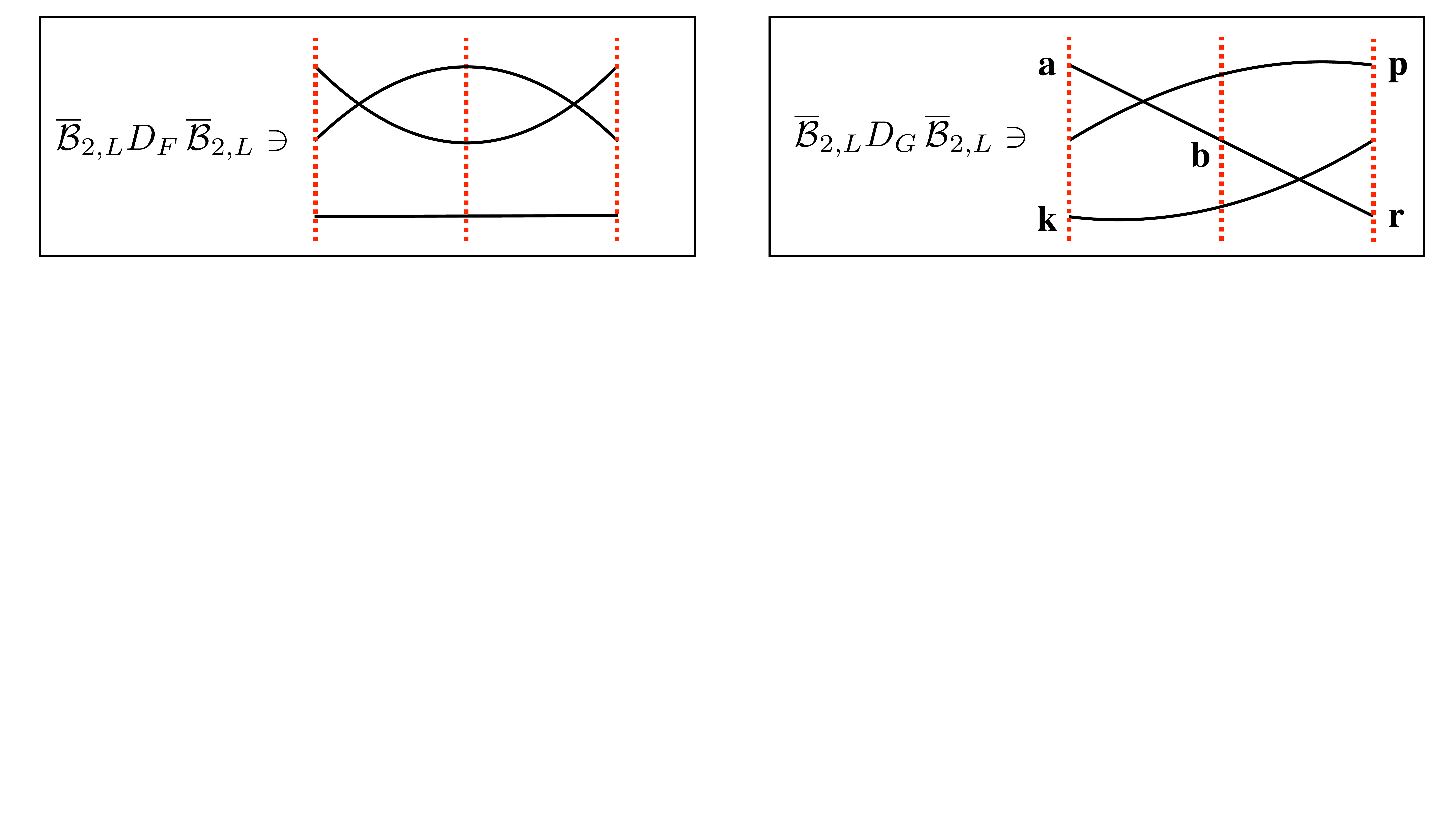}
\vspace{-2.7truein}
\caption{Example of the difference between $D_F$ and $D_G$ when 
connecting two $\overline{\cB}_{2,L}$ segments.
Notation as in Fig.~\ref{fig:topt1}.
In this simple example, the extra factor of $1/2!$ contained in the definition of $D_F$ is needed to
give the symmetry factor associated with the closed loop that crosses the central cut 
in the left-hand diagram.
Momentum labels are discussed in the text in Sec.~\protect\ref{sec:onshell}.
\label{fig:FvsG}
}
\end{center}
\end{figure}

Generalizing to more cuts is straightforward if only $\cB_3$ segments appear, for they can be
connected together with factors of $3 i D_F$ or, equivalently, $i(D_F+D_G)$.
A complication arises, however, when one has adjacent factors of $\overline{\cB}_{2,L}$,
as occurs in the example shown in Fig.~\ref{fig:CL_example}.
In such cases the manner in which indices are contracted matters due to the asymmetry of $\overline{\cB}_{2,L}$.
In Fig.~\ref{fig:CL_example}, the intermediate matrix must be a $D_G$, because the spectator
line is switched. If there is no such switch then the intermediate matrix must be a $D_F$.
This distinction is illustrated in Fig.~\ref{fig:FvsG}.
The relative factor of $2!$ between $D_F$ and $D_G$ is also needed to obtain the correct overall symmetry factor. 
One way to understand this is to note that there are two ways to join a spectator to the interacting pair,
compared to only a single way of joining the spectators.
The conclusion of this discussion is that factors of $\overline{\cB}_{2,L}$ 
{\em must be joined by $i(D_F+D_G)$},
with no freedom to change to any other form. This is why in the cases above where there was such freedom,
we rewrote the result in terms of the combination $D_F+D_G$. 

We thus conclude that the result for $n\ge 1$ relevant cuts can be written
\begin{equation}
C_{3,L}^{(n)} =
\widehat{A}' \,  i (D_F + D_G) \left[ i(\overline{\cB}_{2,L}+\cB_3) i(D_F+D_G)\right]^{n-1} \widehat{A}\,.
\end{equation}
The full correlator is then given by a geometric series
\begin{align}
	C_{3,L}(E,\vec{P}) &= \sum_{n=0}^\iy C_{3,L}^{(n)}(E,\vec{P})
	\\
	&= C_{3,L}^{(0)}(E,\vec{P}) + \sum_{n=1}^\iy \widehat{A}' \,
	i(D_F+D_G) \left[ i( \overline{\cB}_{2,L} + \cB_3 ) i(D_F+D_G) \right]^{n-1} \widehat{A}
	\\
	&= C_{3,\iy}^{(0)}(E,\vec{P})	
	 + \widehat{A}'\, i(D_F + D_G) 
	 \frac{1}{1 - i( \overline{\cB}_{2,L}+ \cB_3) i(D_F + D_G)} \widehat{A} \,.
	 \label{eq:CL_TOPT}
\end{align}
We have obtained a closed-form expression for $C_{3,L}(E,\vec P)$
in which the building blocks are infinite-volume ``TOPT amplitudes''---$\widehat{A}'$,
$\cB_3$, $\cB_2$ (contained in $\overline{\cB}_{2,L}$), and $\widehat{A}$.
Volume dependence enters through the fact that momenta are summed over the set of FV values,
and through the explicit factors of $L^3$ associated with the sums and contained in $\overline{\cB}_{2,L}$.
The simplicity of the result and the straightforward manner of its derivation
are two of the main advantages of using the TOPT approach.



\subsection{On-shell projection}
\label{sec:onshell}

At this stage our expression for $C_{3,L}$ is built from infinite-volume quantities,
but these are, in general, off shell.
To obtain a useful quantization condition we need to rewrite $C_{3,L}$ in terms of fully on-shell quantities,
which in turn can be related to physical amplitudes.
By fully on-shell, we mean that the incoming and outgoing four-momenta are both individually on shell and 
sum to the total four-momentum $P^\mu=(E,\vec{P})$.
Within TOPT the first condition is automatically satisfied, 
as is the conservation of spatial momenta,
so the issue to address is that, in general, the three particles at a relevant cut
do not satisfy $\omega_k+\omega_a+\omega_{b}=E$.
We stress that for the TOPT amplitudes $\cB_3$ and $\overline{\cB}_{2,L}$, one must separately consider
on-shellness for the incoming and outgoing momenta---these amplitudes can, for example,
be on shell on one side but off shell on the other.

The method we use in this subsection is to expand the amplitudes about their on-shell points.
For fixed $E$ and $\vec P$, these expansions are made in the spatial momenta of the external particles.
This necessarily involves consideration of external momenta that do not lie in the discrete set allowed
for finite volumes. Although such external momenta do not enter into the expression for $C_{3,L}$,
Eq.~(\ref{eq:CL_TOPT}), the TOPT amplitudes are well defined for all external momenta.
This is because all internal loops have been converted to infinite-volume integrals, so changes in external
momenta can be propagated through the diagrams into small changes in loop momenta.
Since the integrands are nonsingular by construction, the dependence on external momenta is smooth.
This result was used in HS1 and HS2 in the context of Feynman amplitudes.

We separately analyze cuts involving $D_G$ and $D_F$,
which we refer to as ``G cuts" and ``F cuts," respectively, and then apply the results to $C_{3,L}(E,\vec P)$.

\subsubsection{$G$ cuts}
\label{sec:Gcut}

To evaluate Eq.~\Eqref{eq:CL_TOPT}, we must repeatedly consider quantities of the form\footnote{%
If $X'=\widehat{A}'$ or $X=\widehat{A}$ then the corresponding outer momentum labels are absent.}
\begin{equation}
	\left[X' D_G X\right]_{ka;pr}= \left[X'\right]_{ka;k'a'} \left[D_G\right]_{k'a';p'r'} \left[X\right]_{p'r';pr} \,,\qquad
	X'\in\{ \widehat{A}', \overline\cB_{2,L}, \cB_3 \}\,, \ \
	X \in\{ \widehat{A}, \overline\cB_{2,L}, \cB_3 \}\,.
\label{eq:XDGXp}
\end{equation}
Our aim is to rewrite these in terms of a part in which the right-hand momenta of $X'$ (with indices $k'a'$)
and the left-hand momenta of $X$ (with indices $p'r'$) are both on shell, 
plus a residue that does not have a pole in $E$. 
In the following we refer to the momenta that we will set on shell as the ``inside'' momenta, 
while those that are left off shell (here $ka$ and $pr$) as the ``outside'' momenta.

We consider in detail the case where $X'=X=\overline\cB_{2,L}$, from which the results for other choices
can easily be deduced. Writing this out, we have
\begin{align}
	\left[\overline\cB_{2,L} D_G\, \overline\cB_{2,L}\right]_{ka;pr} 
	&= \left[\overline\cB_{2,L}\right]_{ka;k'a'} 
	\left[D_G\right]_{k'a';p'r'} \left[\overline\cB_{2,L}\right]_{p'r';pr} 
	\label{eq:C2L_DG_C2L}
	\\
	&= \cB_2(E_{2,k},\vec P_{2,k}; \vec{a}; \vec{p}) 
	\frac{1}{2\omega_{b_{kp}}(E-\omega_k-\omega_p-\omega_{b_{kp}})}  
	\cB_2(E_{2,p},\vec P_{2,p}; \vec{k}; \vec{r}) \,, 
	\label{eq:C2_DG_C2}
\end{align}
where in the first line repeated indices are summed.
The choice of momentum labels is illustrated in the right-hand diagram of Fig.~\ref{fig:FvsG}.

Our aim in the following is to expand the two factors of $\cB_2$ about the points at which their
internal momenta are on shell. Consider first the left-hand $\cB_2$.
Its internal (right-hand) momenta are $\vec{p}$ and $\vec{b}_{pk}=\vec P-\vec k-\vec p \equiv \vec b$.
We wish to adjust $\vec{p}$ (which also changes $\vec b$ since $\vec P$ is fixed) until 
$\omega_p+\omega_b=E_{2,k}$.
The way we do so is adapted from the approach used in Sec. IVB of HS1.
We first change variables to those in the center-of-mass frame (CMF) of the scattered pair.
Since the four-momentum of the pair is $(\omega_p+\omega_b,\vec P_{2,k})$,
the boost to this frame has parameters
\begin{equation}
\beta_k = \frac{|\vec P_{2,k}|}{\omega_p+\omega_b}\,,\quad
\gamma_k = \frac{\omega_p+\omega_b}{2 \omega_p^*}\,, \quad
4 \omega_p^{*2} = (\omega_p+\omega_b)^2-\vec P_{2,k}^2\,,
\label{eq:Wu1}
\end{equation}
such that $(\omega_p,\vec p)$ is boosted to $(\omega_p^*,\vec p_k^*)$, 
with 
\begin{equation}
	p^*_{k\parallel} = \gamma_{k} (p_\parallel - \beta_{k} \omega_p) \,,
	\quad \vec{p}^{\,*}_{k\perp} = \vec{p}_\perp\,,
\label{eq:Wu2}
\end{equation}
where parallel and perpendicular components are relative to $\vec P_{2,k}$.
Similarly, $(\omega_b,\vec b)$ is boosted to $(\omega_p^*,-\vec p_k^*)$.
Here the subscripts ``$k$'' are a reminder that the boost depends on $\vec k$.
This boost differs from that used in HS1, because here all particles are on shell while energy is not conserved,
whereas in HS1 the particle with momentum $\vec b$ is off shell and energy is conserved.
The boost used here has the advantage that it is well defined for all choices of $\vec k$, 
since $|\beta_k| < 1$.
We refer to it below as the ``Wu boost.''\footnote{%
We learned of this boost from J.-J. Wu, who used it in the context of the Hamiltonian effective field theory
description of finite-volume effects~\cite{Wuboost20}.
}
We also stress that $\vec p^*_k$ completely fixes $\vec p$
since it determines $E_{2,k}^*=2\omega_p^*$, from which,
for given $\vec P_{2,k}$,  one obtains $\omega_p+\omega_b$ and thus
the inverse boost. Thus we can change variables\footnote{%
The fact that $\cB_2$ is not Lorentz invariant presents no obstruction to this change of variables.}
 in $\cB_2$ from $\vec p$ to $\vec p_k^*=p_k^* \widehat{\vec{p}}_k^*$:
\begin{equation}
\cB_2(E_{2,k},\vec P_{2,k};\vec a;\vec p) \equiv
\cB^*_2(E_{2,k},\vec P_{2,k};\vec a;\vec p_k^*)\,,
\end{equation}
where the asterisk on $\cB_2^*$ simply indicates the same function expressed in terms of the new variables.

The next step is to decompose the angular dependence of $\cB^*_2$  into spherical harmonics,
\begin{equation}
\cB^*_2(E_{2,k},\vec P_{2,k};\vec a;\vec p_k^*)
\equiv
\cB_2^*(E_{2,k},\vec P_{2,k};\vec a;p_k^*)_{\ell m} \sqrt{4 \pi} Y_{\ell m}(\widehat{\vec{p}}_k^*)
\,,
\label{eq:C2sphdecomp}
\end{equation}
where there is an implicit summation on angular-momentum indices, and the
factor of $\sqrt{4\pi}$ follows the conventions of HS1.
We use real spherical harmonics throughout to avoid an overabundance of asterisks.
Since we expect $\cB_2$ to be nonsingular in our kinematic regime, the coefficients
$\cB_2^*(E_{2,k},\vec P_{2,k};\vec a;p_k^*)_{\ell m}$ should be smooth functions of $p_k^*$.
Indeed, if we can Taylor-expand $\cB_2^*(E_{2,k},\vec P_{2,k};\vec a;\vec p_k^*)$ about $\vec p_k^*=0$, it follows that,
in order to avoid singularities at $p_k^*=0$ from the spherical harmonics or the absolute value, 
we can pull out a factor of $p_k^{*\ell}$ from $\cB_2^*$,
\begin{equation}
\cB_2^*(E_{2,k},\vec P_{2,k};\vec a;p_k^*)_{\ell m} \equiv
\cB_2^{**}(E_{2,k},\vec P_{2,k};\vec a;p_k^{*2})_{\ell m}  p_k^{*\ell} \,,
\end{equation}
where $\cB_2^{**}$ is a smooth function of $p_k^{*2}$.
Thus we can rewrite the expansion as
\begin{equation}
\cB_2(E_{2,k},\vec P_{2,k};\vec a;\vec p)
=
\cB_2^{**}(E_{2,k},\vec P_{2,k};\vec a;p_k^{*2})_{\ell m} \cY_{\ell m}(\vec{p}_k^*)
\,,
\quad
\cY_{\ell m}(\vec{p}_k^*) \equiv
\sqrt{4\pi} Y_{\ell m} (\widehat{\vec{p}}_k^*)
p_k^{*\ell}\,.
\end{equation}
The $\cY_{\ell m}$ are simply the harmonic polynomials, rescaled by $\sqrt{4\pi}$.

The final step is to set the amplitude on shell (on its right-hand side) by adjusting $p_k^{*2}$.
If we set\footnote{%
The definitions of $q_{2,k}^*$ and $E_{2,k}^*$ are the same as in HS1, and the on-shellness conditions
also match.}
\begin{equation}
p_k^{*2}=q_{2,k}^{*2} \equiv {E_{2,k}^{*2}}/4 - m^2\,,\ \ {\rm where}\ \
E_{2,k}^{*2} = E_{2,k}^2 - \vec P_{2,k}^2
\label{eq:onshellcond}
\,,
\end{equation}
then we achieve the desired result:
\begin{equation}
p_k^{*2}=q_{2,k}^{*2} \ \ \Rightarrow \ \
4\omega_p^{*2} = E_{2,k}^{*2} \ \ \Rightarrow \ \
(\omega_p\!+\!\omega_b)^2 - \vec P_{2,k}^2 =
E_{2,k}^2 - \vec P_{2,k}^2 \ \ \Rightarrow\ \ 
E_{2,k} = \omega_p\!+\!\omega_b \ \ \Rightarrow\ \ 
E = \omega_k\!+\!\omega_p\!+\!\omega_b\,.
\end{equation}

We note that as $|\vec k|$ increases, $q_{2,k}^{*2}$ becomes negative, 
so that enforcing the on-shell condition of Eq.~\Eqref{eq:onshellcond} 
requires an extrapolation of $\cB_2$ below the two-particle threshold. 
In other words, even though $\vec k$ is such that, for the given values of $E$ and $\vec P$, the
other two particles cannot go on shell, we still must include a contribution from the 
TOPT amplitudes extrapolated below threshold.
This feature is common to all three-particle quantization conditions~\cite{Polejaeva:2012ut}.
However, we do not expect the three-particle levels to be sensitive to  amplitudes far below threshold.
To avoid this region, we introduce a function
$H(\vec k)$ that cuts off the sum over $\vec k$
(and depends implicitly on $E$ and $\vec P$). 
At this stage, the details of this function do not matter, aside from four properties:
(i) it must equal unity for all values of $\vec k$ for which an on-shell three-particle state is kinematically allowed,
(ii) it must remain unity for a finite distance of $\cO(m)$ below threshold,
(iii) it must be smooth in $\vec k$, 
and (iv) it must vanish for large $|\vec k|$.
The first property ensures that the pole terms that can lead to power-law FV dependence are 
fully incorporated into the analysis.
The second ensures that exponentially suppressed volume terms are suppressed by $\exp(-\delta L)$ with
$\delta = \cO(m)$.
The third property is needed to avoid introducing unwanted power-law dependence, as we will see shortly.
The final property truncates the sum over $\vec k$, which is essential to turn the quantization condition
into a practical tool.
Functions with these properties can be constructed easily from the example given in HS1.
We stress that the quantization condition that we derive is valid for any cutoff function satisfying
these properties, up to exponentially suppressed corrections. 
In other words, we do not lose control of power-law volume dependence when we
add the cutoff function by hand.

We now write $\cB_2$ in terms of an on-shell part and a residue
\begin{align}
\cB_2(E_{2,k},\vec P_{2,k};\vec a;\vec p)
&=
\cB_2^{**}(E_{2,k},\vec P_{2,k};\vec a;q_{2,k}^{*2})_{\ell m} \cY_{\ell m}(\vec{p}_k^*) H(\vec k)
+
\delta \cB_2(E_{2,k},\vec P_{2,k};\vec a;\vec p)\,,
\label{eq:C2decomp}
\end{align}
where the residue is
\begin{multline}
\delta \cB_2(E_{2,k},\vec P_{2,k};\vec a;\vec p) 
=
\left[
\cB_2^{**}(E_{2,k},\vec P_{2,k};\vec a;p_{k}^{*2})_{\ell m} 
- \cB_2^{**}(E_{2,k},\vec P_{2,k};\vec a;q_{2,k}^{*2})_{\ell m} 
\right]\cY_{\ell m}(\vec{p}_k^*) H(\vec k)
\\
+
\cB_2(E_{2,k},\vec P_{2,k};\vec a; \vec p) [1 - H(\vec k)]
\,.
\end{multline}
The key point here is that $\delta \cB_2$ cancels the pole in the energy denominator in
Eq.~(\ref{eq:C2_DG_C2}). 
For the first term, the smoothness of $\cB^{**}_{2;\ell m}$ implies that 
\begin{equation}
\left[
\cB_2^{**}(E_{2,k},\vec P_{2,k};\vec a;p_{k}^{*2})_{\ell m} 
- \cB_2^{**}(E_{2,k},\vec P_{2,k};\vec a;q_{2,k}^{*2})_{\ell m} \right] \propto
p_k^{*2}- q_{2,k}^{*2} = - \frac14 (E-\omega_k+\omega_p+\omega_b) (E-\omega_k-\omega_p-\omega_b)\,,
\end{equation}
which explicitly cancels the pole.
For the second term in $\delta \cB_2$, the factor of $1-H(\vec k)$ vanishes for all choices of $\vec k$
for which on-shell kinematics are possible, and in a finite neighborhood thereof, so that the pole
is avoided.
Thus the total summand involving $\delta \cB_2$ is finite and smooth,
allowing loop sums involving momenta crossing the cut to be converted to integrals, as discussed further below.

Our final step for the left-hand amplitude is to rewrite it
in terms of $\cB^*_{2;\ell m}$, i.e.~the angular components of the on-shell amplitude.
Then Eq.~(\ref{eq:C2decomp}) becomes
\begin{equation}
\cB_2(E_{2,k},\vec P_{2,k};\vec a;\vec p)
=
\cB_2^{*}(E_{2,k},\vec P_{2,k};\vec a;q_{2,k}^{*})_{\ell m} \frac{\cY_{\ell m}(\vec{p}_k^*)}{q_{2,k}^{*\ell}}
H(\vec k)
+
\delta \cB_2(E_{2,k},\vec P_{2,k};\vec a;\vec p)
\,.
\label{eq:leftC2part}
\end{equation}
The construction makes clear that the apparent pole at $q_{2,k}^*=0$ is canceled by the behavior of 
$\cB^*_{2;\ell m}$ near threshold.

We now make an analogous decomposition of the right-hand $\cB_2$. The steps are the
same with the roles of $\vec k$ and $\vec p$ interchanged. We thus find
\begin{equation}
\cB_2(E_{2,p},\vec P_{2,p};\vec k;\vec r)
=
H(\vec p) \frac{\cY_{\ell m}(\vec{k}_p^*)}{q_{2,p}^{*\ell}}
\cB_2^{*}(E_{2,p},\vec P_{2,p};q_{2,p}^{*};\vec r)_{\ell m} 
+ \mc{O}(E-\omega_k-\omega_p-\omega_b)
\,.
\label{eq:rightC2part}
\end{equation}
Here we have set the left-hand momenta on shell.

We also find it convenient to rewrite the pole in the relativistic form used in Refs.~\cite{\BHSQC,\dwave,\largera}
\begin{equation}
\frac1{2\omega_b (E-\omega_k-\omega_p-\omega_b)} = \frac1{b^2-m^2} \left[1 
+ \mc{O}(E-\omega_k-\omega_p-\omega_b)\right]\,,
\label{eq:polepart}
\end{equation}
where $b$ here is the four-vector 
\begin{equation}
b=(E-\omega_k-\omega_p,\vec P-\vec k-\vec p)=(\omega_b, \vec b) + (E-\omega_k-\omega_p-\omega_b,\vec 0)
\,.
\end{equation}
This step is not essential, and we could proceed with the derivation with the form
of the pole used in HS1.

Inserting the results from Eqs.~\Eqref{eq:leftC2part}, \Eqref{eq:rightC2part}, and \Eqref{eq:polepart}
into Eq.~(\ref{eq:C2_DG_C2}), we find the pole part
\begin{equation}
\cB_2^{*}(E_{2,k},\vec P_{2,k};\vec a;q_{2,k}^{*})_{\ell m}
G^b_{k\ell m;p\ell' m'}
\cB_2^{*}(E_{2,p},\vec P_{2,p};q_{2,p}^{*};\vec r)_{\ell' m'} 
\,,
\label{eq:fullpolepart}
\end{equation}
where the ``switch matrix'' $G^b$ is\footnote{%
This is the same as the object $G^b$ appearing in HS1, except that we use the relativistic form of the pole and the Wu boost.}
\begin{equation}
{G}^b_{k\ell m;p\ell' m'}
=
 \frac{\cY_{\ell m}(\vec{p}_k^*)}{q_{2,k}^{*\ell}}
\frac{H(\vec k)H(\vec p)}{b^2-m^2}
\frac{\cY_{\ell'm'}(\vec{k}_p^*)}{q_{2,p}^{*\ell'}}\,.
\label{eq:Gb}
\end{equation}
Equation~\Eqref{eq:fullpolepart} has achieved our goal of pulling out a term in which the inner
indices are set on shell. 
Whenever the index set $\{k \ell m\}$ appears instead of, say, $\{ka\}$, this indicates that
an on-shell projection has been carried out following the procedure explained above.

\begin{figure}[tb]
\begin{center}
\vspace{-10pt}
\includegraphics[width=\textwidth]{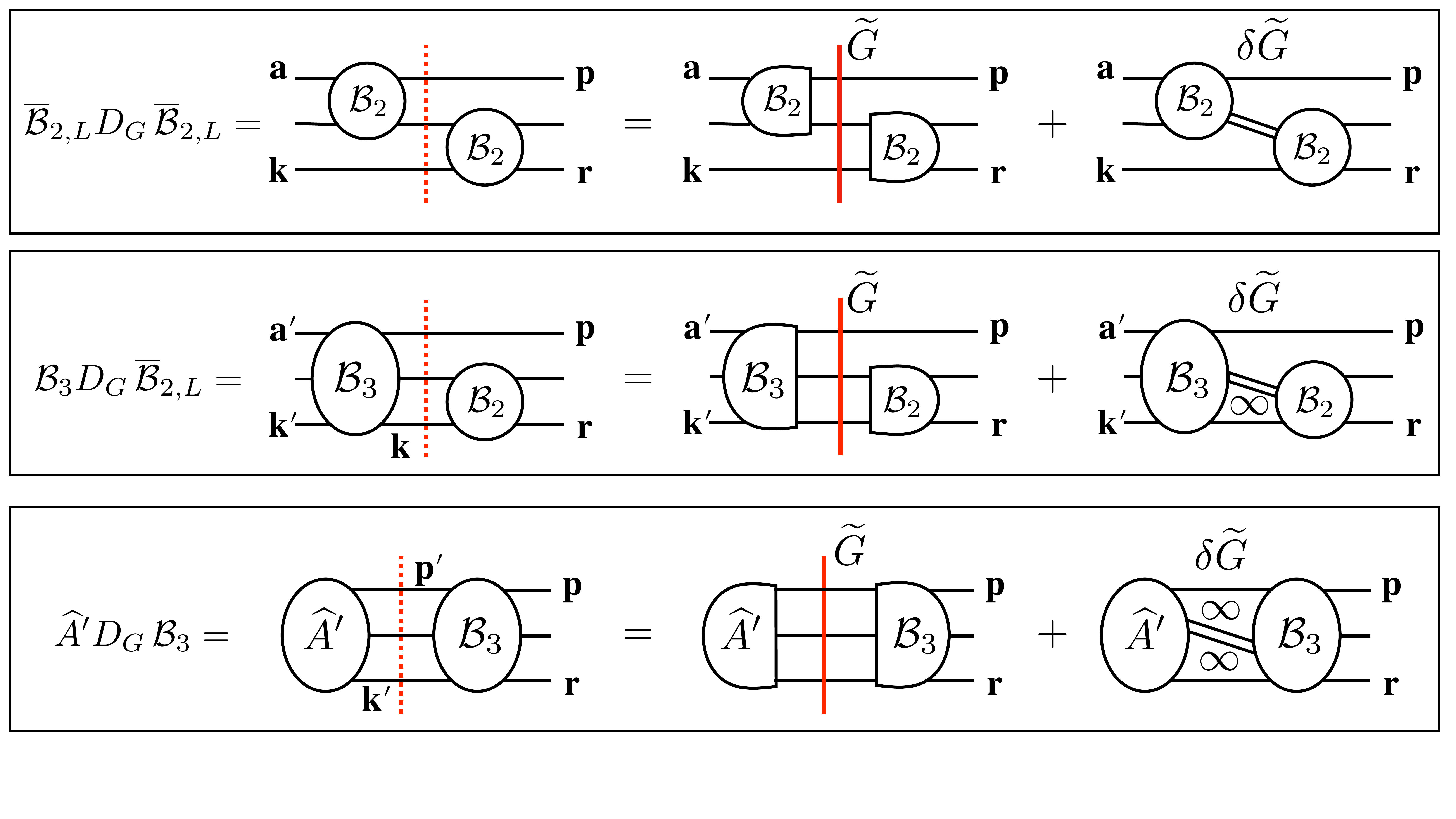}
\vspace{-0.3truein}
\caption{
Diagrammatic illustration of on-shell projection for G cuts, specifically of the results 
Eq.~(\ref{eq:finalC2DGC2}) [upper panel],
Eq.~(\ref{eq:finalC3DGC2}) [middle panel],
and Eq.~\eqref{eq:Gsplit} 
with $X'=\widehat{A}'$ and $X=\cB_3$ [lower panel].
The relevant cut corresponding to $D_G$ is shown by the dashed (red) vertical line,
while the insertion of the matrix $\wt G$ is shown by the solid (red) vertical line.
Rounded ends of kernels are off shell, while straightened ends are on shell.
The angled double solid line connecting amplitudes indicates that the pole in the energy denominator
has been canceled.
A loop containing ``$\infty$" is integrated, while other loops are summed.
\label{fig:Gcuts}
}
\end{center}
\end{figure}

To package the final result in a way that generalizes to other choices of $X'$ and $X$
in Eq.~\Eqref{eq:XDGXp}, we reintroduce the factors of $2\omega L^3$ that cancel if $X'=X=\overline{\cB}_{2,L}$
but do not cancel in general:\footnote{%
Here $\wt{G}$ is related to the matrix $G$ of HS1 by 
$\wt{G}_{k\ell m;p\ell' m'}=(2\omega_k L^3)^{-1} G_{k\ell m;p\ell' m'}$, except that we use the relativistic
form of the pole and the Wu boost.
In fact, at this stage we could change to using the boost of HS1 (and, if desired, the nonrelativistic form
of the pole) in $\wt G$. This only leads to a change in $\delta\wt G$. For completeness, we note that
our $\wt{G}$ differs from the similar quantity of the same name used in
Refs.~\cite{\BHSnum,\dwave,\largera}: our version contains an extra factor of $1/L^3$.
}
\begin{align}
\left[\overline{\cB}_{2,L}\right]_{ka;k'\ell m}
&= 2\omega_k L^3 \delta_{kk'} \cB_2^{*}(E_{2,k},\vec P_{2,k};\vec a;q_{2,k}^{*})_{\ell m}
\label{eq:C2Loffon}
\\
\left[\overline{\cB}_{2,L}\right]_{p' \ell' m';pr}
&= 2\omega_p L^3 \delta_{p'p}
\cB_2^{*}(E_{2,p},\vec P_{2,p};q_{2,p}^{*};\vec r)_{\ell' m'} 
\label{eq:C2Lonoff}
\\
\wt{G}_{k\ell m;p\ell' m'} 
&= \frac1{2\omega_k L^3} {G}^b_{k\ell m;p\ell' m'} \frac1{2\omega_p L^3}
\label{eq:Gt}
\,.
\end{align}
Using these matrices we have
\begin{align}
	\left[\overline\cB_{2,L} D_G \overline\cB_{2,L}\right]_{ka;pr} 
	&= \left[\overline\cB_{2,L} \left(\wt G + \delta\wt G\right) \overline\cB_{2,L}\right]_{ka;pr} \,.
\label{eq:finalC2DGC2}
\end{align}
where the $\wt G$ term contains the pole, while $\delta \wt{G}$ is simply the sum of all nonsingular
contributions. We will not need an explicit form for $\delta \wt{G}$.
Equation~\Eqref{eq:finalC2DGC2}  gives the result in a convenient, but highly compact, notation.
It should always be kept in mind that any object adjacent to a factor of $\wt{G}$ is projected on shell. 
The $\delta\wt{G}$ contribution, however, does not include an on-shell projection.
The result~\Eqref{eq:finalC2DGC2} is shown diagrammatically in the upper panel of Fig.~\ref{fig:Gcuts}, where
the $\delta \wt G$ term is seen to sew together the two $\cB_2$'s into an enlarged, infinite-volume amplitude.

The analysis proceeds similarly if we replace one or both of the $\overline\cB_{2,L}$'s with one
of the other TOPT amplitudes.
As an example, consider 
\begin{equation}
\left[\cB_3 D_G \overline{\cB}_{2,L}\right]_{k'a';pr}
=
\left[\cB_3\right]_{k'a';ka} \left[ D_G\right]_{ka;p'r'}\left[ \overline{\cB}_{2,L}\right]_{p'r';pr}
\,,
\label{eq:finalC3DGC2}
\end{equation}
which is illustrated in the middle panel of Fig.~\ref{fig:Gcuts}.
For notational convenience 
we have interchanged the dummy indices $\{ka\}$ and $\{k'a'\}$ compared to earlier.
The differences from the analysis above are 
(i) that $\vec k\ne \vec k'$, so that the on-shell projection
for $\cB_3$ involves a boost determined by the inner momentum (here $\vec k$);
(ii) the $1/(2\omega_{k} L^3)$ factor in $D_G$ is not canceled;
(iii) $\vec k$ is summed in the final result---with $1/(2\omega_{k} L^3)$ providing
the correct measure factor; and
(iv) in the finite $\delta \wt G$ term, the sum over $\vec k$ can be converted to an integral,
since the pole has been canceled, and this attaches a $\cB_2$ to the $\cB_3$ to create an enlarged
infinite-volume amplitude. 
This step relies also on the smoothness of $H(\vec k)$.
This last point implies that we should view $\delta\wt G$ as an operator that acts differently depending
on the adjacent kernels, and in particular implies integration over all internal loops that cross the original cut.

To give an explicit expression we need to define the version of $\cB_3$ after on-shell projection on the right,
\begin{equation}
\left[\cB_3\right]_{k'a';k \ell m} \equiv \cB^{*}_3(\vec k',\vec a'; \vec k,a_k^*=q_{2,k}^*)_{\ell m} 
 \frac{\cY_{\ell m}(\vec{a}_k^*)}{q_{2,k}^{*\ell}}\,,
\end{equation}
using which we have
\begin{equation}
\left[\cB_3 D_G \overline{\cB}_{2,L}\right]_{k'a';pr}
=
\left[\cB_3 (\wt G + \delta \wt G) \overline{\cB}_{2,L}\right]_{k'a';pr}\,,
\label{eq;C3DGC2}
\end{equation}
i.e. a result of exactly the same form as when $X'=X=\overline\cB_{2,L}$.

The other cases follow analogously, and we do not discuss them in detail. 
Dropping external indices, the general result is simply
\begin{equation}
X' D_G X = X'( \wt G + \delta \wt G) X\,.
\label{eq:Gsplit}
\end{equation}
The only new feature that enters if both $X'$ and $X$ are symmetric kernels is
that both factors of $1/(2\omega L^3)$ are uncanceled, 
so both internal momenta end up summed (for the $\wt G$ term) or integrated
(for the $\delta \wt G$ term). This is illustrated for $X'=\widehat{A}'$ and $X=\cB_3$ in 
the lower panel of Fig.~\ref{fig:Gcuts}.

\subsubsection{$F$ cuts}
\label{sec:Fcut}

Next, we wish to derive an analogous result for 
	$X' D_{F} X$.
One option is to split up the on- and off-shell contributions exactly as for $D_G$; 
we refer to this as the ``$\wt\Sigma_F$ approach," 
as the analog to $\wt{G}$ that arises is a sum over spatial momenta.
Although the $\wt\Sigma_F$ approach is perfectly valid and well defined, we relegate its details to 
Appendix~\ref{app:opt1}, as there is a more standard method---following essentially the same
approach as in HS1---that we now discuss.
We shall refer to the standard method as the ``$\wt{F}$ approach," for reasons that will soon be apparent.

\begin{figure}[tb]
\begin{center}
\vspace{-10pt}
\includegraphics[width=\textwidth]{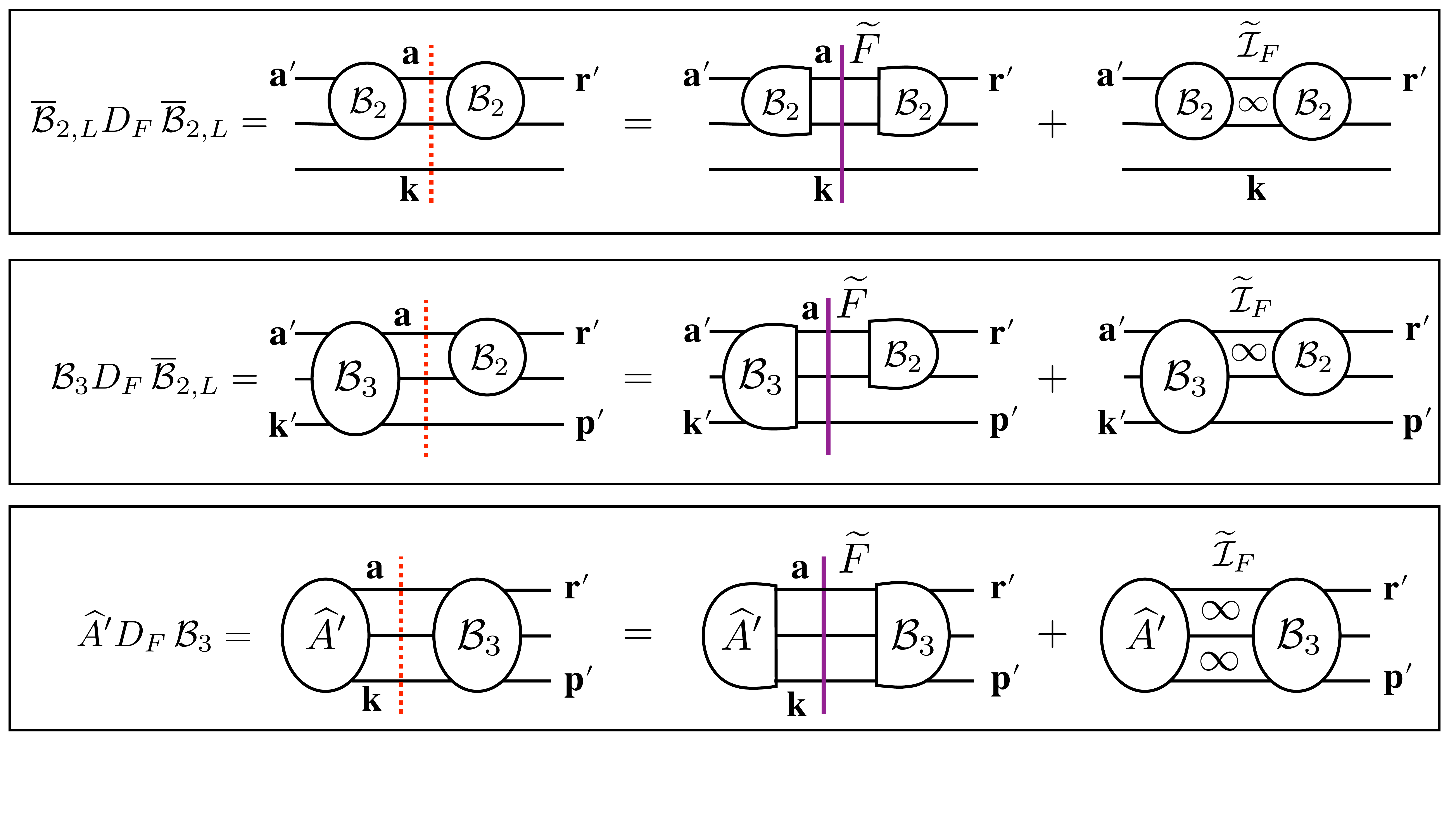}
\vspace{-0.5truein}
\caption{
Diagrammatic illustrations of on-shell projection for F cuts.
Notation as in Fig.~\ref{fig:Gcuts}  except that
the insertion of the matrix $\wt F$ is shown by the solid (purple) vertical line.
The integral over the momentum $\vec a$ (the upper loop) requires a pole prescription, 
for which we use a generalized PV prescription. Momentum labels are
matched to the discussion in the text.
\label{fig:Fcuts}
}
\end{center}
\end{figure}

The results we obtain are illustrated in Fig.~\ref{fig:Fcuts}. As the figures show, an essential  difference
between the G and F cuts is that, for the latter, 
at least one of the momenta crossing the cut is part of an internal loop.
This feature is exemplified by the explicit expression for $X'=X=\overline{\cB}_{2,L}$,
\begin{equation}
	\left[\overline\cB_{2,L} D_F\, \overline\cB_{2,L}\right]_{k a';p r'} 
	= \delta_{k p}2\omega_{k} L^3  \sum_{\vec a} \frac1{2\omega_{a}L^3} 
	 \cB_2(E_{2,k},\vec P_{2,k}; \vec{a}'; \vec{a}) 
	\frac{1}{2\omega_{b_{ka}}(E\!-\!\omega_{k}\!-\!\omega_{a}\!-\!\omega_{b_{ka}})}  
	\cB_2(E_{2,k}, \vec P_{2,k}; \vec{a}; \vec{r'}) \,, 
	\label{eq:C2_DF_C2}
\end{equation}
which is shown diagrammatically in the upper panel in Fig.~\ref{fig:Fcuts}.
The $\vec a$ loop is always present, and, in the $\wt{F}$ approach, is treated first. 
The method is that introduced in Ref.~\cite{\KSS} and extended in HS1:
one replaces the sum in Eq.~\Eqref{eq:C2_DF_C2} using the identity 
$\sum = \left[\sum-\PV\!\int\right] + \PV\!\int$.
Here PV indicates that we are using a generalized principal-value (PV) pole prescription,
in the class introduced in Ref.~\cite{\largera}.
The sum-integral difference projects the inner momenta of the adjacent amplitudes on shell,
as shown by the identity derived in Appendix A of HS1.\footnote{%
Strictly speaking, the argument given in HS1 must be slightly modified to use the boost
introduced above, but the essence is unchanged.}
Using this identity, we find, in our example,
\begin{equation}
\left[\overline\cB_{2,L} D_F\, \overline\cB_{2,L}\right]_{k a';p r'} 
=
\left[\overline\cB_{2,L}\right]_{k a';k' \ell m} \wt F_{k'\ell m;p' \ell' m'}
\left[\overline\cB_{2,L}\right]_{p' \ell' m';p r'} 
+
\left[\overline\cB_{2,L} \wt \cI_F \overline\cB_{2,L}\right]_{k a';p r'} 
\,,
\label{eq:C2DFC2}
\end{equation}
where the partially on-shell amplitudes are defined in Eqs.~\Eqref{eq:C2Loffon} and \Eqref{eq:C2Lonoff},
$\wt F$ is a matrix acting in the on-shell index space\footnote{%
Here $\wt F$ is related to the matrix $F$ of HS1 by
$\wt{F}_{k\ell m;p\ell'm'}  = (2\omega_k L^3)^{-1} F_{k\ell m;p \ell' m'}$. 
The fact that we use a different boost in the on-shell projection
only changes the sum-integral difference by exponentially suppressed terms. 
This is true also if the pole in $D_{ka}$ is changed to the
relativistic form used for $\wt G$, and for changes in the UV regulator.
Our $\wt F$ differs from the quantity of the same name used in Refs.~\cite{\BHSnum,\dwave,\largera} by
having an additional factor of $1/L^3$.
}
\begin{equation}
	\wt{F}_{k\ell m;p\ell'm'} \equiv \delta_{kp} H(\vec k)
	\left[ \frac1{L^3} \sum_{\vec{a}}^\UV - \PV\!\int_{\vec{a}}^\UV \right] 
	\frac{\cY_{\ell m}(\vec{a}_{k}^*)}{q_{2,k}^{*\ell} }
	\frac{L^3 D_{ka}}{2!} \frac{\cY_{\ell'm'}(\vec{a}_{k}^{*})}{q_{2,k}^{*\ell'}} 
	\,,
\label{eq:Ft}
\end{equation}
with $\int_{\vec a} \equiv \int d^3a/(2\pi)^3$, and the action of the integral operator $\wt \cI_F$ is
\begin{equation}
\left[\overline\cB_{2,L} \wt \cI_F \overline\cB_{2,L}\right]_{k a';p r'} 
\equiv
\sum_{\vec k',\vec p',\vec r} 
\left\{ H(\vec k) \,\PV\!\! \int_{\vec{a}}^\UV  + [1-H(\vec k)] \int_{\vec{a}}^\UV \right\}
	\left[\overline{\cB}_{2,L}\right]_{k a';k' a}
	\frac{L^3 D_{k' a}\delta_{k' p'}\delta_{ar}}{2!} \left[\overline{\cB}_{2,L}\right]_{p' r;p r'} 
	\,. 
	\label{eq:I_F}
\end{equation}
The integral operator ties together the two factors of $\cB_2$, as shown in the figure.
Note that the overall factor of $L^3$ cancels that in the $1/(2\omega_a L^3)$ contained in $D_{ka}$.

One subtle feature of Eqs.~\Eqref{eq:Ft} and \Eqref{eq:I_F} is the appearance of factors of $H(\vec k)$.
As for the G cuts, these are introduced so as to cut off the sum over $\vec k$. The identity
(\ref{eq:C2DFC2}) is valid for any choice of $H(\vec k)$ satisfying the properties enumerated earlier.
In particular, the second integral on the right-hand side of Eq.~\Eqref{eq:I_F} originates as a sum over
$\vec a$, but can be converted to an integral because the overall $1-H(\vec k)$ cancels 
the pole in $D_{ka}$.\footnote{%
Note also that, despite appearances, the $H(\vec k)$ dependence of the two integrals in
Eq.~\Eqref{eq:I_F}  {\em do not cancel}, because they are defined with different pole prescriptions:
PV for the first integral, while no prescription is needed for the second (since the pole is avoided).
This is equivalent to using the $i\epsilon$ prescription for the second integral.}

The results for other choices of $X'$ and $X$ take the same form, and are illustrated in the middle and lower
panels of Fig.~\ref{fig:Fcuts}. The only new feature occurs when both $X'$ and $X$ are symmetric amplitudes.
In this case the sum over $\vec k$ is not resolved by Kronecker deltas, and remains as an internal loop.
Here we can use the result that the PV integral over $\vec a$ leads to a smooth function of $\vec k$, despite
the pole in the integrand. This is shown for the standard PV prescription in Appendix B of HS1, 
and holds also for the generalizations of Ref.~\cite{\largera}. Because of this result, and the smoothness of
the TOPT amplitudes, the sum over $\vec k$ in the $\wt \cI_F$ term can be replaced by an integral,
as exemplified by the last term in the lower panel in the figure.
Thus, as for $\delta \wt G$, the integral operator $\wt \cI_F$ acts in a manner that depends on the
adjacent amplitudes.

In summary, the general result (with matrix indices implicit) is
\begin{align}
	X' D_F\,X &= X' \left(\wt{F} + \wt\I_F\right)X\,,
	\label{eq:Fsplit}
\end{align}
where any amplitude adjacent to $\wt F$ is placed on shell with indices $\{k\ell m\}$.

\subsubsection{Application to $C_{3,L}(E,\vec{P})$}

We can use the results Eqs.~\Eqref{eq:Gsplit} and \Eqref{eq:Fsplit} to rewrite our
expression for the three-particle correlator, Eq.~(\ref{eq:CL_TOPT}), so as to isolate on-shell contributions,
\begin{align}
	C_{3,L}-C_{3,\iy}^{(0)} 
	&= \widehat{A}'\, i (\wt{F}+\wt{G} + \wt\I_F + \delta\wt{G})
	 \frac{1}{1 - i (\overline\cB_{2,L}+\cB_{3}) i (\wt{F}+\wt{G} + \wt\I_F + \delta\wt{G}\,)} \widehat{A}
	\\
	&= \delta C_{3,\infty} + 
	\wt{A}'^{\oneu} i(\wt{F}+\wt{G}\,) \frac{1}{1-i\Kalluu i(\wt{F}+\wt{G}\,)} \wt{A}^{\oneu} \,, 
	\label{eq:CL_final_L}
\end{align}
where
\begin{align}
	i\Kalluu &\equiv 
	\frac{1}{1-i (\overline{\cB}_{2,L} + \cB_3) i(\wt\I_F + \delta\wt{G}\,)} i(\overline{\cB}_{2,L} + \cB_3) \,,
	\label{eq:Kalluu_L}
	\\
	\wt{A}'^{\oneu} &\equiv \widehat{A}' \frac{1}{1-i(\wt\I_F + \delta\wt G)i(\overline{\cB}_{2,L} + \cB_3)}\,,
	\\
	\wt{A}^{\oneu} &\equiv \frac{1}{1-i(\overline{\cB}_{2,L} +\cB_3) i(\wt\I_F+\delta\wt G)} \widehat{A}\,,
	\label{eq:ALp}
	\\
	\delta C_{3,\infty} &\equiv \widehat{A}'\; i( \wt\I_F + \delta\wt G) 
	\frac{1}{1-i(\overline{\cB}_{2,L} + \cB_3) i(\wt\I_F + \delta\wt G)} \widehat{A} \,.
	 \label{eq:delta_C3L}
\end{align}
Since $\wt{A}'^{\oneu}$, $\Kalluu$, and $\wt{A}^{\oneu}$ all appear adjacent to $\wt F+\wt G$, they are
all projected into the on-shell $\{k\ell m\}$ index space.
We will refer them as ``on-shell kernels''.

Various aspects of these results deserve further explanation.
The first is our use of tildes.
We have added these in order to distinguish the kernels from quantities in HS1 that have similar names,
but different definitions.

The second new feature is the appearance of superscripts $\oneu$ and ${(u,u)}$.
This notation, borrowed from HS1, indicates asymmetric quantities.\footnote{%
An important caveat, however, is that the precise nature of the asymmetry here differs from that in
HS1. We discuss this further below.
}
The asymmetry here arises from the presence of the asymmetric amplitude $\overline{\cB}_{2,L}$.
When we expand out the geometric series in Eqs.~\Eqref{eq:Kalluu_L}-\Eqref{eq:ALp},
the external amplitude can either be a $\overline{\cB}_{2,L}$ or a symmetric amplitude. 
For example, we can rewrite Eq.~(\ref{eq:ALp}) as
\begin{equation}
\wt{A}^{\oneu} 
= 
\widehat{A}
+
i(\overline{\cB}_{2,L} +\cB_3) i(\wt\I_F+\delta\wt G)
 \frac{1}{1-i(\overline{\cB}_{2,L} +\cB_3) i(\wt\I_F+\delta\wt G)} \widehat{A}\,.
\label{eq:ALpnew}
\end{equation}
The presence of the $\overline{\cB}_{2,L}$ in the second term on the right-hand side implies that
this is an asymmetric quantity, since $\vec k$ is always associated with the spectator momentum
when connecting to $\overline{\cB}_{2,L}$.

\begin{figure}[tb]
\begin{center}
\vspace{-10pt}
\includegraphics[width=\textwidth]{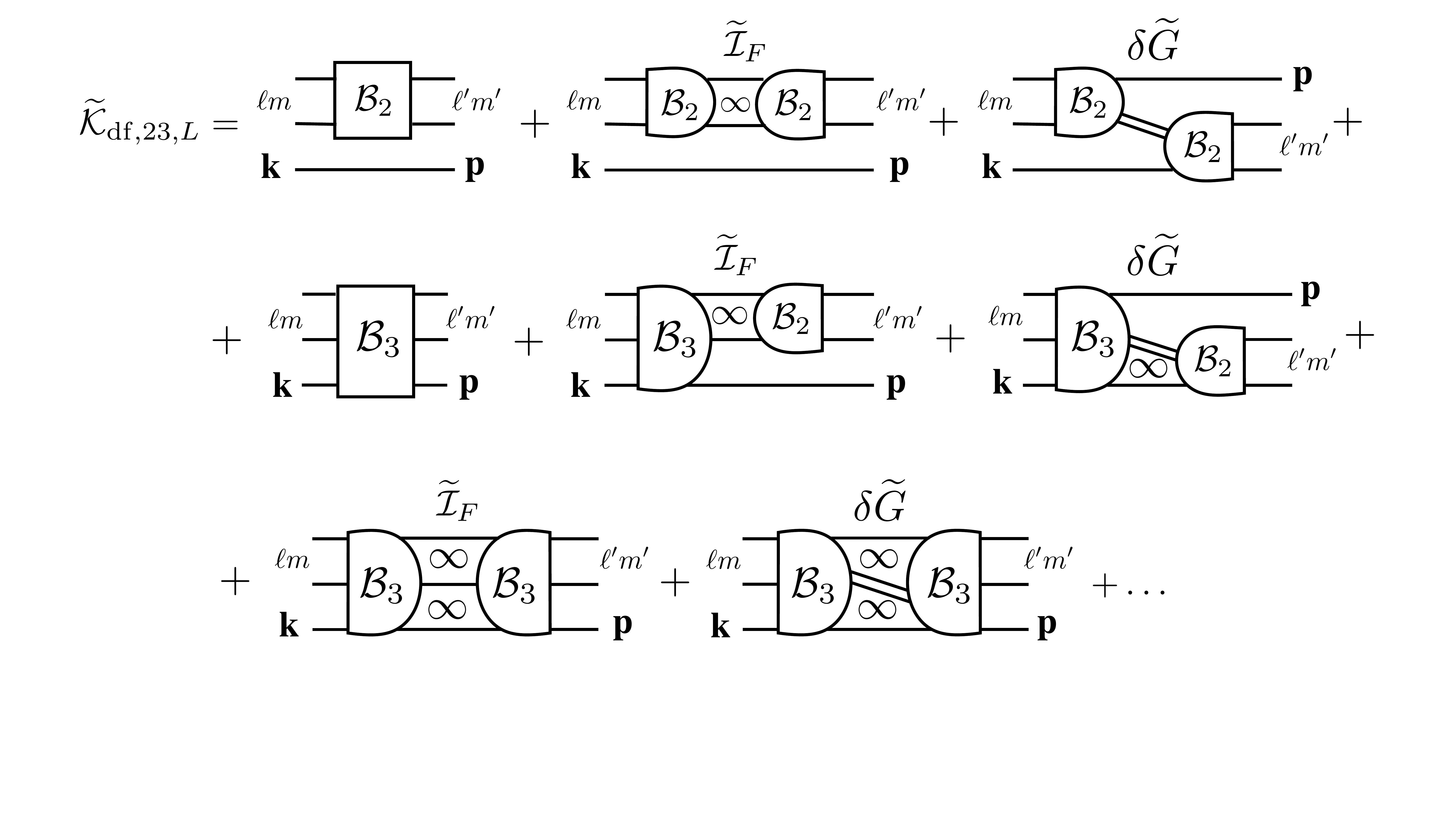}
\vspace{-0.8truein}
\caption{
Diagrams contributing to $[\Kalluu]_{k \ell m;p\ell' m'}$. Notation as in Figs.~\ref{fig:Gcuts} and \ref{fig:Fcuts}.
Factors of $i$ are implicit.
\label{fig:Kdf23L}
}
\end{center}
\end{figure}

The third new aspect is the subscript ``$\df,23$'' on $\Kalluu$, as well as the use of the name $\cK$.
To explain these features, we expand the geometric series in Eq.~(\ref{eq:Kalluu_L}), 
leading to the contributions shown diagrammatically in Fig.~\ref{fig:Kdf23L}. 
These are exactly the set of diagrams that give rise, in TOPT, to the $2\to 2$ amplitude (with
a spectator) combined with the $3\to3$ amplitude, except that we have replaced the relevant cuts with
integral operators. This is similar to what one does when defining a K matrix, namely removing the
imaginary parts that arise from unitary cuts. 
In particular, since all the integrals that appear either use a PV prescription or avoid the pole, $\Kalluu$ is real.
Because of this similarity, we refer to it as a K matrix,
and, indeed, the connection to standard K matrices can, in part, be made more precise, as we show below.
We use the subscript ``23'' to indicate that it contains amplitudes for both two- and three-particle scattering. 
Finally, ``df'' stands for ``divergence free," which is to say that, by construction, it contains no singularities
due to three-particle cuts. This use of ``df'' is taken over from HS1.

The final issue concerns the volume dependence of the kernels.
We find that $\wt{A}'^\oneu$, $\wt{A}^\oneu$, and $\delta C_{3,\infty}$ are infinite-volume quantities,
and that $\Kalluu$ can be simply related to infinite-volume quantities,
both results holding up to exponentially suppressed corrections.
These results, derived in the next subsection, will allow us to make all $L$ dependence explicit.

\subsubsection{Volume (in)dependence of kernels}

The most complicated of the kernels is $\Kalluu$, and we address this first. 
As is clear from Fig.~\ref{fig:Kdf23L}, the $2\to2$ part of $\Kalluu$ is given by the geometric series
\begin{align}
i\overline{\cK}_{2,L} &\equiv
 i\overline{\cB}_{2,L} \frac1{1-i\wt\cI_F i\overline{\cB}_{2,L}} 
= i\overline{\cB}_{2,L} + i\overline{\cB}_{2,L}\, i \wt \cI_F\, i\overline{\cB}_{2,L} + \cdots \,.
\label{eq:K2Ldef}
\end{align}
The off-shell version of this quantity, i.e.~with indices $\{ka,pr\}$, will be a key building block in the final expression. 
The factors of $2\omega_k L^3$ cancel in pairs, leaving a single overall factor of this type, allowing us to write
\begin{align}
\left[\overline{\cK}_{2,L}\right]_{ka;pr} &=  2\omega_k L^3  \left[\cK_2\right]_{ka;pr}\,,
\label{eq:K2Loffoff}
\\
\left[\cK_2\right]_{ka;pr} &\equiv \delta_{k p} \cK_2(E_{2,k},\vec P_{2,k}; \vec a;\vec r)\,,
\label{eq:K2off}
\end{align}
where $\cK_2(E_2,\vec P_2;\vec a;\vec r)$ 
is the infinite-volume {$2\to2$ quantity obtained by sewing together any number of $\cB_2$ kernels 
with the two-particle version of $\wt\cI_F$.
As the name suggests, it is related to a two-particle K matrix. 
Indeed, we show in Appendix~\ref{app:K2} that the on-shell restriction is given by
\begin{align}
\left[\overline{\cK}_{2,L}\right]_{k\ell m;p\ell' m'} &=  2\omega_k L^3\left[\cK_2\right]_{k\ell m;p\ell' m'}\,,
\label{eq:K2Lonon}
\\
\left[\cK_2\right]_{k\ell m;p\ell' m'} &= \delta_{kp}\delta_{\ell \ell'}\delta_{m m'} \cK_2^{(\ell)}(q_{2,k}^*)\,,
\label{eq:K2on}
\end{align}
where $\cK_2^{(\ell)}$ is the $\ell$th partial-wave amplitude in the two-particle CMF,
with $q_{2,k}^*$ the magnitude of the momentum of each particle.
$\cK_2^{(\ell)}$ has a known relation to the corresponding partial wave
of the scattering amplitude, $\cM_2^{(\ell)}$, given in Eq.~(\ref{eq:result2}).
$\cK_2$ becomes the standard K matrix if we use the standard PV scheme 
and set $H(\vec k)=1$ for all $\vec k$.

Returning to $\Kalluu$, we now reorder the terms in the geometric series \Eqref{eq:Kalluu_L} by first
summing subsets of diagrams involving sequences of $\overline{\cB}_{2,L}$'s and $\widetilde \cI_F$'s
into $\overline{\cK}_{2,L}$'s. This is illustrated in the first panel of Fig.~\ref{fig:K2_D}.
We next sum sequences of the resulting $\overline{\cK}_{2,L}$'s connected by factors of $\delta \wt G$,
leading to the $3\to3$ quantity
\begin{equation}
i\wt \cD_{3,L}^\uu \equiv i\overline{\cK}_{2,L} i \delta \wt G i\overline{\cK}_{2,L} \frac1{1 - i \delta \wt G i\overline{\cK}_{2,L}}\,,
\label{eq:Duu}
\end{equation}
as shown in the lower panel  of Fig.~\ref{fig:K2_D}.
Factors of $2\omega L^3$ cancel except for an inverse such factor for every internal loop,
which would be absorbed if we could convert each loop sum into an integral.
This requires, however, that the summand is smooth.
Here we face a new issue: while the (double-line) connectors between adjacent $\cK_2$'s are nonsingular,
$\cK_2$ itself can have singularities as a function of the loop momentum. 
For example, in the second contribution to $\wt\cD_{3,L}^\uu$ shown in the figure, the four-momentum 
$(E_{2,a'},\vec P_{2,a'})$ passing through the lower $\cK_2$
clearly depends on the loop momentum $\vec a'$.
We know that the on-shell $\cK_2$ has poles for real momenta whenever there is a nearby narrow resonance,
and, following the arguments of Ref.~\cite{\BHSK}, we expect this to extend to the off-shell K matrix that enters
here. There can also be subthreshold poles in $\cK_2$, given our particular definition~\cite{\BHSnum}.
Thus there is, in general, a barrier to converting the sum into an integral, and, for this reason,
the derivation of HS1 works only assuming the absence of singularities in $\cK_2$.
However, it has subsequently been understood that
by generalizing the definition of the PV prescription, one can define a class of two-particle K matrices,
and that by adjusting the parameters of the prescription, one can find definitions 
that are nonsingular for any given physical scattering amplitude~\cite{\largera}.
We assume henceforth that such a prescription has been used, 
and thus that $\wt\cD_{3,L}^\uu$ is an infinite-volume quantity.

\begin{figure}[tb]
\begin{center}
\vspace{-10pt}
\includegraphics[width=\textwidth]{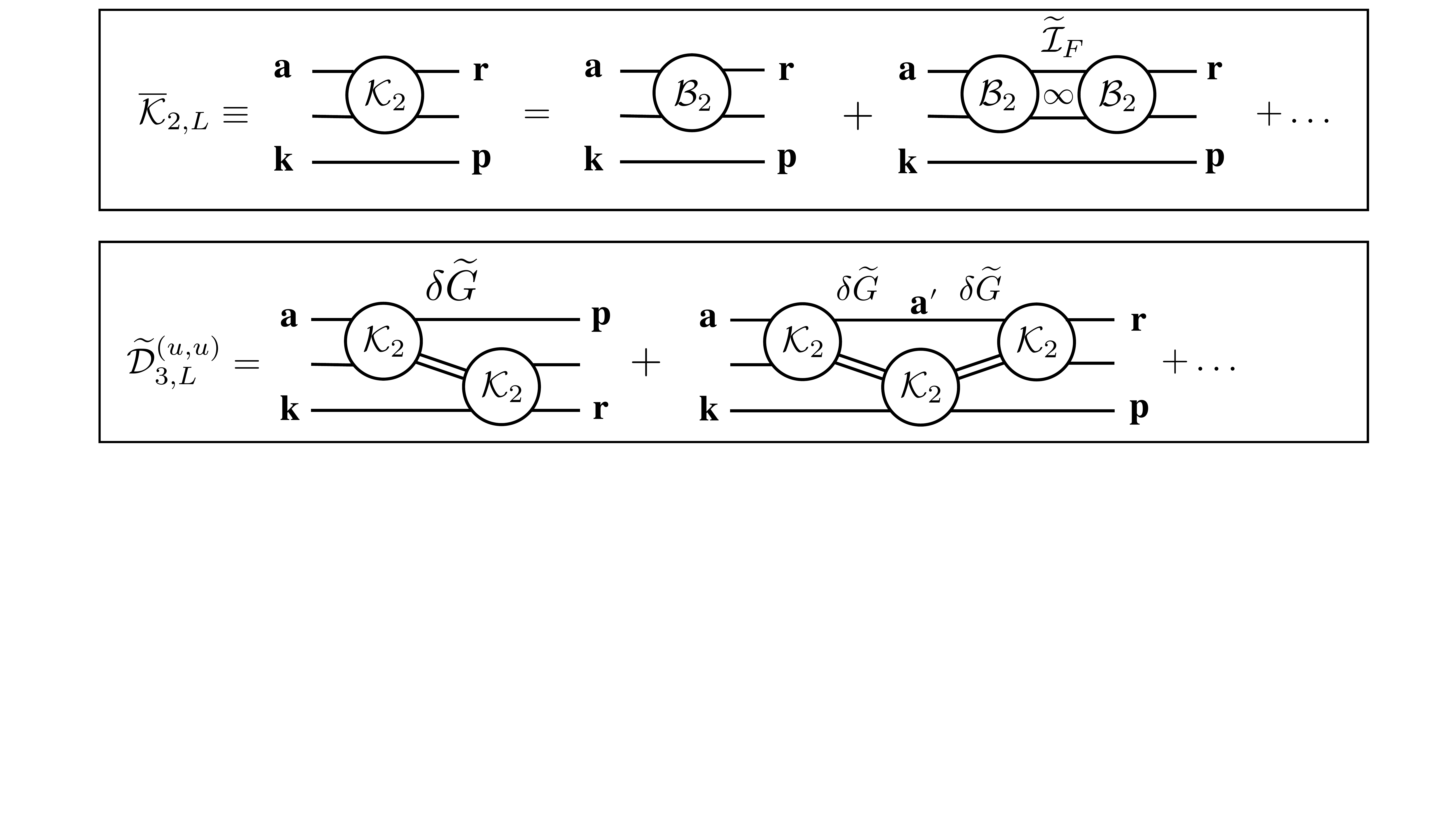}
\vspace{-1.8truein}
\caption{
Diagrams contributing to $\overline{\cK}_{2,L}$ and $\cD_{3,L}^\uu$. 
Notation as in Figs.~\ref{fig:Gcuts} and \ref{fig:Fcuts}.
Factors of $i$ are implicit.
\label{fig:K2_D}
}
\end{center}
\end{figure}

To combine these ingredients we use algebraic manipulations that recur frequently in this work, 
and which we derive in Appendix~\ref{app:algebra}.
These lead to
\begin{align}
i\Kalluu &= i\overline{\cK}_{2,L} + i\Kdft^\uu\,,
\label{eq:Kdf23Ldecomp}
\\
i\Kdft^\uu &\equiv i\wt\cD_{3,L}^\uu + \left(1 + i \wt\cD_{23,L}^\uu i\wt \cI_{FG} \right)
i\cB_3 
\frac1{1- \left(i\wt\cI_{FG} +i\wt\cI_{FG} i\wt \cD_{23,L}^\uu i\wt \cI_{FG}\right) i\cB_3} 
\left(1+ i\wt\cI_{FG} i \wt \cD_{23,L}^\uu \right)\,,
\label{eq:Kdf3uures}
\\
\wt \cI_{FG} &\equiv \wt\cI_F+\delta \wt G\,,
\\
i\wt\cD^\uu_{23,L} &\equiv i\overline{\cK}_{2,L} + i\wt \cD_{3,L}^\uu = i\overline{\cK}_{2,L} \frac1{1 - i \delta \wt G i\overline{\cK}_{2,L}} \,.
\end{align}
In $\Kdft^\uu$, there is an additional loop sum associated with each factor of $\wt\cI_{FG}$
adjacent to a $\wt\cD^\uu_{23,L}$,
with the summand including a factor of $\cK_2$. Using our generalized PV prescription,
 all such sums can be converted to integrals, and this absorbs all remaining factors of $2\omega L^3$.
Thus $\Kdft^\uu$ is an infinite-volume quantity.

Using similar expansions for $\wt{A}'^\oneu$, $\wt{A}^\oneu$, and $\delta C_{3,\infty}$,
we find that the factors of $L^3$ in 
$\overline{\cB}_{2,L}$, $\wt \cI_F$, and $\delta\wt G$ either cancel or can be used to convert sums 
into integrals, again assuming a PV prescription such that $\cK_2$ is smooth.
Thus these three kernels are also infinite-volume quantities.

\subsubsection{Summary}

We close this subsection by taking stock of what has been achieved.
We started from the closed-form expression for the three-particle correlator, Eq.~(\ref{eq:CL_TOPT}), which
is composed of infinite-volume amplitudes, but has the disadvantage that these amplitudes are off shell.
After some technical effort, which involved generalizing results from HS1 so that they applied to TOPT amplitudes, 
we obtained two simple equations, \Eqref{eq:Gsplit} and \Eqref{eq:Fsplit}, 
that allow the correlator to be expressed in terms of on-shell kernels,
as shown explicitly in Eq.~\Eqref{eq:CL_final_L}.
In a final step, we determined the volume (in)dependence of these kernels.
These steps lead to the following result for the correlation function,
 \begin{align}
	C_{3,L}(E,\vec{P}) = \wt C_{3,\iy} (E,\vec{P}) + 
	\wt{A}'^{\oneu} i(\wt{F}+\wt{G}) 
	\frac{1}{1-i\left(2\omega L^3\K_2 + \Kdft^\uu\right) i(\wt{F}+\wt{G}) } \wt{A}^{\oneu} \,, 
	\label{eq:CL_final}
\end{align}
where contributions with no $L$ dependence are collected into\footnote{%
Our notation (with the subscript $\infty$) is slightly misleading
because $\wt C_{3,\iy}$ is not the complete infinite-volume limit of $C_{3,L}$,
since the other term on the right-hand side of Eq.~\Eqref{eq:CL_final}, 
which contains all the volume dependence, 
has a nonvanishing infinite-volume limit.}
\begin{align}
	\wt C_{3,\iy} \equiv C_{3,\iy}^{(0)} + \delta C_{3,\iy} \,,
\end{align}
and we have introduced the diagonal matrix
\begin{equation}
\left[2\omega L^3\right]_{k\ell m;p\ell' m'} = \delta_{kp} \delta_{\ell \ell'} \delta_{m m'} \, 2\omega_kL^3\,.
\label{eq:2omL3}
\end{equation} 
All $L$ dependence is now explicit, 
entering through the quantities $\wt{F}$, $\wt{G}$, and $2\omega L^3$.

Our result can be compared to Eq.~(250) of HS1, rewritten to match our notation:
\begin{align}
C_{3,L}  &= C_{3,\infty} + A'  i F_3 \frac1{1- i \Kdf i F_3} A\,,
\label{eq:HSQC}
\\
F_3 &= \wt F \left[ \frac13 - \frac1{1/(2\omega L^3 \cK_2) + \wt F + \wt G} \, \wt F \right]\,.
\label{eq:F3}
\end{align}
This shows the trade-off that we have made: by using an asymmetric form of the three-particle K matrix,
our final expression is simpler, containing only the combination $\wt F + \wt G$ and no factors of $1/3$.
Another gain is that we have explicit expressions for all quantities in terms of the underlying TOPT amplitudes,
in contrast to HS1, where the definitions of the kernels are constructive and not explicit.



\subsection{New form of the quantization condition}
\label{sec:QC}

To make contact with the FV energy spectrum of the theory,
we exploit the fact that $C_{3,L}(E,\vec{P})$ has a simple pole whenever $E$ lies in the FV spectrum.
Since $\wt C_{3,\iy},\wt{A}'^\oneu,\wt{A}^\oneu$ are all smooth infinite-volume quantities, 
any singularity in $C_{3,L}$ must arise from the quantity lying between 
$\wt{A}'^\oneu$ and $\wt{A}^\oneu$ in Eq.~\eqref{eq:CL_final}.
This quantity is a matrix in the $\{k\ell m\}$ index space, 
and must have a diverging eigenvalue for $C_{3,L}$ to have a pole.
Equivalently, the determinant of its inverse should vanish,
\begin{align}
	\det\left[\wt{F}+\wt{G}\right]^{-1} \det\left[1 - i\left(2\omega L^3\K_2 + \Kdft^\uu\right) i(\wt{F}+\wt{G}) \right] = 0 \,.
\end{align}
The energies where $\det[\wt{F}+\wt{G}]^{-1}=0$ are the free three-particle energies where 
$E=\omega_k+\omega_a+\omega_{b_{ka}}$ for some choice of FV momenta $\vec{k},\vec{a}\in\frac{2\pi}{L}\Z^3$.
For general $\cK_2$ and $\Kdft^\uu$, we expect that the product of the two determinants will not
vanish at these energies, because the second determinant will diverge.\footnote{%
As is well known from numerical investigations, if one truncates the partial-wave expansions
of  $\cK_2$ and $\Kdft^\uu$, then there will be solutions to the quantization condition 
at free energies~\cite{\BHSnum,\dwave,\largera}.}
Physically this corresponds to the fact that a general interaction will shift all FV energies from their free values.
We therefore conclude that for a given $\vec{P}$, an energy $E$ can only be in the finite-volume spectrum of the interacting theory if
\begin{align}
	\det\left[1 + \left(2\omega L^3\K_2 + \Kdft^\uu\right) (\wt{F}+\wt{G}) \right] = 0 \,.
	 \label{eq:QC_Ft}
\end{align}
This is our alternate form of the three-particle quantization condition.

This result has a superficially similar form to that from HS1, which follows from Eq.~\Eqref{eq:HSQC},
\begin{equation}
\det \left[1 + \Kdf F_3 \right] = 0\,,
\label{eq:QCHS1}
\end{equation}
but many of the details are different.
For example, in Eq.~\eqref{eq:QC_Ft}, the infinite-volume K matrices appear together and separate from the FV quantities $\wt F$ and $\wt G$, whereas $F_3$ in Eq.~\eqref{eq:QCHS1} is a relatively complicated function of $\cK_2$, $\wt F$, and $\wt G$.}
We return to the relation between the two approaches in Sec.~\ref{sec:relation}.

\section{TOPT expression for $\cM_{3,L}$}
\label{sec:M3L}


In order to understand the relation between quantization conditions, we need to first extend the
developments of the previous section from the correlator $C_{3,L}$ to the finite-volume $3\to3$ amplitude
$\cM_{3,L}$. This extension also allows us to determine the infinite-volume 
relation between our asymmetric K matrix $\Kdfuu$ and the full $3\to3$ amplitude $\cM_3$.
This latter relation is somewhat off the main line of development of this paper, so we relegate it
to Appendix~\ref{app:inteqs}.

\begin{figure}[tb]
\begin{center}
\vspace{-10pt}
\includegraphics[width=\textwidth]{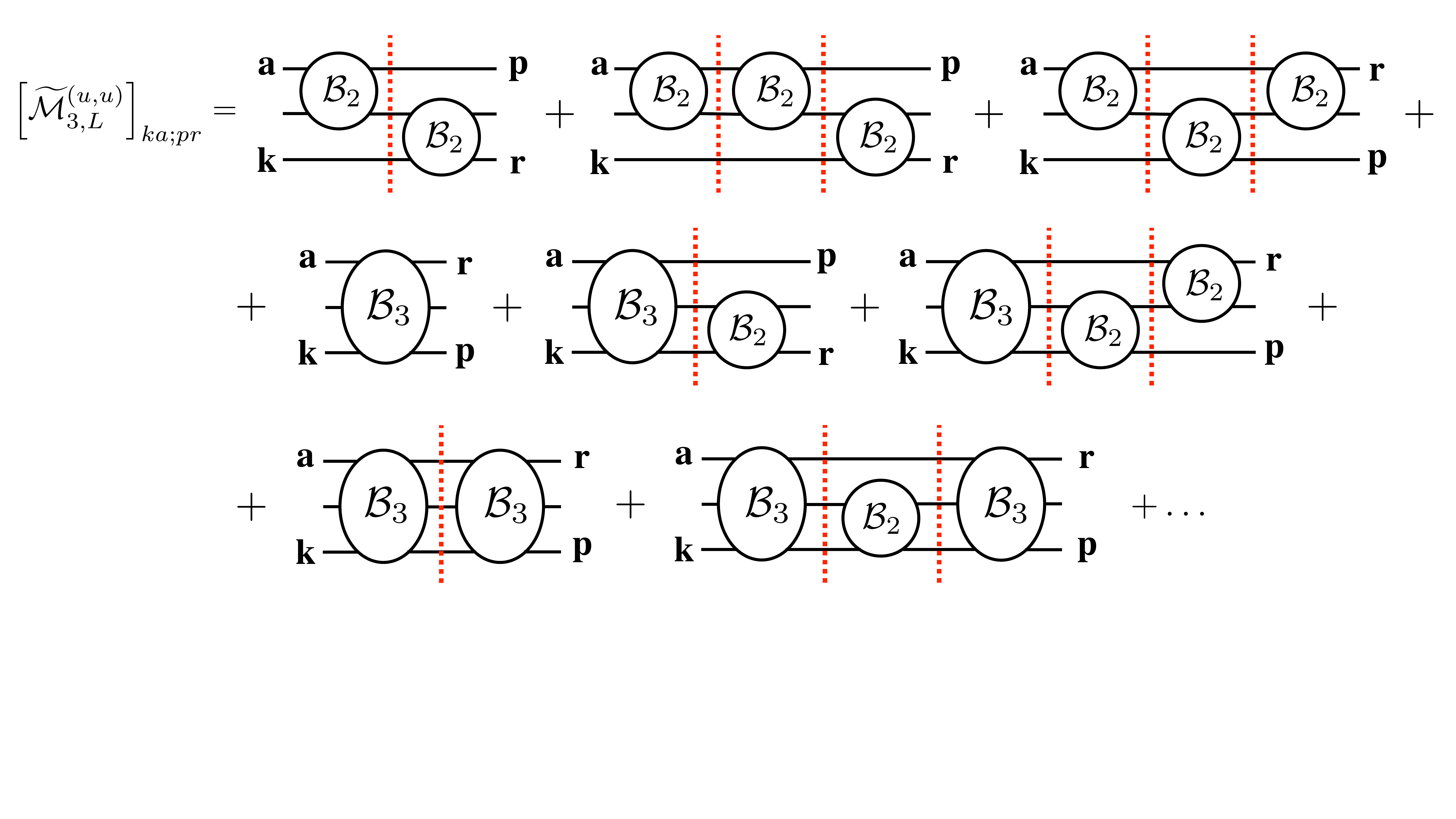}
\vspace{-1.2truein}
\caption{
Diagrams contributing to $\cM_{3,L}^\uu$ in TOPT.
Notation as in Fig.~\ref{fig:topt1}.
The asymmetric feature of this amplitude is that the momenta $\vec k$ and $\vec p$ are always
assigned to a spectator line, if one is present.
\label{fig:M3Luu}
}
\end{center}
\end{figure}

$\cM_{3,L}$ is defined as the amputated, connected, $3\to3$ finite-volume amplitude.
It is in general off shell, and thus a matrix in $\{ka\}$ space. 
It is simpler to begin by considering an asymmetric version of the amplitude, $\wt \cM_{3,L}^\uu$, 
defined so that, if there is an external factor of $\overline{\cB}_{2,L}$, the spectator propagator
is always labeled with one of the external momenta (typically called $\vec k$ or $\vec p$).
This definition is illustrated in Fig.~\ref{fig:M3Luu}.\footnote{%
We include a tilde on $\wt \cM_{3,L}^\uu$ since it is different from the similar quantity $\cM_{3,L}^\uu$
defined in HS2, with the latter having an asymmetry based on the Feynman skeleton expansion.
We stress, however, that the symmetrized version $\cM_{3,L}$ is the same as in HS2 (when evaluated
on shell).}
As we have seen several times above, results are simplified if we combine the 
asymmetric three-particle amplitude with the corresponding two-particle quantity, here $\overline{\cM}_{2,L}$.
(This is defined in Eq.~\Eqref{eq:M2L}, and  is simply $\cM_{2,L}$ packaged with an inverse spectator propagator.)
The TOPT result for this combination is a geometric series,
\begin{equation}
i\left(\overline{\cM}_{2,L} + \wt\cM_{3,L}^\uu\right)
= 
 \frac{1}{1 - i( \overline{\cB}_{2,L}+ \cB_3) i(D_F + D_G)} i( \overline{\cB}_{2,L}+ \cB_3)\,.
 \label{eq:M23uu}
 \end{equation}
 To obtain the full $\cM_{3,L}$ we symmetrize by summing over the different attachments of the external momenta
 \begin{equation}
\left[ \cM_{3,L}\right]_{ka; k' a'} =
\sum_{x\in P} \sum_{x' \in P'}\left[ \wt\cM_{3,L}^\uu\right]_{x;x'} \,,
\qquad
P=\{ka,ab,bk\}\,,\ \
P'=\{k'a',a'b',b'k'\}\,.
\label{eq:symm}
\end{equation}
Only three terms are needed on both sides (rather than the $3!$ one might have expected)
because $\cB_2$ is symmetric under $a\leftrightarrow b$ interchange.
No overall factor of $1/9$ is needed because this factor is built into $\cB_3$---see the discussion above 
Eq.~\Eqref{eq:C3symmetry}. 

The remaining steps are essentially a repeat of those we used for $C_{3,L}$. 
We find that Eq.~\Eqref{eq:CL_final_L} is replaced by\footnote{
Strictly speaking, this way of writing the result only holds if all quantities are on shell, so that
all matrices are square. Indeed, we have only considered above the on-shell form of $\Kalluu$.
However, its definition, given in Eqs.~\Eqref{eq:Kdf23Ldecomp} and \Eqref{eq:Kdf3uures}, 
can be extended off shell.
The same is true for the result here, if one expands out the geometric series and evaluates it term by term,
and also for Eq.~(\ref{eq:M3Lb}).
These extensions are convenient, but not essential for the following discussion. }
\begin{equation}
i \left(\overline{\cM}_{2,L} + \wt \cM_{3,L}^\uu\right) = i\Kalluu \frac1{1 - i (\wt F+\wt G) \, i \Kalluu}
\,,
\label{eq:M23LtoKdf}
\end{equation}
Using the algebraic result (\ref{eq:appalgfinal}), as well as
the decomposition of $\Kalluu$, Eq.~\Eqref{eq:Kdf23Ldecomp},
and the result  for $\overline{\cM}_{2,L}$, Eq.~\Eqref{eq:M2LresF},
we can extract the expression for $\wt \cM_{3,L}^\uu$:
\begin{align}
i\wt\cM_{3,L}^\uu &= i\cD_L^\uu + i\wt\cM_{\df,3,L}^\uu
\label{eq:M3Lb}
\end{align}
where
\begin{align}
i\cD^{(u,u)}_L &\equiv i\overline{\cM}_{2,L} i\wt G i\overline{\cM}_{2,L}\frac1{1-i \wt G i\overline{\cM}_{2,L} }\,,
\label{eq:DLuu}
\\
i\wt\cM_{\df,3,L}^\uu &\equiv
\left[1 + i \cD^\uu_{23,L} i(\wt F+\wt G)\right]
i\wt\cT^\uu_L 
\left[1 + i(\wt F+\wt G) i \cD_{23,L}^\uu \right]\,,
\label{eq:Mdf3Lb}
\end{align}
with
\begin{align}
i\wt\cT^\uu_L &\equiv i\Kdfuu  \frac1{1- \left[1 +i (\wt F+ \wt G) i \cD_{23,L}^\uu\right]  i(\wt F+\wt G) i\Kdfuu} \,,
\label{eq:Tuu}
\\
i\cD^\uu_{23,L} &\equiv i\overline{\cM}_{2,L} +  i \cD_L^\uu  
=
i\overline{\cM}_{2,L} \frac1{1-i \wt G i\overline{\cM}_{2,L} }
\,.
\label{eq:D23Luu}
\end{align}
Here $\cD_L^\uu$ is the same as the quantity of the same name appearing in HS2, since
the asymmetry arises from the external $\overline{\cM}_{2,L}$, which is the same in both approaches.\footnote{%
This equality holds only on shell, which is all that we require in the following subsection.}
These results can also be expressed in terms of $\overline{\cK}_{2,L}$ instead of $\overline{\cM}_{2,L}$,
\begin{align}
 i\wt\cM_{\df,3,L}^\uu &=
\frac{1}{1 - i\overline{\cK}_{2,L}i(\wt F + \wt G)}
i\Kdfuu  \frac1{1 -  i(\wt F+\wt G)\frac{1}{1 - i\overline{\cK}_{2,L}i(\wt F + \wt G)} i\Kdfuu} 
\frac{1}{1 - i(\wt F + \wt G)i\overline{\cK}_{2,L}}\,,
\\
i\cD^\uu_{23,L} &=
 \frac{1}{1 - i\overline{\cK}_{2,L}i(\wt F + \wt G)}i\overline\cK_{2,L}\,.
\end{align}
These forms are used in our companion paper~\cite{BS2}.

The results (\ref{eq:M3Lb})-(\ref{eq:D23Luu}) 
make all volume dependence of $\wt\cM_{3,L}^\uu$ explicit: it enters
through $\wt F$, $\wt G$, and $\overline{\cM}_{2,L}$. 
We also note that this result holds both for the off-shell amplitude and its on-shell limit.

We end this subsection with a side remark.
As was pointed out in HS2, the quantization condition can be obtained from the off-shell
$\wt\cM_{3,L}^\uu$ instead of $C_{3,L}$, since the former is an (amputated) three-particle correlator.
This is made particularly clear by the fact that,
by comparing Eqs.~\Eqref{eq:CL_TOPT} and (\ref{eq:M23uu}),
we can explicitly relate the two quantities:
\begin{equation}
C_{3,L} -C_{3,\infty}^{(0)} = \widehat{A}' i (D_F+D_G) \widehat{A}
+ 
\widehat{A}' i (D_F+D_G) i \left(\overline{\cM}_{2,L} + \wt\cM_{3,L}^\uu\right) i (D_F+D_G)  \widehat{A}\,.
\end{equation}

\section{Relation to quantization condition of HS1}
\label{sec:relation}



In this section we show that  our new quantization condition, Eq.~\Eqref{eq:QC_Ft},
can be rewritten in the HS1 form of Eq.~\Eqref{eq:QCHS1},
and  that the two approaches therefore lead to equivalent results.
We refer to this transformation as the ``symmetrization'' of the quantization condition, since
the HS1 form is written in terms of a symmetric three-particle K matrix.
As a side benefit, we obtain the algebraic relation between
our asymmetric amplitude $\Kdft^\uu$ and the symmetric quantity of HS1, $\Kdf$.

\subsection{Recap of result for $\cM_{3,L}^\uu$ from HS2}
\label{sec:M3LuuHS}

The connection to the HS1 QC is provided by studying the result for $\cM_{3,L}^\uu$,
the asymmetric finite-volume three-particle amplitude introduced in HS2.
This is defined as for our $\wt \cM_{3,L}^\uu$, except that its
asymmetry is based on the skeleton expansion in terms of 2PIs and 3PIs
Bethe-Salpeter (B-S) kernels built from Feynman diagrams.
Specifically, if the external legs connect to a 2PIs B-S kernel, then the spectator propagator
associated with that kernel
is connected to the spectator momentum of $\cM_{3,L}^\uu$. 
This is the analog of our definition of $\wt \cM_{3,L}^\uu$ (see Fig.~\ref{fig:M3Luu})
except that we use an expansion in terms of TOPT amplitudes, 
and for us $\cB_2$ plays the role of the 2PIs B-S kernel.
Since $\cB_2$ contains only a subset of the time orderings that contribute to the 2PIs B-S kernel,
more contributions are symmetrized in $\wt \cM_{3,L}^\uu$ than in $\cM_{3,L}^\uu$.
We stress, however, that, after complete symmetrization, both objects lead to the same
amplitude, $\cM_{3,L}$.
This is an example of the fact that there are many different ways to define asymmetric amplitudes,
all of which symmetrize to the same quantity.

The result for $\cM_{3,L}^\uu$ is given in Eq.~(67) of HS2. 
Converting the expressions to our notation, we have
\begin{align}
i\cM_{3,L}^\uu &=
 i\cD_L^\uu + \big(\cL_L^{(u)}\big) \big(i\Kdf\big)
\frac1{1 - \big(iF_3\big) \big( i\Kdf\big)}
\big(\cR_L^{(u)}\big)\,,
\label{eq:M3LuuHS}
\end{align}
where
\begin{align}
\big(\cL_L^{(u)}\big)  &= \big(1\ 0\big) + i \cD_{23,L}^\uu i\wt F \big(1\ 1\big)\,,
\label{eq:LuHS}
\\
\big(i\Kdf\big) &= 
\begin{pmatrix} i\KdfuuHS & i\cK_{\df,3}^{(u,s+\tilde s)} \\
i\cK_{\df,3}^{(s+\tilde s,u)} & i\cK_{\df,3}^{(s+\tilde s,s+\tilde s)} \end{pmatrix} \,,
\label{eq:KdfHS}
\\
\big( iF_3 \big) &= \begin{pmatrix} 1 \\ 1 \end{pmatrix}
i F_3
\big(1\ 1\big)
=
\begin{pmatrix} 1 \\ 1 \end{pmatrix}
[1/3 + i\wt F \cD_{23,L}^\uu] i\wt F
\big(1\ 1\big)
\,,
\label{eq:F3HS}
\\
\big(\cR_L^{(u)}\big) &= \begin{pmatrix}1\\ 0 \end{pmatrix}
+\begin{pmatrix}1\\1\end{pmatrix}
i\wt F i \cD_{23,L}^\uu\,.
\label{eq:RuHS}
\end{align}
Here $\cD_L^\uu$ and $\cD_{23,L}^\uu$, 
defined in Eqs.~\Eqref{eq:DLuu} and \Eqref{eq:D23Luu}, respectively,
are the same as in our result for $\wt\cM_{3,L}^\uu$, Eq.~\Eqref{eq:M3Lb}.
We note that the expression for $F_3$ given on the right-hand side of Eq.~(\ref{eq:F3HS}) 
is an alternative way of writing the earlier form, Eq.~\Eqref{eq:F3}.

The quantities in round braces in
Eqs.~(\ref{eq:M3LuuHS})-(\ref{eq:RuHS}) live in a two-dimensional space: 
$(\cL_{L}^\oneu)$ being a row vector, $(i\Kdf)$ and $(iF_3)$ being matrices,
and $(\cR_L^\oneu)$ being a column vector.
The $1$ in the denominator of Eq.~(\ref{eq:M3LuuHS} indicates the identity matrix in this space.
The indices in this space are $\oneu$ and $(s+\tilde s)$,
as exemplified by the expression for $(i\Kdf)$ in Eq.~(\ref{eq:KdfHS}).
These indices were introduced in HS1 to denote the different ways in which
the spectator-momentum label is attached to diagrams.
The precise definition is given in Appendix~\ref{app:asymkernel}. 
As an example of this matrix notation, the symmetrized $\Kdf$ (which is
differentiated from the matrix version by the absence of surrounding parentheses) is given by
\begin{equation}
i\Kdf \equiv \begin{pmatrix}1 \ 1 \end{pmatrix} 
\big( i\Kdf \big) \begin{pmatrix}1 \\ 1 \end{pmatrix}
=
i\KdfuuHS + i\cK_{\df,3}^{(u,s+\tilde s)} +
i\cK_{\df,3}^{(s+\tilde s,u)} + i\cK_{\df,3}^{(s+\tilde s,s+\tilde s)}\,.
\label{eq:symmKdf}
\end{equation}
This is the quantity that appears in the quantization condition of HS1, Eq.~(\ref{eq:HSQC}).

A noteworthy feature of the result (\ref{eq:M3LuuHS}) is that
$\cM_{3,L}^\uu$ is not given in terms only of the symmetrized $\Kdf$.
This is because of the $(1 \ 0)$ term in $(\cL_L^\oneu)$,
and the corresponding term in $(\cR_L^\oneu)$,
which project onto asymmetric components of $(\Kdf)$.
If one symmetrizes, and considers $\cM_{3,L}$,
then it is possible to write the result in terms of the symmetric $\Kdf$,
as shown by Eq.~(68) of HS2. 

The dependence of $\cM_{3,L}^\uu$ brings up a potential conflict.
On the one hand, 
we expect that we can obtain the HS1 quantization condition
from $\cM_{3,L}^\uu$, since any three-particle correlation 
function should have poles at the spectral energies.
We know that the resulting quantization condition, Eq.~(\ref{eq:HSQC}), 
contains the symmetric $\Kdf$.
On the other hand,
$\cM_{3,L}^\uu$ depends also on asymmetric components of $\Kdf$, as just noted.
The resolution is that the central portion of the second term on the right-hand side of
Eq.~(\ref{eq:M3LuuHS}) can be rewritten as
\begin{align}
\big(i\Kdf\big) \frac1{1 - \big(iF_3\big) \big( i\Kdf\big)}
&=
\big(i\Kdf\big) + 
\big(i\Kdf\big)  \frac1{1 - \big(iF_3\big) \big( i\Kdf\big)} \big(iF_3\big)\big(i\Kdf\big) 
\label{eq:QCc}
\\
&=
\big(i\Kdf\big) + 
\big(i\Kdf\big) \begin{pmatrix} 1 \\ 1 \end{pmatrix}  
\frac1{1 - iF_3 i\Kdf} iF_3 \begin{pmatrix}1 \ 1 \end{pmatrix} \big(i\Kdf\big) \,.
\label{eq:QCd}
\end{align}
This shows that the geometric series leading to the poles does contain the symmetric $\Kdf$.

One final technical point needs to be mentioned. In HS1 and HS2, the versions of 
$\wt F$ and $\wt G$ differ from those used here in three ways:
(i) they use a different boost to the interacting pair CMF (which only changes $\wt G$);
(ii) $\wt F$ uses the original PV prescription rather than the generalized one used here;
and (iii) $\wt G$ is defined with the nonrelativistic energy denominator.
However, the derivations of HS1 and HS2 go through essentially unchanged if
one uses our versions of $\wt F$ and $\wt G$, and, in particular, the expressions given 
in Eqs.~(\ref{eq:M3LuuHS})-(\ref{eq:RuHS}) remain valid.


\subsection{Asymmetrizing $\cM_{3,L}^\uu$}
\label{sec:KtoK}

The expressions for $\wt\cM_{3,L}^\uu$ and $\cM_{3,L}^\uu$,
given in Eqs.~\Eqref{eq:M3Lb} and (\ref{eq:M3LuuHS}), 
have a similar structure, but differ in many details.
In this subsection we bring the almost symmetric result for $\cM_{3,L}^\uu$ into an asymmetric form
similar to that of $\wt\cM_{3,L}^\uu$.

In Appendix~\ref{app:asym} we derive the following three ``asymmetrization'' identities 
(valid up to exponentially suppressed corrections, 
and for both the Wu boost and the boost of HS1)
\begin{align}
X^\oneu \wt F \left( 1\ 1\right)  &= X^\oneu (\wt F+ \wt G -
\overrightarrow{\cI}_G  )\left(1\ 0\right) 
\label{eq:subL}
\\
\begin{pmatrix}1\\1\end{pmatrix}  \wt F  X^\oneu 
&=
\begin{pmatrix}1 \\ 0 \end{pmatrix} (\wt F+\wt G - \overleftarrow{\cI}_G  )X^\oneu 
\,,
\label{eq:subR}
\\
\begin{pmatrix}1\\1\end{pmatrix} \frac{\wt F}3  \left(1\ 1\right)
&=
\begin{pmatrix}1\\0\end{pmatrix}  (\wt F+\wt G + \otimes_G )  \left(1\ 0\right)\,.
\label{eq:subLR}
\end{align}
where $X^\oneu$ is a generic asymmetric amplitude, e.g. $\overline{\cM}_{2,L}$ or $\cD_{23,L}^\uu$,
and there is an implicit matrix of amplitudes such as $(\Kdf)$ on the right of Eq.~\Eqref{eq:subL},
on the left of Eq.~\Eqref{eq:subR}, and on both sides of Eq.~\Eqref{eq:subLR}.
The integral operators $\overrightarrow{\cI}_G$, $\overleftarrow{\cI}_G$,  and $\otimes_G$ are defined
in the appendix. The first two are similar to $\wt\cI_F$,  but their action is directional, as indicated by the arrows.
The effect of all three operators is to sew together the adjacent amplitudes leading to new infinite-volume quantities.

Using the three identities, we can rewrite the expression for $\cM_{3,L}^\uu$, Eq.~(\ref{eq:M3LuuHS}), 
solely in terms of $\KdfuuHS$:
\begin{align}
i\cM_{\df,3,L}^\uu &\equiv i\cM_{3,L}^\uu - i \cD_L^\uu
\\
&= \left[1 + i \cD^\uu_{23,L} i(\wt F+\wt G - \overrightarrow{\cI}_G) \right]
i \cT_L^\uu
\big[1 + i (\wt F+ \wt G - \overleftarrow{\cI}_G)i \cD^\uu_{23,L}  \big]\,,
\label{eq:Mdf3Luu}
\\
i \cT_L^\uu &= i\KdfuuHS \frac1{1 - \left[ i (\wt F+\wt G+\otimes_G)
+ i (\wt F+\wt G - \overleftarrow{\cI}_G)i \cD_{23,L}^\uu i(\wt F+\wt G - \overrightarrow{\cI}_G) \right] i\KdfuuHS} \,.
\label{eq:TLuu}
\end{align}
In Appendix~\ref{app:KtoK}, we show that this result can be reorganized into 
\begin{equation}
i \cM_{\df,3,L}^\uu = 
\left[1 + i \cD^\uu_{23,L} i(\wt F+\wt G)\right]
i\Kdfuun  \frac1{1- \left[1 +i (\wt F+ \wt G) i \cD_{23,L}^\uu\right]  i(\wt F+\wt G) i\Kdfuun} 
\left[1 + i(\wt F+\wt G) i \cD_{23,L}^\uu \right]
\,,
\label{EQ:M3LC}
\end{equation}
where the primed version of the HS1 asymmetric amplitude is
\begin{equation}
i\Kdfuun \equiv
\left(1 - i\overline{\cK}_2 i\overrightarrow{\cI}_G \right)
i\KdfuuHS 
\frac1{1 -  
\left[i\otimes_G + i\overleftarrow{\cI}_G i\overline{\cK}_2 i\overrightarrow{\cI}_G\right]
 i\KdfuuHS} 
\left(1- i\overleftarrow{\cI}_G i\overline{\cK}_2\right)\,.
\label{EQ:KTOKON2}
\end{equation}
We observe that the form of Eq.~\Eqref{EQ:M3LC} is identical to that of the result for
$\wt\cM_{3,L}^\uu$, Eq.~\Eqref{eq:Mdf3Lb}, with $\Kdfuun$ playing the role of $\Kdfuu$.

An interesting implication of this result is that the HS1 quantization condition can be rewritten
in the form derived here, Eq.~\Eqref{eq:QC_Ft}, but with $\Kdfuu$ replaced by $\Kdfuun$:
\begin{equation}
	\det\left[1 + \left(2\omega L^3\K_2 + \Kdfuun\right) (\wt{F}+\wt{G}) \right] = 0 \,.
	 \label{eq:QC_Ftn}
\end{equation}
To show this, we use the result, noted above, that the quantization condition can be derived
from the poles in $\cM_{3,L}^\uu$. 
To obtain an expression for $\cM_{3,L}^\uu$,
we start from Eq.~(\ref{EQ:M3LC}), and reverse the steps leading from Eq.~\Eqref{eq:M23LtoKdf} to
Eq.~\Eqref{eq:Mdf3Lb}, obtaining
\begin{equation}
i \left(\overline{\cM}_{2,L} + \cM_{3,L}^\uu\right) = i (2\omega L^3 \cK_2 + \Kdfuun)
\frac1{1 - i (\wt F+\wt G) \, i (2\omega L^3 \cK_2 + \Kdfuun)}
\,,
\end{equation}
the denominator of which leads immediately to the quantization condition Eq.~\eqref{eq:QC_Ftn}.

We now have two quantization conditions of exactly the same form, Eqs.~(\ref{eq:QC_Ft})
and (\ref{eq:QC_Ftn}), but containing different asymmetric three-particle K matrices, 
$\Kdfuu$ and $\Kdfuun$ respectively. This does not, however, imply that these two K matrices are the same.
One way of seeing this is to note that asymmetrization is not unique:
there are many ways to divide a symmetric amplitude into asymmetric components. 
This is because, for a given asymmetric diagram (in either the Feynman or TOPT approach),
one can choose to assign it directly to the asymmetric amplitude, or to first symmetrize and then assign.
When using the identities \Eqref{eq:subL}-\Eqref{eq:subLR}, the left-hand sides involve only the 
symmetric part of $\Kdf$, while the right-hand sides involve only the $(u)$ parts. Since the latter are
ambiguous, the identity must be satisfied for all possible choices of asymmetric amplitude.
In other words, the operators appearing on the right-hand sides, e.g.~$\wt F+\wt G-\overrightarrow{\cI}_G$,
must have (an infinite number of) zero modes. These observations do not impact the derivation just given,
in which we {\em choose} a particular asymmetrization. However, they imply that we could have made
another choice, in which case the resulting $\Kdfuun$ would have been different while the form of the 
resulting quantization condition, Eq.~(\ref{eq:QC_Ftn}), would have been unchanged.

\subsection{Symmetric form of the new quantization condition}
\label{sec:symmQC}

Having understood how asymmetrization turns the HS1 quantization condition into our new form, we now
follow the inverse path and bring our quantization condition into HS1 form. 

What we need to do is to rewrite our result for $\wt \cM_{\df,3,L}^\uu$, Eq.~(\ref{eq:Mdf3Lb}),
in the form given in Eq.~\Eqref{eq:Mdf3Luu}, for then we can use the asymmetrization identities in reverse
and obtain the HS2 form, Eq.~\Eqref{eq:M3LuuHS}.
In order to follow these steps we must first invert Eq.~(\ref{EQ:KTOKON2}).
We can do so by discretizing momentum space so that all relations are matrix
equations. Then we obtain
\begin{align}
i\Kdfuunn &= i \wt Z^\uu
\frac1{1 + 
\left[i\otimes_G + i\overleftarrow{\cI}_G i\overline{\cK}_2 i\overrightarrow{\cI}_G\right]
i \wt Z^\uu}\,,
 \\
 i \wt Z^\uu &\equiv \frac1{1-i \overline{\cK}_2 i \overrightarrow{\cI}_G}
i \Kdfuu \frac1{1- i \overleftarrow{\cI}_G \overline{\cK}_2}\,.
\end{align}
The next step is to obtain the other components of this new version of the K matrix,
namely $\wt\cK_{\df,3}^{\prime(s+\tilde s,u)}$, etc.
This is done using relations from HS1, which are recalled in Appendix~\ref{app:asym}.
Then the steps above lead to the analog of Eq.~(\ref{eq:M3LuuHS}), 
\begin{align}
i\wt\cM_{\df,3,L}^\uu &=
  \big(\cL_L^{(u)}\big) \big(i\wt\cK_{\df,3}'\big)
\frac1{1 - \big(iF_3\big) \big( i\wt \cK_{\df,3}'\big)}
\big(\cR_L^{(u)}\big)\,,
\label{eq:M3Luun}
\end{align}
with $(\cL^\oneu_L)$, $(F_3)$, and $(\cR^\oneu_L)$ unchanged from above.
Finally, we can use the arguments at the end of Sec.~\ref{sec:M3LuuHS} to determine the
quantization condition from $\wt\cM_{\df,3,L}^\uu$, obtaining
\begin{equation}
\det\left[1 + \wt \cK_{\df,3}' F_3\right]\,,
\label{eq:QCnn}
\end{equation}
where the symmetrized K matrix is
\begin{equation}
\wt\cK_{\df,3}'
=
\wt\cK_{\df,3}^{\prime(u+s+\tilde s,u+s+\tilde s)} \,.
\end{equation}
Thus we find that the symmetrized version of our new quantization condition is of exactly the same form
as that of HS1.

We now argue that the 
symmetrized K matrix obtained here and that of HS1 are the same, i.e.\footnote{%
We stress that this result will hold only if the same choice of boost in $\wt G$ 
and cutoff function $H(\vec k)$
is used in both cases.}
\begin{equation}
\wt\cK_{\df,3}'  = \Kdf\,.
\label{eq:equivalence}
\end{equation}
To do so, we use the result from HS2 that, 
by symmetrizing the infinite-volume limit of Eq.~(\ref{eq:M3LuuHS}),
the full $\cM_3$ can be written in terms of integral equations containing $\Kdf$ alone---the 
asymmetric form $\KdfuuHS$ is not needed, unlike for $\cM_{\df,3}^\uu$. 
Similarly, by symmetrizing the infinite-volume limit of Eq.~(\ref{eq:M3Luun}), 
$\cM_3$ is given by {\em exactly the same expression} in terms of $\wt \cK_{\df,3}'$.
Now we assume that this relation is invertible and one-to-one.
If so, the two symmetrized K matrices must be equal.
Another way of stating this claim is that we are effectively assuming that there are no
redundant parts of the symmetrized K matrices, which we view as plausible given the explicit
construction presented here. This is in contrast to the asymmetrized K matrices, which we know are
ambiguous.

The equality of $\wt\cK_{\df,3}'$ and $\Kdf$ is also consistent with the quantization condition
having the same form in both cases. However, this is not a sufficient argument to demonstrate equality
of the K matrices, since the asymmetric form of the quantization conditions also agree, and yet we know
that the asymmetric K matrices differ.

The result (\ref{eq:equivalence}) connects the Feynman-diagram-based method of HS1/HS2 and the
TOPT approach followed here. It also provides an explicit expression for $\Kdf$ that goes beyond
the constructive definition given in HS1. 

We close by emphasizing two points.
First, we stress that the steps leading to the result (\ref{eq:M3Luun}) for $\wt \cM_{\df,3,L}^\uu$
do not depend on the derivations of HS1 and HS2. 
While we have made use of results from these works as motivation
for the logical progression of our approach, in the end the steps needed are simply algebraic.
Second, although the symmetric K matrices are the same, this does not mean that $\cM_{3,L}^\uu$
and $\wt\cM_{3,L}^\uu$ are the same, because they depend also on $\KdfuuHS$ and $\Kdft^\uu$,
respectively, and these differ. This is an important consistency check, as we know that 
$\cM_{3,L}^\uu$ and $\wt \cM_{3,L}^\uu$ are in fact different.

\section{Conclusions}
\label{sec:conc}


In this paper we provide a more direct and explicit path to the relativistic, model-independent,
three-particle quantization condition of HS1~\cite{\HSQCa}.
Although it is reassuring to check the result of the complicated and lengthy derivation of HS1, this is not
our fundamental motivation.
Instead, we expect that our method will simplify the generalization 
to nondegenerate particles, and hope to present results for this shortly.
This is a complementary generalization to that achieved recently
in Ref.~\cite{\isospin}, which considers degenerate, but  potentially distinguishable,
spinless particles. We expect it to be profitable to combine the two approaches.

As part of our derivation, we have shown that the three-particle quantization condition can be written in a simpler
form in terms of asymmetric amplitudes 
[see Eqs.~\Eqref{eq:QC_Ft} and \Eqref{eq:QC_Ftn}].
We do not necessarily
propose this as a practical alternative to the HS1 form, because the asymmetric amplitudes
will require, at any order in the threshold expansion of Refs.~\cite{\BHSnum,\dwave},
a larger number of parameters for a general description than the symmetric form appearing 
in the HS1 quantization condition.\footnote{%
This can be seen explicitly in chiral perturbation theory
from the leading-order result for $\KdfuuHS$.}
The parametrizations must therefore be redundant, a result that is presumably related to the
ambiguity in defining an asymmetric amplitude.
However, we do expect that the new, asymmetric form of the quantization condition has
theoretical implications.  Indeed, in a companion paper we show that it allows one to
derive a form of the quantization condition in terms of the R matrix of Refs.~\cite{\Maiisobar,\isobar}
and thus obtain a generalization of the result of the FVU approach to all partial waves~\cite{BS2}.

A more distant goal of our approach is to allow the generalization to more than three particles.
In this regard we note that the complications associated with the possibility of on-shell intermediate
states in three-particle scattering, which led in HS1 to the introduction of the divergence-free three-particle
K matrix, $\Kdf$, are dealt with very simply and automatically in the TOPT approach used here.
This gives us some hope that the additional complications that arise with more than three particles
will be manageable.
In this regard, the alternative approach for dealing with F cuts that is sketched in Appendix~\ref{app:opt1}
may be helpful as it does not require a choice of PV prescription.

An issue that we have commented on only peripherally in the text is whether the various three-particle K matrices
that we have considered are Lorentz invariant.
This is not relevant to the derivation, 
but is important for practical applications, since invariance restricts the number of terms that contribute.
We summarize the status here. First, we stress that all the quantization
conditions presented here---both the original HS1 form and our asymmetric form, and for both choices
of boost---hold for all kinematically allowed choices of $P^\mu=(E, \vec P)$. 
In particular they hold for choices of $P^\mu$ for which the three particles are relativistic, which is why
we call the quantization conditions relativistic.
This is, however, a separate issue from the relativistic invariance of the K matrices.
Here the status is that, if we define $\wt G$ using the boost of HS1 and the relativistic form of the pole (which is
possible for both the HS1 K matrices and the new TOPT versions introduced here), then
the original symmetric $\Kdf$ and the asymmetric version 
$ \cK_{\df,3}'^\uu$ are both Lorentz invariant.\footnote{%
The latter result follows because $\cK_{\df,3}'^\uu$ is related to the Lorentz invariant
quantity $\cM_3^\uu$ by integral equations of the same form as given in Appendix~\ref{app:inteqs}, 
which are boost invariant.}
These appear, respectively, in the original HS1 quantization condition, 
Eq.~\Eqref{eq:QCHS1},
and the asymmetric form found here,
Eq.~\Eqref{eq:QC_Ftn}.
On the other hand, the asymmetric K matrix $\Kdfuu$ appearing in the TOPT version of
the asymmetric quantization condition, Eq.~\Eqref{eq:QC_Ft},
is not invariant (see Appendix~\ref{app:inteqs}).
However, the symmetrized quantity $\wt \cK_{\df,3}'$ is invariant, since it equals $\Kdf$.

\section*{Acknowledgments}

We thank Max Hansen and Jia-Jun Wu for comments and discussions.
This work was supported in part 
by the U.S. Department of Energy contract No.~DE-SC0011637.

\appendix

\section{Technical comments on time-ordered perturbation theory}
\label{app:TOPT}


In this appendix we address two technical issues concerning the application of TOPT described in the main text.
These are, first, the use of the physical, renormalized mass in energy denominators---and more generally, our apparent neglect of self-energy diagrams---and, second, the presence of an additional class of diagrams with relevant (three-particle) cuts.
Both issues have been partially addressed previously in Ref.~\cite{\BHSQC}, and our discussion here
leans heavily on the analysis in that work.

We begin with the first issue, which we first restate in more detail.
In the discussion in the main text, the kinematic factor associated with each cut
involves the physical mass $m$ rather than the bare mass.
In particular, the factors of $\omega_k$ that appear in both energy denominators and propagator factors
are given by $\omega_k=\sqrt{m^2 + \vec k^2}$.
This appears to ignore the fact that the full propagator in any RFT has 
a more complicated analytic form than a simple pole, due to the usual iteration of self-energy diagrams.
In fact, we are not ignoring self-energy diagrams, but instead dealing with them first
in the context of a Feynman diagram decomposition, and then converting to TOPT to give the rules
described in Sec.~\ref{sec:TOPT}.

To explain our approach,
we begin by writing the quantity under consideration, i.e.~$C_{3,L}$ or $\wt \cM_{3,L}^\uu$,
in terms of the Feynman diagrams that follow
from the Lagrangian of our generic relativistic effective field theory.
Following HS1, we organize these diagrams into a skeleton expansion in terms of 
Bethe-Salpeter kernels and appropriately defined dressed propagators.
The only subtlety here is that for diagrams in which all of the momentum is carried by a single propagator,
the self-energy diagrams that dress this propagator must be 3PIs, instead of the usual 1PIs. 
This allows all possible contributions
with three-particle intermediate states to be made explicit.
This is explained in the text surrounding Eq.~(49) of HS1, 
and the distinction between 1PIs and 3PIs self-energies is illustrated (in the context of a theory
without the $\mathbb Z_2$ symmetry) in Fig.~4 of Ref.~\cite{\BHSQC}.

\begin{figure}[tb]
\begin{center}
\vspace{-10pt}
\includegraphics[width=\textwidth]{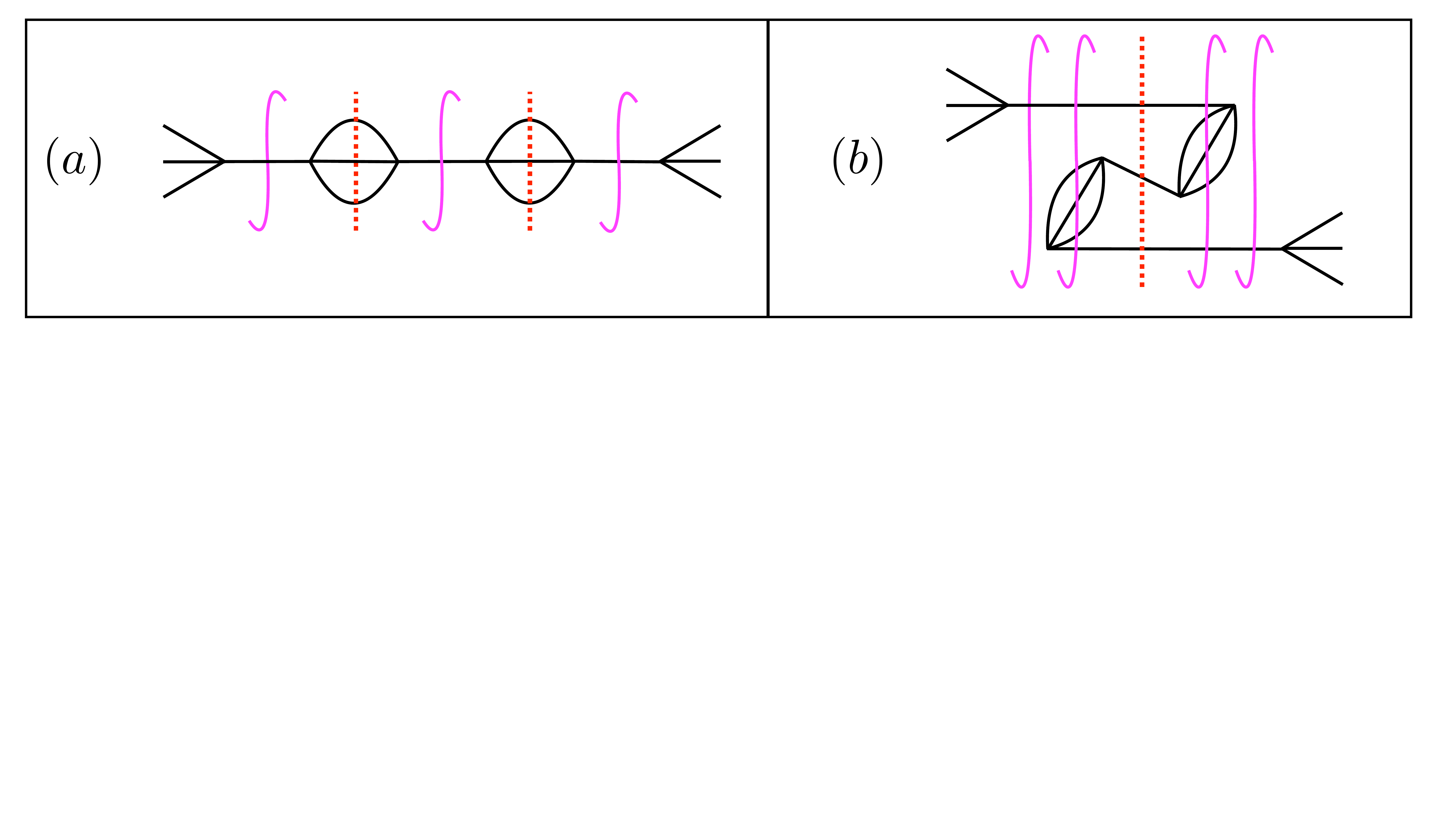}
\vspace{-2.5truein}
\caption{
Examples of TOPT diagrams for $\wt\cM_{3,L}^\uu$
in which all the momentum flows through a single propagator.
Notation as in Fig.~\ref{fig:topt1}.
The two panels show different time orderings of the same Feynman diagram,
and involve self-energy insertions containing three propagators.
The time-ordering shown in (a) has two genuine three-particle cuts, lying between which is a contribution to $\cB_3$.
Time-ordering (b) has a fake three-particle cut that cancels when all time orderings are included.
If the propagators carrying all the momentum are collapsed to point-like vertices, which is valid for $p^2 \gg m^2$
as discussed in the text,
then diagrams of type (a) remain, while those with the Z-type time ordering shown in (b) are removed,
since such a time ordering is no longer possible.
\label{fig:topt_app}
}
\end{center}
\end{figure}

At this stage HS1 use TOPT in a qualitative way to explain why all the self-energy diagrams 
in both types of dressed propagators (1PIs- and 3PIs-dressed) 
can be evaluated in infinite volume [see footnote 18 of HS1].
We now follow Ref.~\cite{\BHSQC} and use a diagram-by-diagram regularization,
in which each Feynman diagram is accompanied by counterterms chosen such 
that it satisfies the renormalization conditions given in Eq.~(14) of Ref.~\cite{\BHSQC}.
In words, these conditions ensure that all self-energy diagrams, 
and their first derivatives with respect to $p^2$, vanish on shell (when evaluated in infinite volume).
Each self-energy diagram thus behaves as $(p^2-m^2)^2$ close to the on-shell point,
where we are using the result that Feynman diagrams yield Lorentz-invariant expressions.
It then follows that, in the usual geometric series that builds up the fully dressed propagator,
only the leading term---a single, undressed propagator---has a pole, and this is of unit residue
and at the position of the physical mass. All other contributions to the dressed propagator are either
momentum-independent constants or vanish as powers of $p^2-m^2$.
For example, a sequence with an undressed propagator followed by a self-energy and another undressed
propagator has the leading behavior $(p^2-m^2)^{-1} (p^2-m^2)^2 (p^2-m^2)^{-1}$, i.e.~a constant.
Such contributions correspond in position space to delta functions or derivatives thereof, and thus can
be collapsed to pointlike interactions. 
(Examples of this collapse, albeit in a slightly different context, are given in Appendix~B.2 of Ref.~\cite{\BHSQC}.)
Any tadpole loops that result
(propagators beginning or ending at the same vertex) can also be collapsed, since, as discussed in
Appendix B.1 of Ref.~\cite{\BHSQC}, they have nonsingular summands that cannot enter into a cut.
The end result of these manipulations is that we are left to evaluate the
subset of diagrams in which there are no self-energy contributions or tadpole loops,
except for self-energy diagrams involving three-particle cuts if they are on 
a single propagator that carries all the momentum. 
[An example of such a diagram is Fig.~\ref{fig:topt_app}(a), viewed as a Feynman diagram.]
When evaluating this reduced class of diagrams we must use modified vertices, due to the collapse of
propagators and tadpole loops, but the key point is that all propagators that remain have their free form
in terms of the physical mass.

At this stage we can break each Feynman diagram into its constituent time orderings, following 
the method explained, for example, in Ref.~\cite{Sterman:1994ce}. This leads to the rules 
described in Sec.~\ref{sec:TOPT}, with all factors of $\omega$ containing the physical mass.
The only subtlety is the need to break up counterterms for vertex diagrams
into Lorentz noncovariant parts so that each TOPT diagram is finite. This does not present problems,
as discussed in Appendix~B.5 of Ref.~\cite{\BHSQC}.
Thus we have resolved the first issue.

We now turn to the second issue, which concerns 
a class of diagrams that leads to ``fake'' three-particle cuts.
By fake, we mean that they will be canceled when all time orderings are added.
Diagrams in this class all have the momentum carried by a single propagator, 
and involve the self-energy diagrams that allow three-particle cuts.
These are the self-energy diagrams that were not part of collapsed dressed propagators in the analysis above.
In TOPT, such diagrams can  have genuine three-particle cuts, 
as shown for example in Fig.~\ref{fig:topt_app}(a),
as well ``Z-type'' configurations that have fake cuts, as in Fig.~\ref{fig:topt_app}(b).
We know the latter cuts must cancel, because if we sum over all time orderings, we will end
up with a result having singularities (higher-order poles) only at $p^2=m^2$.

The simplest way of dealing with this issue is to restrict $E^*$ to lie far above $m$, so that we do not
approach the single-particle pole. 
For example, we could consider $E^*>  E_0^*=2 m$, so that $p^2 > E_0^{*2}=4 m^2$ and $p^2-m^2 > 3 m^2$.
 In that case, the single propagator can be Taylor-expanded about $p^2=E_0^{*2}$, and thus collapsed to
 a series of momentum-dependent vertices.
 This completely removes the Z-type time-orderings, while retaining those that lead to genuine relevant cuts.

\section{Relating $\overline\cK_{2,L}$ to $\cK_2$}
\label{app:K2}


In this appendix we derive Eqs.~\Eqref{eq:K2Ldef}-\Eqref{eq:K2on} in the main text, i.e.~we show
that the two-particle matrix contained in $\overline{\cK}_{2,L}$ is indeed (a variant of) the K matrix.
A secondary purpose is to explain the definition of the generalized PV pole prescription.

Our approach is to consider the two-particle finite-volume amplitude $\cM_{2,L}$, which is given by the sum
of all amputated $2\to2$ diagrams. Since our notation is set up for three-particle correlators, we package
$\cM_{2,L}$ in an analogous manner to that used for $\overline{\cB}_{2,L}$ [Eq.~\Eqref{eq:C2L}],
\begin{equation}
\left[\overline{\cM}_{2,L} \right]_{ka;pr} \equiv \delta_{kp} 2 \omega_k L^3 \cM_{2,L}(E_{2,k},\vec P_{2,k}; \vec a;\vec r)\,,
\quad E_{2,k}\equiv E-\omega_k\,,\ \ \vec P_{2,k}\equiv \vec P-\vec k\,.
\label{eq:M2L}
\end{equation}
This amplitude is off shell in general. It is given in TOPT by
\begin{equation}
i\overline{\cM}_{2,L} = i\overline{\cB}_{2,L} \frac1{1-i D_F i \overline{\cB}_{2,L}}\,,
\label{eq:M2Lres}
\end{equation}
which, using the on-shell projection result Eq.~(\ref{eq:Fsplit}), as well as the definition of
$\overline{\cK}_{2,L}$, Eq.~\Eqref{eq:K2Ldef}, can be rewritten as
\begin{equation}
i\overline{\cM}_{2,L} = i\overline{\cK}_{2,L} \frac1{1-i \wt  F i \overline{\cK}_{2,L}}\,.
\label{eq:M2LresF}
\end{equation}
If we project external indices on shell, so that all matrices are square,
we can invert this result to obtain
\begin{equation}
\left(\overline{\cM}_{2,L}^{\rm on}\right)^{-1} = 
\left(\overline{\cK}_{2,L}^{\rm on}\right)^{-1} + \wt  F\,,
\label{eq:result1}
\end{equation}
where the ``on" labels indicate that both amplitudes must be completely on shell for the equation to hold.

The next step is to take the infinite-volume limit in such a way that the left-hand side goes over
to the (inverse of the) on-shell infinite-volume scattering amplitude. To obtain this limit,
we first remove extraneous common factors (introduced by carrying along the spectator)
by multiplying Eq.~\Eqref{eq:result1} by 
the matrix $2\omega L^3$ [defined in Eq.~\Eqref{eq:2omL3}]
and dropping the $\delta_{kp}$ that is common to all three terms.
We then take the $L\to\infty$ limit holding $E_{2,k}$ and $\vec P_{2,k}$ fixed, which ensures that
in the CMF of the scattering pair, the momentum of each particle in the pair
is held fixed at $q_{2,k}^*$.
Following the prescription used in HS2, we make this limit well defined by 
reintroducing the factors of $i\epsilon$ into the energy denominators contained in the factors of $D_F$
in Eq.~\Eqref{eq:M2Lres}, and only then turning sums into integrals.
The result is
\begin{equation}
\delta_{\ell \ell'} \delta_{m m'} \left[\cM_2^{(\ell)}(q_{2,k}^*)\right]^{-1}
=
\delta_{\ell \ell'} \delta_{m m'} \left[\cK_2^{(\ell)}(q_{2,k}^*)\right]^{-1}
+
 \delta_{\ell \ell'}\delta_{m m'}  \wt\rho_\PV^{(\ell)}(q_{2,k}^{*2})
\,,
\label{eq:result2}
\end{equation}
where $\cM_2^{(\ell)}$ is the $\ell$th partial wave of $\cM_2$,
\begin{equation}
\wt\rho_\PV^{(\ell)}(q_{2,k}^{*2}) \equiv 
H(\vec k) \left[\wt\rho(q_{2,k}^{*2}) + \frac1{32\pi^2}I_{\PV}^{(\ell)}(q_{2,k}^{*2}) \right]\,,
\label{eq:rhoPVdef}
\end{equation}
with the phase space factor given by 
\begin{equation}
\wt\rho(q_{2,k}^{*2}) \equiv \frac1{16\pi E_{2,k}^*} \left\{
\begin{array}{lr} -i {|q_{2,k}^{*}|} &\quad q_{2,k}^{*2} > 0 \\
\hphantom{-i}| q_{2,k}^{*}| &\quad q_{2,k}^{*2} \le 0 \end{array}\right.\,,
\label{eq:rhodef}
\end{equation}
and $I_{\PV}^{(\ell)}$ is an arbitrary real, smooth function.
Here we have assumed that $H(\vec k)$ is, in fact, a function of $q_{2,k}^{*2}$, as is the case 
in all numerical work to date~\cite{\BHSnum,\dwave,\largera,\HHanal}.
The second term on the right-hand side of Eq.~(\ref{eq:result2})
is obtained using Eqs.~(22)-(26) of HS1 
(where the standard PV prescription is defined in the context of $\wt F$), 
together with Eq.~(3.5) of Ref.~\cite{\largera} 
(where the generalized PV prescription is defined),
which together lead to
\begin{equation}
2\omega_k L^3 \wt F_{k\ell m;p\ell' m'} = \delta_{kp} 
\left[ F^{i\epsilon}_{\ell m;\ell' m'}(\vec k) + \delta_{\ell \ell'}\delta_{m m'} 
\wt\rho_\PV^{(\ell)}(q_{2,k}^{*2})\right]\,,
\label{eq:Finf}
\end{equation}
where
\begin{align}
	F^{i\eps}_{\ell m;\ell'm'}(\vec k) &\equiv \frac{H(\vec k)}{2!} \left[ \frac{1}{L^3}\sum_{\vec a}^\UV - \int_{\vec{a}}^\UV \right] \frac1{2\omega_a} \frac{\cY_{\ell m}(\vec{a}_{k}^{*})}{q_{2,k}^{*\ell}} \frac1{2\omega_b (E-\omega_k-\omega_a-\omega_b+i\eps)} \frac{\cY_{\ell'm'}(\vec{a}_{k}^{*})}{q_{2,k}^{*\ell'}}
\end{align}
is the quantity defined in Eq.~(24) of HS1.
Note that $F^{i\epsilon} \to 0$ in the ``$i\epsilon$'' $L\to\infty$ limit.

\section{Algebraic matrix manipulations}
\label{app:algebra}


In the main text, we encounter several times 
[see for example, Eqs.~\Eqref{eq:Kalluu_L} and \Eqref{eq:M23LtoKdf}]
matrix expressions of the form
\begin{align}
m_2+m_3 &= (c_2+c_3) \frac1{1 - (f+g) (c_2+c_3)}\,,
\label{eq:appalg1}
\\
&= \frac1{1 -(c_2+c_3)(f+g)} (c_2+c_3) \,,
\\  
m_2 &= c_2 \frac1{1- f c_2} 
\quad \Rightarrow \quad m_2^{-1} = c_2^{-1}-f
\,,
\end{align}
from which we wish to determine an expression for $m_3$.
For the sake of clarity and completeness,
we collect here the algebraic steps that lead to the form used in the main text.
We stress that these and similar steps have been repeatedly used in previous RFT papers,
i.e.~in HS1, HS2, and Refs.~\cite{\BHSQC,\BHSK,\largera,\isospin}.

As a first step, we define $d_{23}$ as $m_2+m_3$ evaluated when $c_3\to 0$:
\begin{equation}
d_{23} \equiv c_2 \frac1{1-(f+g) c_2}\,.
\end{equation}
This can be rewritten as
\begin{equation}
d_{23}^{-1} = c_2^{-1} - f - g = m_2^{-1} - g\ \ \Rightarrow \ \
d_{23} = m_2 \frac1{1-g m_2} = m_2 + d_3\,, \ \ d_3 \equiv m_2 g m_2 \frac1{1-g m_2}\,.
\end{equation}
In words, $m_2$ is obtained by summing all the $c_2$ terms joining by factors of $f$, and
$d_{23}$ is then obtained by putting in factors of $g$ in all possible ways. 

Our aim is to pull out the $c_3$ dependence of $m_3$ from Eq.~(\ref{eq:appalg1}). The steps are
\begin{align}
m_2+m_3 -d_{23}&= (1+ c_3 c_2^{-1})\frac1{c_2^{-1} - (f+g)(1+c_3 c_2^{-1})} - d_{23}
\\
&= (1+ c_3 c_2^{-1})\frac1{d_{23}^{-1} - (f+g)c_3 c_2^{-1}}- d_{23}
\\
&= (1+ c_3 c_2^{-1})\frac1{1 - d_{23}(f+g)c_3 c_2^{-1}} d_{23} -d_{23}
\\
&= c_3 c_2^{-1}\frac1{1 - d_{23}(f+g)c_3 c_2^{-1}} d_{23} 
+ 
d_{23} (f+g) c_3 c_2^{-1}\frac1{1 - d_{23}(f+g)c_3 c_2^{-1}} d_{23} 
\\
&= 
[1+d_{23} (f+g)] c_3 c_2^{-1}\frac1{1 - d_{23}(f+g)c_3 c_2^{-1}} d_{23} 
\\
&=
\frac1{1-c_2 (f+g)} c_3\frac1{1 - c_2^{-1}d_{23}(f+g)c_3 } c_2^{-1} d_{23} \,.
\end{align}
This can be further simplified using
\begin{equation}
c_{2}^{-1} d_{23} = \frac1{1 - (f+g) c_2}
= 1 + (f+g) d_{23}\,.
\end{equation}
A useful way of rewriting the final result is
\begin{equation}
m_3 = d_3 +
[1+d_{23} (f+g)] c_3 \frac1{1 - [1+ (f+g) d_{23} ](f+g)c_3 } [1+(f+g) d_{23}]\,,
\label{eq:appalgfinal}
\end{equation}
which is used to obtain, for example, Eqs.~\Eqref{eq:Kdf3uures} and \Eqref{eq:M3Lb}.

Clearly this derivation relies on the existence of the various inverse matrices that appear,
and thus, in particular, it assumes that the matrices are square.

\section{Asymmetrization identities}
\label{app:asym}


In this appendix we derive the identities needed in Sec.~\ref{sec:KtoK} to asymmetrize
the HS2 amplitude $\cM_{3,L}^\uu$. These results are extensions of the symmetrization
identities derived in HS1 [see Eqs.~(163) and (198) of that work and surrounding discussions].

\subsection{General asymmetric kernels}
\label{app:asymkernel}

Here we review the notation developed in HS1 to describe general asymmetric kernels,
e.g.~$\KdfuuHS$, as well as collecting some of their key properties. To lighten the notation we denote
a generic asymmetric kernel as $Z^\oneu$, where we only make explicit the symmetry status of
one ``side'' of the kernel.
The meaning of the superscripts $\oneu$ and $\uu$ have been explained in the main text,
with the key point being that, in the $\{k\ell m\}$ index space describing three on-shell particles,
the momentum label $k$ is always associated with the spectator.

We only consider amplitudes that are fully on shell,
denoting the four-momenta of the three particles as $k$ (spectator), $p$, and $b=P-k-p$ (the interacting pair).
The discussion in Sec.~\ref{sec:Gcut} explains the procedure for on-shell projection that defines
$Z_{k\ell m}^\oneu$ (where again we show only one set of indices).
Using Eq.~\Eqref{eq:C2sphdecomp}, the full momentum dependence of $Z^\oneu$ is given by
\begin{equation}
Z^\oneu(\vec k, \widehat{\vec{p}}_k^*) = Z^\oneu_{k\ell m} \sqrt{4 \pi} Y_{\ell m} (\widehat{\vec{p}}_k^*)\,,
\end{equation}
where $\vec p_k^*$ is obtained by the boost of Eqs.~\Eqref{eq:Wu1} and \Eqref{eq:Wu2}
(which is equivalent to the boost of HS1 since the particles are on shell).
Since the kernel is on shell, it depends only on the direction of $\vec p_k^*$
and not its magnitude (for given $\vec k$).
Because we are considering identical particles, $Z^\oneu$ is invariant under
$p\leftrightarrow b$ interchange. Since this interchange is effected in our variables by
changing the sign of $\widehat{\vec{p}}_k^*$, $Z^\oneu$ satisfies
\begin{equation}
Z^\oneu(\vec k, \widehat{\vec{p}}_k^*) = Z^\oneu(\vec k, \widehat{\vec{b}}_k^*) 
=Z^\oneu(\vec k, -\widehat{\vec{p}}_k^*)
\ \ \Leftrightarrow\ \
Z^\oneu_{k \ell m} = 0 \ {\rm if} \ \ell\ \textrm{is odd}\,.
\label{eq:Zu}
\end{equation}

We next define asymmetric kernels with superscript ${(s)}$.
Here the momentum $k$ is assigned to one of the interacting pair, while $p$
is assigned to the spectator:
\begin{equation}
Z^\ones(\vec k,\widehat{\vec{p}}_k^*) 
= Z^\ones_{k\ell' m'} \sqrt{4\pi} Y_{\ell' m'}(\widehat{\vec{p}}_k^*)
\equiv Z^\oneu(\vec p,\widehat{\vec{k}}_p^*)
\,.
\label{eq:Zs}
\end{equation}
We stress that there is a one-to-one relation
between $\{\vec k, \widehat{\vec{p}}_k^*\}$  and $\{\vec p,\widehat{\vec{k}}_p^*\}$, i.e.
one set of variables uniquely determines the other.

In the third option, both $k$ and $p$ are assigned to the interacting pair.
Since this configuration is obtained from the $Z^\ones$ assignment by interchanging $p$ and $b$,
we have
\begin{equation}
Z^\onets(\vec k,\widehat{\vec{p}}_k^*) 
= Z^\onets_{k\ell' m'} \sqrt{4\pi} Y_{\ell' m'}(\widehat{\vec{p}}_k^*)
\equiv Z^\ones(\vec k, -\widehat{\vec{p}}_k^*)
\ \ \Rightarrow\ \
Z^\onets_{k\ell m} = (-1)^\ell Z^\ones_{k \ell m}\,.
\label{eq:Zts}
\end{equation}
In addition, using the one-to-one relation between $\{\vec k, \widehat{\vec{p}}_k^*\}$
and $\{\vec b, \widehat{\vec{p}}_b^*\}$, and the symmetry of $Z^\oneu$ under $p\leftrightarrow b$,
we have
\begin{equation}
Z^\onets(\vec k, \widehat{\vec{p}}_k^*) = Z^\oneu(\vec b,\widehat{\vec{p}}_b^*)\,.
\label{eq:Ztstou}
\end{equation}

We will also need the result from HS1 that 
$\wt F$ vanishes if $\ell'+\ell$ is odd:
\begin{equation}
(-1)^{\ell'}\wt F_{k'\ell' m';k \ell m} (-1)^\ell  
= 
\wt F_{k'\ell' m';k \ell m}\,.
\end{equation}
This holds for all boosts that agree on shell, and thus for the Wu boost we use in this work.
Together with the results in Eqs.~(\ref{eq:Zu}) and (\ref{eq:Zts}), this implies the following useful set of
equalities,
\begin{equation}
X^\oneu \wt F Z^\onets=X^\oneu \wt F Z^\ones\,, \ \ 
X^\onets \wt F Z^\oneu=X^\ones \wt F Z^\oneu\,, \ \
X^\ones \wt F Z^{\ones} = X^\onets \wt F Z^{\onets}\,, \ \
X^\onets \wt F Z^{\ones} = X^\ones \wt F Z^{\onets}\,,
\label{eq:tstos}
\end{equation}
where $X$ is another kernel.

The symmetric on-shell amplitude is obtained by adding all three attachments
\begin{equation}
Z \equiv Z^{(u+s +\tilde s)} = Z^\oneu+Z^\ones+Z^\onets,
\label{eq:onshellsym}
\end{equation}
where we are using the convention that adding superscripts corresponds to adding the underlying amplitudes.
Equation~\Eqref{eq:onshellsym} is the on-shell version of the off-shell symmetrization definition given in 
Eq.~\Eqref{eq:symm}.

\subsection{Deriving Eqs.~(\protect\ref{eq:subL}) and (\protect\ref{eq:subR})}

The first two asymmetrization identities of Sec.~\ref{sec:KtoK} 
have essentially been derived in Eq.~(163)-(165) of HS1.
Here we need a slightly more explicit form, so we repeat the essential steps.

We begin with Eq.~\Eqref{eq:subL}.
For concreteness, we act the identity on the vector of amplitudes 
$\left(Z^\oneu,Z^{(s+\tilde s)}\right)$. Then the identity to be demonstrated can be rewritten as
\begin{equation}
\overrightarrow{\Delta} \equiv X^\oneu [\wt F Z^\ones + \wt F Z^\onets - \wt G Z^\oneu] =
- X^\oneu \overrightarrow{\cI}_G Z^\oneu
\,,
\label{eq:DeltaL}
\end{equation}
where $\overrightarrow{\Delta}$ is simply a shorthand for the left-hand side of this equation,
with the arrow pointing in the direction of the amplitudes that are being asymmetrized.
Using results from Eq.~(\ref{eq:tstos}), this can be written as
\begin{align}
\overrightarrow{\Delta} &= X^\oneu (2 \wt F Z^\ones- \wt G Z^\oneu)
= X^\oneu (2 \wt \Sigma_F Z^\ones - 2 \wt I_F Z^\ones - \wt G Z^\oneu)
\,,
\label{eq:DeltaL2}
\end{align}
where the second form is obtained by splitting $\wt F$, Eq.~\Eqref{eq:Ft},  
into its sum and integral part, $\wt F = \wt \Sigma_F - \wt I_F$. 
The explicit form for $\wt\Sigma_F$ is given in Eq.~\Eqref{eq:tilde-sigmaF}.
Note that the integral
$\wt I_F$ differs from the integral {\em operator} $\wt \cI_F$ of Eq.~\Eqref{eq:I_F},
the latter being denoted by a calligraphic symbol.
We regulate the UV by inserting a factor of $H(\vec p)$, and choose the relativistic form
of the pole term, both choices that only change $\wt F$ by exponentially suppressed terms.
Then, using the definition of $\wt G$, Eq.~\Eqref{eq:Gt}, we find 
\begin{multline}
X^{(u)} \left[ 2 \wt \Sigma_F Z^{(s)} - \wt G Z^{(u)} \right] =
\sum_{\vec k}
\frac1{2\omega_k L^3} \sum_{\vec p} \frac1{2\omega_p L^3}
\sum_{\ell m} X^{(u)}_{k\ell m} 
\frac{\cY_{\ell m}(\vec p_k^{\,*})}{q_{2,k}^{*\ell}}
\frac{H(\vec k) H(\vec p)}{b^2-m^2}
\\
\times\left\{\sum_{\ell' m'}
\left(\frac{p_k^*}{q_{2,k}^*}\right)^{\ell'} 
\sqrt{4\pi}Y_{\ell' m'}(\widehat {\vec p}_k^*) Z_{k\ell'm'}^{(s)}
-
\sum_{\ell' m'}
\left(\frac{k_p^*}{q_{2,p}^*}\right)^{\ell'}
\sqrt{4\pi} Y_{\ell' m'}(\widehat {\vec k}_p^*) Z_{p\ell'm'}^{(u)}
\right\} \,.
\label{eq:sumFG}
\end{multline}
The key observation is now that the expression in curly braces vanishes when
$b^2=m^2$, i.e.~when all three particles are on shell.
For then $p_k^*=q_{2,k}^*$ and $k_p^*=q_{2,p}^*$,
so the sums over $\ell'$ and $m'$ can be done, leading to 
$Z^\ones(\vec k, \widehat{\vec{p}}_k^*)-Z^\oneu(\vec p,\widehat{\vec{k}}_p^*)$,
which vanishes because of Eq.~(\ref{eq:Zs}).
Because of this cancellation, the sum over $\vec p$ can be replaced by an integral.
This integral requires no pole prescription, but if we wish to separate the two terms in curly braces,
then we must choose a prescription, and we use the generalized PV prescription.
Then the first term in curly braces gives $2 \wt I_F$, which cancels the $-2\wt I_F$ term
in Eq.~(\ref{eq:DeltaL2}). What remains is
\begin{align}
\overrightarrow{\Delta} &= - \sum_{\vec k}
\frac1{2\omega_k L^3} \PV\!\int_{\vec p} \frac1{2\omega_p}
X^{(u)}_{k\ell m}  G^b_{k\ell m;p\ell' m'}  Z_{p\ell'm'}^{(u)} \,,
\label{eq:DeltaL4}
\end{align}
where $G^b$ is defined in Eq.~\Eqref{eq:Gb}.
What happens to the sum over $\vec k$ depends on the form of $X^\oneu$.
If $X^\oneu=\overline{\cM}_{2,L}$, which contains a Kronecker delta,
$\vec k$ is set equal to the external spectator momentum.
If $X^\oneu$ is a three-particle amplitude such as $\cD^\uu_L$, then $\vec k$ is an internal index
and the sum over it can be converted to an integral,  since the PV integration over $\vec p$ leads
to a smooth function.
In either case, $X^\oneu$ and $Z^\oneu$ are sewed together by an integral operator.
We define $\overrightarrow{\cI}_G$ to be this integral operator, leading to the right-hand side
of the identity Eq.~\Eqref{eq:DeltaL}. It is similar to the operator $\wt\cI_F$, and thus we use a similar name.

To summarize, in the difference $\overrightarrow{\Delta}$,
the terms cancel exactly on shell, allowing the sums to be replaced by 
(PV-regulated) integrals. Once this is done, $\wt F$ vanishes, since it is a sum-integral
difference. Thus one simply ends up with an integral over the $-\wt G$ contribution.
We note that the argument holds for both choices of boost to the CMF of the scattered pair considered
in the main text, i.e.~the Wu boost and the boost used in HS1.
One needs only to use the same boost in $\wt G$ and $\overrightarrow{\cI}_G$.

The argument for the second identity, Eq.~\Eqref{eq:subR}, 
is essentially the horizontal reflection of that for Eq.~\Eqref{eq:subL},
and we do not repeat the steps. The only change is that the directionality is reversed,
leading to the integral operator $\overleftarrow{\cI}_G$, which asymmetrizes to the left.

\subsection{Derivation of Eq.~(\ref{eq:subLR})}

To derive Eq.~\Eqref{eq:subLR}, we make it concrete by applying $(X^\oneu, X^{(s+\tilde s)})$
on the left and $(Z^\oneu,Z^{(s+\tilde s)})$ on the right. 
Since $X$ and $Z$ are stand-ins for $\Kdf$, they are three-particle amplitudes for which the
spectator momentum labels are summed, and not constrained by a Kronecker delta.
We denote the difference between the left-hand side of Eq.~\Eqref{eq:subLR}
and the $\wt F+\wt G$ term on the right-hand side by $\overleftrightarrow{\Delta}$.
Our aim is to show that this is an integral operator.
We begin by breaking it into four parts
\begin{align}
3\overleftrightarrow{\Delta} &\equiv
X^{(u+s+\tilde s)} \wt F Z^{(u+s+\tilde s)} -3  X^\oneu (\wt F+\wt G) Z^\oneu
\\
&=
\overleftrightarrow{\Delta}_1 + \overleftrightarrow{\Delta}_2 
+ \overleftrightarrow{\Delta}_3 + \overleftrightarrow{\Delta}_4\,,
\end{align}
where
\begin{align}
\overleftrightarrow{\Delta}_1 &=
X^\oneu  \wt F Z^{(s+\tilde s)}  - X^{\oneu} \wt G Z^{\oneu}
\,,
\label{eq:DeltaLR1}
\\
\overleftrightarrow{\Delta}_2 &=
X^{(s+\tilde s)} \wt F Z^{\oneu}  - X^{\oneu} \wt G Z^{\oneu}
\,,
\label{eq:DeltaLR2}
\\
\overleftrightarrow{\Delta}_3 &=
X^\ones \wt F Z^{\ones}
+
X^\onets \wt F Z^{\onets}
- 
2 X^\oneu \wt F Z^{\oneu}
\,,
\label{eq:DeltaLR3}
\\
\overleftrightarrow{\Delta}_4 &=
X^\onets \wt F Z^{\ones}
+
X^\ones \wt F Z^{\onets}
- 
X^\oneu \wt G Z^{\oneu}
\,.
\label{eq:DeltaLR4}
\end{align}
The first two parts can be evaluated using the identities Eq.~(\ref{eq:subL}) and (\ref{eq:subR}), leading to
\begin{equation}
\overleftrightarrow{\Delta}_1 = - X^\oneu\overrightarrow{\cI}_G Z^\oneu
\ \ {\rm and} \ \
\overleftrightarrow{\Delta}_2 = - X^\oneu\overleftarrow{\cI}_G Z^\oneu\,.
\label{eq:Delta12res}
\end{equation}
For the remaining two parts a new analysis is needed.

For $\overleftrightarrow{\Delta}_3$,  using the third result in Eq.~(\ref{eq:tstos}), we obtain
\begin{align}
\overleftrightarrow{\Delta}_3 &=
2 X^\ones \wt F Z^{\ones}
- 
2 X^\oneu \wt F Z^{\oneu}\,.
\end{align}
Separating the $\wt F$'s into sum and integral parts, we have
\begin{align}
\overleftrightarrow{\Delta}_3 &\equiv
 \overleftrightarrow{\Delta}_{3\Sigma}
- \overleftrightarrow{\Delta}_{3I}\,,
\label{eq:DeltaLR3B}
\\
\overleftrightarrow{\Delta}_{3\Sigma} &=
2 X^\ones \wt \Sigma_F Z^{\ones}
- 
2 X^\oneu \wt \Sigma_F Z^{\oneu}\,,
\label{eq:DeltaLR3Sigma}
\\
\overleftrightarrow{\Delta}_{3I} &=
2 X^\ones  \wt I_F Z^{\ones}
- 
2 X^\oneu \wt I_F Z^{\oneu}\,.
\label{eq:DeltaLR3I}
\end{align}
The integral part can be converted into a double integral because of the 
smoothness of the first PV-regulated integral,
\begin{equation}
\overleftrightarrow{\Delta}_{3I} =
\int_{\vec k} \PV\! \int_{\vec p} 
\left\{X^\ones_{k\ell' m'} F_{\ell' m';\ell m} ^b(\vec k,\vec p)
Z^{\ones}_{k\ell m}
- 
 X^\oneu_{k\ell' m'} F_{\ell' m';\ell m} ^b(\vec k,\vec p)
Z^{\oneu}_{k\ell m}\right\}\,,
\label{eq:DeltaLR3IB}
\end{equation}
where (again regulating the $\vec p$ integral in the UV with $H(\vec p)$, but here keeping the original
form of the pole)
\begin{equation}
F^b_{\ell' m';\ell m}(\vec k, \vec p) =
	\frac{\cY_{\ell'm'}(\vec{p}_{k}^*)}{q_{2,k}^{*\ell'}} 
	\frac{H(\vec k) H(\vec p)}
	{2\omega_k 2\omega_p 2\omega_b (E-\omega_k-\omega_p-\omega_b)} 
	\frac{\cY_{\ell m}(\vec{p}_{k}^{*})}{q_{2,k}^{*\ell}} 
	\,.
\label{eq:Fb}
\end{equation}

The sum part $\overleftrightarrow{\Delta}_{3\Sigma}$ has the same form as \Eqref{eq:DeltaLR3IB}
except that both integrals are replaced by sums. 
Naively we might expect the two terms to cancel, since the difference between $(s)$ and $(u)$ quantities
is just a $\vec k \leftrightarrow \vec p$ relabeling.
To investigate this we interchange the dummy variables $\vec k$ and $\vec p$ for the second term in the sum,
resulting in\footnote{%
An alternative approach, used, for example, in the analysis around Eq.~(196) of HS1, is to use the
sum-integral identity in reverse to write the original expression for $\overleftrightarrow{\Delta}_3$
in terms of {\em off-shell} amplitudes, which are more easily manipulated. We do not follow this approach,
however, since it requires accounting for the difference between off-shell amplitudes calculated
using Feynman diagrams and TOPT. Instead, we work entirely with on-shell amplitudes.}
\begin{equation}
\overleftrightarrow{\Delta}_{3\Sigma} =
\frac1{L^6} \sum_{\vec k,\vec p}
\left\{X^\ones_{k\ell' m'} F_{\ell' m';\ell m}^b(\vec k,\vec p)
Z^{\ones}_{k\ell m}
- 
 X^\oneu_{p\ell' m'} F_{\ell' m';\ell m}^b(\vec p,\vec k)
Z^{\oneu}_{p\ell m}\right\}\,.
\label{eq:DeltaLR3S}
\end{equation}
The summand has a pole at each of the free three-particle energies,
with residue
\begin{equation}
\frac1{8 \omega_k \omega_p\omega_b}
\left[X^\ones(\vec k,\widehat{\vec{p}}_k^*) Z^\ones(\vec k,\widehat{\vec{p}}_k^*)
-
X^\oneu(\vec p,\widehat{\vec{k}}_p^*) Z^\oneu(\vec p,\widehat{\vec{k}}_p^*)\right]\Bigg|_{\textrm{on shell}}\,,
\end{equation}
which vanishes due to Eq.~\Eqref{eq:Zs}.
Thus the sum can be converted into an integral. We choose to do the $\vec p$ integral first
using the generalized PV prescription, leading to
\begin{equation}
\overleftrightarrow{\Delta}_{3\Sigma} =
\int_{\vec k} \PV\!\int_{\vec p}
\left\{X^\ones_{k\ell' m'} F_{\ell' m';\ell m}^b(\vec k,\vec p)
Z^{\ones}_{k\ell m}
- 
 X^\oneu_{p\ell' m'} F_{\ell' m';\ell m}^b(\vec p,\vec k)
Z^{\oneu}_{p\ell m}\right\}\,.
\label{eq:DeltaLR3SigmaB}
\end{equation}
As far as we can see, this difference does not vanish. What we have achieved, however, is to convert
the sum part of $\overleftrightarrow{\Delta}_3$ into an integral.
Subtracting from this the result from Eq.~(\ref{eq:DeltaLR3IB}), the terms involving $(s)$ quantities cancel,
leading to
\begin{align}
\overleftrightarrow{\Delta}_{3} &= \int_{\vec k} \PV\! \int_{\vec p}
\left\{ X^\oneu_{k\ell' m'} F^b_{\ell' m';\ell m}(\vec k,\vec p)  Z^{\oneu}_{k\ell m}
- 
X^\oneu_{p\ell' m'} F^b_{\ell' m';\ell m}(\vec p,\vec k) Z^{\oneu}_{p\ell m} \right\}
\\
&\equiv X^\oneu \overleftrightarrow{\cI}_F Z^\oneu \,.
\label{eq:Delta3res}
\end{align}
We have not found a useful way to simplify this further, but this result is sufficient for our purposes.
The key point is that it involves an integral operator that acts on the $(u)$ components of the amplitudes.
\bigskip

Finally, we consider $\overleftrightarrow{\Delta}_{4}$, Eq.~(\ref{eq:DeltaLR4}), which can be analyzed
using a combination of the methods used above. We only give an outline of the calculation.
First, using Eq.~(\ref{eq:tstos}), we see that the first two terms are the same, so that 
\begin{equation}
\overleftrightarrow{\Delta}_4 =
X^\onets 2 \wt F Z^{\ones} -  X^\oneu  \wt G Z^{\oneu}
=
X^\onets 2\wt \Sigma_F Z^{\ones} -  X^\oneu  \wt G Z^{\oneu} - X^\onets 2 \wt I_F Z^\ones
\,.
\label{eq:DeltaLR4B}
\end{equation}
The first two terms involve double sums over $\vec k$ and $\vec p$. For the $\wt G$ term we replace
the sum over $\vec k$ with that over $\vec b$, which is simply a change of variables, and then rename
$\vec k$ as $\vec b$ and vice versa. Then the residue of the pole in $E$ is
\begin{equation}
\frac1{8 \omega_k \omega_p\omega_b}
\left[X^\onets(\vec k,\widehat{\vec{p}}_k^*) Z^{\ones}(\vec k,\widehat{\vec{p}}_k^*)
-
X^\oneu(\vec b,\widehat{\vec{p}}_b^*) Z^{\oneu}(\vec p,\widehat{\vec{b}}_p^*)\right]\Bigg|_{\textrm{on shell}}
\,.
\end{equation}
The identities Eqs.~(\ref{eq:Zu}) and (\ref{eq:Ztstou}) imply that this vanishes.
Thus, once again, we can replace the sums by (PV-regulated) integrals.
This sends $\wt\Sigma_F\to\wt I_F$, canceling the existing $\wt I_F$ term
(which, as above, can be converted to a double integral) and leaving only the integral
over the $\wt G$ term.

In this way we find
\begin{equation}
\overleftrightarrow{\Delta}_4 =
-  X^\oneu 
\int_{\vec k} \, \PV\! \int_{\vec p} Z^\oneu_{b\ell' m'} G^b_{b\ell' m;p \ell m} Z^{\oneu}_{p\ell m}
\equiv X^\oneu \overleftrightarrow{\cI}_G Z^\oneu
\,.
\label{eq:Delta4res}
\end{equation}
We use a bidirectional arrow since the order of integrals here is irrelevant (as long as the first one is
done using the PV prescription).  This can be shown by starting from the form involving $X^\ones \wt F Z^{(\tilde s)}$.

Pulling together the results for the four components, given in
Eqs.~(\ref{eq:Delta12res}), (\ref{eq:Delta3res}), and (\ref{eq:Delta4res}),
we have
\begin{align}
\overleftrightarrow{\Delta} &= - \frac13 X^\oneu 
\left(\overrightarrow{\cI}_G + \overleftarrow{\cI}_G - \overleftrightarrow{\cI}_F
+ \overleftrightarrow{\cI}_G\right) Z^\oneu
\\
&\equiv
X^\oneu \otimes_G Z^{\oneu}\,,
\end{align}
where in the second line we have introduced a compact notation.
All that matters for the argument in the main text is that this is a known integral joining operator
acting on the asymmetric $\oneu$ kernels.
As before, this result holds independent of the choice of boost.

\section{Relating $\Kdfuu$ to $\cM_3$}
\label{app:inteqs}


The aim in this appendix is to take an appropriate infinite-volume limit 
of  Eq.~\Eqref{eq:M3Lb} and obtain integral equations relating $\cM_{3}$ to $\Kdfuu$.
All quantities in this subsection will be on shell,
so that our $\cM_{3,L}$, $\cM_{2,L}$, and $\cD_L^\uu$ are strictly the same as those in HS2.
Since our result for $\wt \cM_{3,L}^\uu$ is similar to that for the corresponding quantity $\cM_{3,L}^\uu$
in HS2 [see Eq.~(68) of that work],
we can take over much  of the analysis essentially without change.

The infinite-volume limit  that is required has been described in Appendix~\ref{app:K2}.
It sends 
\begin{equation}
\left[\cM_{3,L}\right]_{k\ell m;p \ell' m'} \to \cM_3(\vec k, \vec p)_{\ell m; \ell' m'}\,,
\end{equation}
where now $\vec k$ and $\vec p$ are continuous variables.
Analogous limits hold for $\wt \cM_{3,L}^\uu$, $\cD_L^\uu$, and $\wt \cT_L^\uu$.
As noted earlier, $\Kdfuu$ is already an infinite-volume quantity, so the only change is to replace
discrete with continuous momenta.
For the other quantities, we have
\begin{align}
\lim_{L\to\infty} [\overline{\cM}_{2,L}]_{k\ell m;p\ell'm'} &= 
\overline\delta(\vec k-\vec p) \cM_2(\vec k)_{\ell m;\ell' m'}\,,
\\
\overline\delta(\vec k-\vec p) &\equiv 2 \omega_k (2\pi)^3 \delta^3(\vec k-\vec p)
\\
\cM_2(\vec k)_{\ell m;\ell' m'} &= \delta_{\ell \ell'} \delta_{m m'}\cM_2^{(\ell)}(q_{2,k}^*)\,,
\end{align}
where $\cM_2^{(\ell)}$ is the $\ell$th partial wave of $\cM_2$,
while from Eq.~(\ref{eq:Finf}) we read off that
\begin{align}
\lim_{L\to\infty} 2\omega_k L^3 \wt F_{k\ell m;p \ell' m'} 2 \omega_p L^3
&= 
\overline\delta(\vec k-\vec p) 
\wt\rho_\PV(\vec k)_{\ell m;\ell' m'}\,,
\\
\wt\rho_\PV(\vec k)_{\ell m;\ell' m'} &= \delta_{\ell \ell'} \delta_{m m'}  \wt\rho_\PV^{(\ell)}(q_{2,k}^{*2}) 
\,,
\end{align}
where $\wt \rho_\PV^{(\ell)}$, given in Eq.~(\ref{eq:rhoPVdef}), is a smooth function,
and, finally, 
\begin{equation}
\lim_{L\to\infty} 
2\omega_k L^3 \, \wt G_{k\ell m;p \ell' m'} 2\omega_p L^3 \equiv
G^\infty_{\ell m; \ell' m'}(\vec k,\vec p)
= 
 \frac{\cY_{\ell m}(\vec{p}_k^*)}{q_{2,k}^{*\ell}}
\frac{H(\vec k)H(\vec p)}{b^2-m^2 + i\epsilon}
\frac{\cY_{\ell'm'}(\vec{k}_p^*)}{q_{2,p}^{*\ell'}}\,.
\label{eq:Ginfty}
\end{equation}
The latter function, taken from HS2, is simply $G^b$, Eq.~(\ref{eq:Gb}), but with the $i\epsilon$ added back,
and the discrete momentum indices converted to continuous arguments.

When the $L\to\infty$ limit is taken in this way, it is straightforward to see that
the factors of $(2\omega L^3)^{-1}$ coming with $\wt F$ and $\wt G$
convert all momentum sums into integrals with Lorentz-invariant measure
\begin{equation}
\sum_{\vec k} \frac1{2\omega_k L^3}  \xrightarrow{L\to\infty}
 \int \frac{d^3k}{2\omega_k (2\pi)^3} \equiv \int_{\vec k} \frac1{2\omega_k}\,.
\end{equation}
Matrix equations involving geometric series then become integral equations.
In particular, Eq.~\Eqref{eq:DLuu} becomes 
\begin{equation}
i\cD^\uu(\vec k,\vec p) = i\cM_2(\vec k) 
\int_s \frac1{2\omega_s} iG^\infty(\vec k,\vec s) 
\left[\overline\delta(\vec s-\vec p) 
 i\cM_2(\vec p)
+ i \cD^\uu(\vec s,\vec p)\right]\,,
\end{equation}
where angular-momentum indices are implicit. 
This is identical to Eq.~(85) of HS2.
The core geometric series in Eq.~\Eqref{eq:M3Lb} becomes an integral equation for $\wt \cT^\uu$,
\begin{multline}
i\wt\cT^\uu(\vec k,\vec p) = i\Kdfuu(\vec k,\vec p) +
\int_{\vec r,\vec s,\vec t} \frac1{2\omega_r 2\omega_s2\omega_t}
\left[\overline\delta(\vec k-\vec r)  
i\wt\rho_\PV(\vec k)
+ iG^\infty(\vec k,\vec r)\right]
\\
\times \left[\overline\delta(\vec r - \vec s) 
 i\cM_2(\vec r)
+ i \cD^\uu(\vec r, \vec s)\right]
\left[\overline\delta(\vec s-\vec t) 
i\wt\rho_\PV(\vec s)
+ iG^\infty(\vec s,\vec t)\right]
i\wt\cT^\uu(\vec t,\vec p) \,.
\end{multline}
This differs from the corresponding equation in HS2 [Eq.~(91) of that work]
due to the asymmetry of  our $\Kdfuu$, and the presence here
of factors of $\wt F+\wt G$ in place of $\wt F$.
Finally, the factors on either side of $\wt\cT_L^\uu$ in Eq.~\Eqref{eq:M3Lb} become integral operators.
That on the left becomes
\begin{equation}
\wt\cL^\uu(\vec k, \vec s) = \overline\delta(\vec k-\vec s) 
+ \int_{\vec r} \frac1{2\omega_r} 
\left[\overline\delta(\vec k-\vec r) 
 i\cM_2(\vec k)
+ i \cD^\uu(\vec k, \vec r)\right]
\left[\overline{\delta}(\vec r-\vec s) 
i\wt\rho_\PV(\vec r)
+ iG^\infty(\vec r,\vec s)\right]
\,,
\end{equation}
while that on the right, $\wt\cR^\uu$, is given by the horizontal reflection. 
These also differ from their analogs in HS2 ($\cL^\uu$ and $\cR^\uu$) by the presence here of contributions
resulting from $\wt F+\wt G$ in place of $\wt F$.

Putting these pieces together, we obtain the final expression for $\wt\cM_{3}^\uu$,
\begin{equation}
i\wt\cM_3^\uu(\vec k, \vec p) = i\cD^\uu(\vec k,\vec p)
+ \int_{\vec r,\vec s} \frac1{2\omega_r2\omega_s}
i\wt\cL^\uu(\vec k,\vec r) i\wt\cT^\uu(\vec r, \vec s) i\wt\cR^\uu(\vec s,\vec p)
\,.
\end{equation}
This is then symmetrized to obtain $\cM_3(\vec k,\vec p)$.
The symmetrization of on-shell quantities is given by the on-shell limit of
Eq.~\Eqref{eq:symm}, and has been discussed extensively in HS2 
[see discussion around Eqs.~(35)-(37) of that work].

One property of this result is that, despite the apparently Lorentz invariant
form of the relations derived in this appendix, $\Kdfuu$  is not Lorentz invariant.
Here by $\Kdfuu$ we refer to the form obtained after multiplying $\Kdfuu(\vec k,\vec p)_{\ell' m';\ell m}$ 
by spherical harmonics and summing over angular momentum indices,
i.e.~the quantity that depends only on the external momenta.
It cannot be Lorentz invariant because $\wt \cM_3^\uu$  is not invariant, as its asymmetry is
defined in terms of TOPT kernels, and these are frame dependent. 

In summary, the relation between our asymmetric kernel $\Kdfuu$ and $\cM_3$ is of a similar form
to that between $\Kdf$ and $\cM_3$ obtained in HS2. The main point is that such a relation exists,
so that our new quantization condition has the same logical status as that of HS1.
We expect that solving the integral equations numerically (usually done by going back to the
matrix form) will be of similar difficulty.

\section{Deriving Eqs.~(\protect\ref{EQ:M3LC}) and (\protect\ref{EQ:KTOKON2})}
\label{app:KtoK}


In this appendix we provide some details of the derivation of Eqs.~(\ref{EQ:M3LC}) and (\ref{EQ:KTOKON2}).
We make extensive use of the following identities:
\begin{align}
i\cD_{23,L}^\uu &=
\frac1{1-i\overline{\cK}_2i(\wt F+\wt G)} i\overline{\cK}_2 
\label{eq:identity1}
\\
&= \left[ 1 + \frac1{1-i\overline{\cK}_2i(\wt F+\wt G) }i\overline{\cK}_2 i(\wt F+\wt G)\right] i\overline{\cK}_2\,,
\label{eq:identity1a}
\\
i(\wt F+\wt G) \left[1 + i\cD_{23,L}^\uu 
i(\wt F+\wt G)\right]
&= i(\wt F+\wt G)\frac1{1-i\overline{\cK}_2i(\wt F+\wt G)}
= 
\frac1{1-i(\wt F+\wt G) i\overline{\cK}_2}i(\wt F+\wt G)
\,,
\label{eq:identity2}
\end{align}
where $\cD_{23,L}^\uu$ is defined in Eq.~\Eqref{eq:D23Luu}.

We start from Eqs.~(\ref{eq:Mdf3Luu}) and (\ref{eq:TLuu}).
Using the identities \Eqref{eq:identity1a} and \Eqref{eq:identity1} in turn, we find
\begin{align}
i\cD_{23,L}^\uu (- i\overrightarrow{\cI}_G) i\Kdfuun
&= \left[1 + \frac1{1-i\overline{\cK}_2 i(\wt F+\wt G)}i\overline{\cK}_2 i(\wt F+\wt G)\right] 
i\overline{\cK}_2  (-i \overrightarrow{\cI}_G) i \Kdfuun
\\
&= \left[1 + i \cD_{23,L}^\uu i(\wt F+ \wt G)\right] 
i\overline{\cK}_2  (-i \overrightarrow{\cI}_G) i \Kdfuun\,,
\end{align}
so that the factor on the left-hand end of Eq.~(\ref{eq:Mdf3Luu}) can be written 
(when acting to the right on $\Kdfuun$)
\begin{align}
\left[1 + i\cD_{23,L}^\uu i(\wt F+ \wt G - \overrightarrow{\cI}_G)\right] 
&=
\left[1 + i\cD_{23,L}^\uu  i(\wt F+ \wt G)\right] 
\left( 1 - i\overline{\cK}_2 i \overrightarrow{\cI}_G \right) \,.
\label{eq:left}
\end{align}
The factor on the right-hand end of Eq.~(\ref{eq:Mdf3Luu}) gives the horizontal reflection of this expression.

To simplify $\cT_L^\uu$, Eq.~(\ref{eq:TLuu}), we first use \Eqref{eq:identity1a} to obtain
\begin{align}
 i\overleftarrow{\cI}_G i \cD_{23,L}^\uu i \overrightarrow{\cI}_G
&=
 i\overleftarrow{\cI}_G   \overline{\cK}_2  i \overrightarrow{\cI}_G
+
 i\overleftarrow{\cI}_G \overline{\cK}_2
\frac1{1- i(\wt F+\wt G) i\overline{\cK}_2} i(\wt F+\wt G) 
i\overline{\cK}_2 i \overrightarrow{\cI}_G\,,
\end{align}
and then expand out the term that lies between factors of $\KdfuuHS$ using (\ref{eq:identity2})
\begin{align}
   i( \wt F+ \wt G + \otimes_G)  & +
i (\wt F+\wt G - \overleftarrow{\cI}_G )
i\cD_{23,L}^\uu
i( \wt F+ \wt G - \overrightarrow{\cI}_G)
\\
\begin{split}
&=
  i \otimes_G  + i(\wt F+ \wt G) \frac1{1- i\overline{\cK}_2 i(\wt F+\wt G) }
 \\&\qquad
-  i(\wt F+ \wt G) \frac1{1-i\overline{\cK}_2 i(\wt F+ \wt G)} i\overline{\cK}_2  i\overrightarrow{\cI}_G
\\
&\qquad -i \overleftarrow{\cI}_G i\overline{\cK}_2 \frac1{1-i(\wt F+\wt G) i\overline{\cK}_2} i(\wt F+\wt G)
\\
&\qquad 
+  i\overleftarrow{\cI}_G   \overline{\cK}_2  i \overrightarrow{\cI}_G
+
 i\overleftarrow{\cI}_G \overline{\cK}_2
\frac1{1- i(\wt F+\wt G) i\overline{\cK}_2} i(\wt F+\wt G) 
i\overline{\cK}_2 i \overrightarrow{\cI}_G
\end{split}
\\
\begin{split}
&=
i \otimes_2 
+  i\overleftarrow{\cI}_G   \overline{\cK}_2  i \overrightarrow{\cI}_G
\\
&\qquad + 
\left(1 - i\overleftarrow{\cI}_G i\overline{\cK}_2\right)
\left[1+ i(\wt F+\wt G)i\cD_{23,L}^\uu\right] i(\wt F+\wt G)
\left(1 - i\overline{\cK}_2 i\overrightarrow{\cI}_G \right)
\,.
\end{split}
\label{eq:middle}
\end{align}

Inserting Eq.~(\ref{eq:left}), its reflection, and Eq.~(\ref{eq:middle})
into Eqs.~(\ref{eq:Mdf3Luu}) and (\ref{eq:TLuu}) and reorganizing leads to Eqs.~(\ref{EQ:M3LC})
and (\ref{EQ:KTOKON2}).

\section{$\wt\Sigma_F$ approach}
\label{app:opt1}


In Sec.~\ref{sec:Fcut}, we chose to deal with off-shell F cuts using the ``$\wt F$ approach," which is essentially the standard strategy used in HS1 and subsequent RFT works.
In this appendix we sketch an alternative method, which we refer to as the $\wt\Sigma_F$ approach.

The $\wt\Sigma_F$ and $\wt F$ approaches share the same goal, namely to rewrite quantities of the form
\begin{align}
	[X'D_FX]_{p'r';pr} = [X']_{p'r';k'a'} [D_F]_{k'a';ka} [X]_{ka;pr} \,, \qquad
	X'\in\{\widehat A', \overline\cB_{2,L}, \cB_3\}\,, ~~
	X\in\{\widehat A, \overline\cB_{2,L}, \cB_3\}
\end{align}
in terms of a part in which the ``middle'' indices are projected on-shell and a remainder.
In the $\wt\Sigma_F$ approach, we use exactly the same strategy  that we used to deal with G cuts
in Sec.~\ref{sec:Gcut}.
The end result is an F-cut analog of Eq.~\eqref{eq:Gsplit}:
\begin{align}
	X' D_{F} X &= X' \left( \wt{\Sigma}_F + \delta\wt F \right) X \,. 
	\label{eq:SigmaF_split}
\end{align}
Here 
\begin{align}
	\big[\wt\Sigma_F\big]_{k' \ell' m'; k\ell m}  &\equiv 
	\frac{\delta_{k'k}}{2!} H(\vec{k}) \sum_{\vec{a}}^\UV
	 \frac{\cY_{\ell'm'}(\vec{a}_{k}^{*})}{q_{2,k}^{*\ell'}} 
	\frac1{2\omega_k L^3} \frac{1}{b_{ka}^2-m^2} \frac1{2\omega_a L^3} 
	\frac{\cY_{\ell m}(\vec{a}_{k}^{*})}{q_{2,k}^{*\ell}}
\label{eq:tilde-sigmaF}
\end{align}
is the analog of $\wt G$,\footnote{%
Following the G-cut procedure exactly actually gives two factors of $H(\vec k)$ in $\wt\Sigma_F$ (one from each endcap), but this is overkill since the spectator is shared by both endcaps. Using $H(\vec k)$ instead of $[H(\vec k)]^2$ in $\wt\Sigma_F$ (and consequently the $\delta\wt F$ term) is simply a matter of preference.}
while $\delta\wt F$ plays the analogous role to $\delta\wt G$.
In particular, $\delta\wt F$ accounts for all nonsingular off-shell contributions 
(with its exact definition depending on the choice of $X'$ and $X$), 
and we can therefore treat it as an infinite-volume quantity by 
replacing all internal sums in $X'\delta\wt FX$ with integrals.

Two important differences between the $\wt F$ and $\wt\Sigma_F$ approaches are now clear.
The first concerns the UV cutoff: $\wt \Sigma_F$ depends on the cutoff, while $\wt F$ does not
(up to exponentially-suppressed terms). 
We stress, however, that $\wt\Sigma_F$ is well defined 
for all choices of cutoff function, due to our use of the Wu boost.

The second difference is that the integrals in $\delta\wt F$ do not require a pole prescription,
since the integrand is smooth. This is in contrast to the integral in $\wt F$ (denoted $\wt \cI_F$ above),
which requires a PV prescription.
This difference is the main advantage of the $\wt\Sigma_F$ approach.

\subsection{Quantization condition}

Inserting Eqs.~\eqref{eq:SigmaF_split} and \eqref{eq:Gsplit} into Eq.~\eqref{eq:CL_TOPT}, we find
\begin{align}
	C_{3,L}-C_{3,\iy}^{(0)}
	&= \widehat{A}'\, i \big(\wt\Sigma_F+\wt{G} + \delta \wt F + \delta\wt{G}\big)
	 \frac{1}{1 - i\big(\overline\cB_{2,L}+\cB_{3}\big) i\big(\wt\Sigma_F+\wt{G} + \delta\wt F + \delta\wt{G}\big)} \widehat{A}
	\\
	&= \delta C^{\Sigma_F}_{3,\iy} + 
	A'^{\,\Sigma_F,(u)} i\big(\wt{\Sigma}_F+\wt{G}\big) 
	\frac{1}{1-i\cK_{\df,23,L}^{\Sigma_F,\uu} i\big(\wt{\Sigma}_F + \wt{G}\big)} A^{\Sigma_F,(u)} \,, 
	\label{eq:C3L_SigmaF}
\end{align}
where all quantities in the last term are evaluated on shell, with
\begin{align}
	i\cK_{\df,23,L}^{\Sigma_F,\uu} &\equiv \frac{1}{1-i\big(\overline\cB_{2,L}
	+\cB_{3}\big) i\big(\delta\wt F + \delta\wt G\big)} i\big(\overline\cB_{2,L}+\cB_{3}\big)
	\\
	A'^{\,\Sigma_F,(u)} &\equiv \widehat{A}' 
	\frac{1}{1-i\big(\delta\wt F + \delta\wt G\big) i\big(\overline\cB_{2,L}+\cB_{3}\big)}
	\\
	A^{\Sigma_F,(u)} &\equiv 
	\frac{1}{1-i\big(\overline\cB_{2,L}+\cB_{3}\big) i\big(\delta\wt F + \delta\wt G\big)} \widehat{A}
	\\
	\delta C^{\Sigma_F}_{3,\iy} &\equiv 
	\widehat{A}' i\big(\delta\wt F + \delta\wt G\big) 
	\frac{1}{1-i(\overline\cB_{2,L}+\cB_{3}) i\big(\delta\wt F + \delta\wt G\big)} \widehat{A} \,.
\end{align}
We note that $\cK_{\df,23,L}^{\Sigma_F,\uu}$ can be split up as
\begin{align}
	\cK_{\df,23,L}^{\Sigma_F,\uu} &= \overline{\cK}{}^{\Sigma_F}_{2,L} + \cK_{\df,3}^{\Sigma_F,\uu} \,,
\end{align}
where
\begin{align}
	i\overline{\cK}{}^{\Sigma_F}_{2,L} &\equiv \frac{1}{1- i\overline\cB_{2,L} i\delta\wt F} i\overline\cB_{2,L}
	= 2\omega L^3 i\cK^{\Sigma_F}_2 \,,
	\\
	\left[\cK^{\Sigma_F}_2\right]_{k'\ell'm';k\ell m} 
	&\equiv \delta_{k'k} \left[\cK^{\Sigma_F}_2(\vec k) \right]_{\ell'm';\ell m} \,.
\end{align}
Here $\cK^{\Sigma_F}_2$ and $\K_{\df,3}^{\Sigma_F,\uu}$ are the respective $\wt\Sigma_F$-approach analogs of the K matrices $\cK_2$ and $\Kdfuu$ in the $\wt F$ approach, with their definitions only differing by using $\delta\wt F$ in place of $\wt\cI_F$.

From Eq.~\eqref{eq:C3L_SigmaF}, we obtain the quantization condition in the $\wt\Sigma_F$ approach:
\begin{align}
	\det\left[ 1 + \left(2\omega L^3\cK_2^{\Sigma_F} + 
	\cK{}_{\df,3}^{\Sigma_F,\uu}\right)\left({\wt\Sigma}_F+\wt{G}\right) \right] = 0 \,.
	\label{eq:QC3_SigmaF}
\end{align}
For this to be useful, we need to relate the infinite-volume
quantities $\cK_2^{\Sigma_F}$ and $\cK{}_{\df,3}^{\Sigma_F,\uu}$ to scattering amplitudes,
and we do so in the next two subsections.

\subsection{Relating $\cK_2^{\Sigma_F}$ to $\cM_2$}

From Appendix \ref{app:K2} and the equations above, we have
\begin{align}
	i\overline\cM_{2,L} &= i\overline\cB_{2,L} \frac1{1-iD_Fi\overline\cB_{2,L}}
	\\
	&= i\overline\cK{}^{\Sigma_F}_{2,L} \frac1{1-i\wt\Sigma_F i\overline\cK{}^{\Sigma_F}_{2,L}} \,,
\end{align}
which gives a simple inverse relation between the on-shell FV amplitudes:
\begin{align}
	\left(\overline{\cM}_{2,L}^{\rm on}\right)^{-1} &= \Big(\overline\cK^{\Sigma_F,\rm on}_{2,L}\Big)^{-1} 
	+ \wt\Sigma_F \,,
\end{align}
where again the ``on" labels indicate that the amplitudes must be completely on shell for the equation to hold.
To obtain the corresponding infinite-volume relation between $\cM_2$ and $\cK_2^{\wt\Sigma_F}$, we follow the same steps as we did in Appendix \ref{app:K2}: eliminate the common spectator term by multiplying by $2\omega L^3$ and dropping the $\delta_{k'k}$, take the $L\to\iy$ limit holding $E_{2,k}$ and $\vec P_{2,k}$ (and therefore $q_{2,k}^*$) fixed by reintroducing the $i\eps$ term in $D_F$, and convert all sums to integrals.
The result is
\begin{align}
	\delta_{\ell'\ell}\delta_{m'm} \left[ \cM_2^{(\ell)}(q_{2,k}^*) \right]^{-1} &= 
	\bigg\{\Big[ \cK_2^{\Sigma_F}(\vec k) \Big]^{-1}\bigg\}_{\ell'm';\ell m} + 
	\left[I_F^{i\eps}(\vec k)\right]_{\ell'm';\ell m} \,,
	\label{eq:M2_K2SigmaF}
\end{align}
where
\begin{align}
	\left[I_F^{i\eps}(\vec k)\right]_{\ell'm';\ell m}  &  \equiv \frac{H(\vec k)}{2!} \int_{\vec{a}}^\UV \frac1{2\omega_a} \frac{\cY_{\ell'm'}(\vec{a}_{k}^{*})}{q_{2,k}^{*\ell'}} \frac1{b_{ka}^2-m^2+i\eps} \frac{\cY_{\ell m}(\vec{a}_{k}^{*})}{q_{2,k}^{*\ell}} \,.
\end{align}

A new feature that arises here is  that $I^{i\eps}_F$ is not diagonal in $\ell$ and $m$.
This is because, when using the Wu boost, the transformation to the pair CMF does not lead
to an integrand that, aside from the harmonic polynomials, is a rotational scalar.
It follows from Eq.~(\ref{eq:M2_K2SigmaF})
that $\cK_2^{\Sigma_F}$ must also have off-diagonal terms.
This is not a problem in principle, but is a cumbersome feature of this approach.

\subsection{Relating $\cK{}_{\df,3}^{\Sigma_F,\uu}$ to $\cM_3$}

We provide only a sketch of the derivation of this relation, since the analysis follows closely that given
in Appendix~\ref{app:inteqs}. We start from Eq.~(\ref{eq:M23uu}) and substitute
Eqs.~\eqref{eq:SigmaF_split} and \eqref{eq:Gsplit} to obtain
\begin{equation}
i \left(\overline{\cM}_{2,L} + \wt \cM_{3,L}^\uu\right) = i\cK{}_{\df,23,L}^{\Sigma_F,\uu}
\frac1{1 - i (\wt \Sigma_F+\wt G) \, i\cK{}_{\df,23,L}^{\Sigma_F,\uu}}
\,.
\end{equation}
Following the steps in the main text, we can extract from this a result for $\wt\cM_{3,L}^\uu$
identical to Eqs.~\Eqref{eq:M3Lb}-\Eqref{eq:D23Luu} except for the substitutions
$\wt \cK_{\df,3}^\uu \to \cK{}_{\df,3}^{\Sigma_F,\uu}$
and $\wt F \to \wt \Sigma_F$.
We then take the infinite-volume limit as in Appendix~\ref{app:inteqs} and obtain the same
set of equations with
$\wt \cK_{\df,3}^\uu \to \cK{}_{\df,3}^{\Sigma_F,\uu}$
and $\wt \rho_\PV(\vec k) \to I_F^{i\eps}(\vec k)$.

\bibliography{ref} 

\begin{thebibliography}{57}
\expandafter\ifx\csname natexlab\endcsname\relax\def\natexlab#1{#1}\fi
\expandafter\ifx\csname bibnamefont\endcsname\relax
  \def\bibnamefont#1{#1}\fi
\expandafter\ifx\csname bibfnamefont\endcsname\relax
  \def\bibfnamefont#1{#1}\fi
\expandafter\ifx\csname citenamefont\endcsname\relax
  \def\citenamefont#1{#1}\fi
\expandafter\ifx\csname url\endcsname\relax
  \def\url#1{\texttt{#1}}\fi
\expandafter\ifx\csname urlprefix\endcsname\relax\def\urlprefix{URL }\fi
\providecommand{\bibinfo}[2]{#2}
\providecommand{\eprint}[2][]{\url{#2}}

\bibitem[{\citenamefont{Beane et~al.}(2008)\citenamefont{Beane, Detmold, Luu,
  Orginos, Savage, and Torok}}]{Beane:2007es}
\bibinfo{author}{\bibfnamefont{S.~R.} \bibnamefont{Beane}},
  \bibinfo{author}{\bibfnamefont{W.}~\bibnamefont{Detmold}},
  \bibinfo{author}{\bibfnamefont{T.~C.} \bibnamefont{Luu}},
  \bibinfo{author}{\bibfnamefont{K.}~\bibnamefont{Orginos}},
  \bibinfo{author}{\bibfnamefont{M.~J.} \bibnamefont{Savage}},
  \bibnamefont{and} \bibinfo{author}{\bibfnamefont{A.}~\bibnamefont{Torok}},
  \bibinfo{journal}{Phys. Rev. Lett.} \textbf{\bibinfo{volume}{100}},
  \bibinfo{pages}{082004} (\bibinfo{year}{2008}), \eprint{0710.1827}.

\bibitem[{\citenamefont{Detmold et~al.}(2008)\citenamefont{Detmold, Savage,
  Torok, Beane, Luu, Orginos, and Parre\~no}}]{Detmold:2008fn}
\bibinfo{author}{\bibfnamefont{W.}~\bibnamefont{Detmold}},
  \bibinfo{author}{\bibfnamefont{M.~J.} \bibnamefont{Savage}},
  \bibinfo{author}{\bibfnamefont{A.}~\bibnamefont{Torok}},
  \bibinfo{author}{\bibfnamefont{S.~R.} \bibnamefont{Beane}},
  \bibinfo{author}{\bibfnamefont{T.~C.} \bibnamefont{Luu}},
  \bibinfo{author}{\bibfnamefont{K.}~\bibnamefont{Orginos}}, \bibnamefont{and}
  \bibinfo{author}{\bibfnamefont{A.}~\bibnamefont{Parre\~no}},
  \bibinfo{journal}{Phys. Rev.} \textbf{\bibinfo{volume}{D78}},
  \bibinfo{pages}{014507} (\bibinfo{year}{2008}), \eprint{0803.2728}.

\bibitem[{\citenamefont{Detmold and Smigielski}(2011)}]{Detmold:2011kw}
\bibinfo{author}{\bibfnamefont{W.}~\bibnamefont{Detmold}} \bibnamefont{and}
  \bibinfo{author}{\bibfnamefont{B.}~\bibnamefont{Smigielski}},
  \bibinfo{journal}{Phys. Rev. D} \textbf{\bibinfo{volume}{84}},
  \bibinfo{pages}{014508} (\bibinfo{year}{2011}), \eprint{1103.4362}.

\bibitem[{\citenamefont{Detmold}(2013)}]{Detmold:2013wda}
\bibinfo{author}{\bibfnamefont{W.}~\bibnamefont{Detmold}},
  \bibinfo{journal}{Eur. Phys. J. A} \textbf{\bibinfo{volume}{49}},
  \bibinfo{pages}{83} (\bibinfo{year}{2013}).

\bibitem[{\citenamefont{Detmold and Nicholson}(2013)}]{Detmold:2013gua}
\bibinfo{author}{\bibfnamefont{W.}~\bibnamefont{Detmold}} \bibnamefont{and}
  \bibinfo{author}{\bibfnamefont{A.~N.} \bibnamefont{Nicholson}},
  \bibinfo{journal}{Phys. Rev. D} \textbf{\bibinfo{volume}{88}},
  \bibinfo{pages}{074501} (\bibinfo{year}{2013}), \eprint{1308.5186}.

\bibitem[{\citenamefont{Mai and D{\"o}ring}(2019)}]{Mai:2018djl}
\bibinfo{author}{\bibfnamefont{M.}~\bibnamefont{Mai}} \bibnamefont{and}
  \bibinfo{author}{\bibfnamefont{M.}~\bibnamefont{D{\"o}ring}},
  \bibinfo{journal}{Phys. Rev. Lett.} \textbf{\bibinfo{volume}{122}},
  \bibinfo{pages}{062503} (\bibinfo{year}{2019}), \eprint{1807.04746}.

\bibitem[{\citenamefont{H{\"o}rz and Hanlon}(2019)}]{Horz:2019rrn}
\bibinfo{author}{\bibfnamefont{B.}~\bibnamefont{H{\"o}rz}} \bibnamefont{and}
  \bibinfo{author}{\bibfnamefont{A.}~\bibnamefont{Hanlon}},
  \bibinfo{journal}{Phys. Rev. Lett.} \textbf{\bibinfo{volume}{123}},
  \bibinfo{pages}{142002} (\bibinfo{year}{2019}), \eprint{1905.04277}.

\bibitem[{\citenamefont{Blanton et~al.}(2020)\citenamefont{Blanton,
  Romero-L\'opez, and Sharpe}}]{Blanton:2019vdk}
\bibinfo{author}{\bibfnamefont{T.~D.} \bibnamefont{Blanton}},
  \bibinfo{author}{\bibfnamefont{F.}~\bibnamefont{Romero-L\'opez}},
  \bibnamefont{and} \bibinfo{author}{\bibfnamefont{S.~R.}
  \bibnamefont{Sharpe}}, \bibinfo{journal}{Phys. Rev. Lett.}
  \textbf{\bibinfo{volume}{124}}, \bibinfo{pages}{032001}
  (\bibinfo{year}{2020}), \eprint{1909.02973}.

\bibitem[{\citenamefont{Mai et~al.}(2020)\citenamefont{Mai, D{\"o}ring, Culver,
  and Alexandru}}]{Mai:2019fba}
\bibinfo{author}{\bibfnamefont{M.}~\bibnamefont{Mai}},
  \bibinfo{author}{\bibfnamefont{M.}~\bibnamefont{D{\"o}ring}},
  \bibinfo{author}{\bibfnamefont{C.}~\bibnamefont{Culver}}, \bibnamefont{and}
  \bibinfo{author}{\bibfnamefont{A.}~\bibnamefont{Alexandru}},
  \bibinfo{journal}{Phys.\ Rev.\ D} \textbf{\bibinfo{volume}{101}},
  \bibinfo{pages}{054510} (\bibinfo{year}{2020}), \eprint{1909.05749}.

\bibitem[{\citenamefont{Culver et~al.}(2020)\citenamefont{Culver, Mai, Brett,
  Alexandru, and D{\"o}ring}}]{Culver:2019vvu}
\bibinfo{author}{\bibfnamefont{C.}~\bibnamefont{Culver}},
  \bibinfo{author}{\bibfnamefont{M.}~\bibnamefont{Mai}},
  \bibinfo{author}{\bibfnamefont{R.}~\bibnamefont{Brett}},
  \bibinfo{author}{\bibfnamefont{A.}~\bibnamefont{Alexandru}},
  \bibnamefont{and} \bibinfo{author}{\bibfnamefont{M.}~\bibnamefont{D{\"o}ring}},
  \bibinfo{journal}{Phys. Rev. D} \textbf{\bibinfo{volume}{101}},
  \bibinfo{pages}{114507} (\bibinfo{year}{2020}), \eprint{1911.09047}.

\bibitem[{\citenamefont{Beane et~al.}(2020)}]{Beane:2020ycc}
\bibinfo{author}{\bibfnamefont{S.}~\bibnamefont{Beane}} \bibnamefont{et~al.}
  (\bibinfo{year}{2020}), \eprint{2003.12130}.

\bibitem[{\citenamefont{Romero-L\'opez et~al.}(2018)\citenamefont{Romero-L\'opez,
  Rusetsky, and Urbach}}]{Romero-Lopez:2018rcb}
\bibinfo{author}{\bibfnamefont{F.}~\bibnamefont{Romero-L\'opez}},
  \bibinfo{author}{\bibfnamefont{A.}~\bibnamefont{Rusetsky}}, \bibnamefont{and}
  \bibinfo{author}{\bibfnamefont{C.}~\bibnamefont{Urbach}},
  \bibinfo{journal}{Eur. Phys. J.} \textbf{\bibinfo{volume}{C78}},
  \bibinfo{pages}{846} (\bibinfo{year}{2018}), \eprint{1806.02367}.

\bibitem[{\citenamefont{L{\"u}scher}(1986)}]{Luscher:1986n2}
\bibinfo{author}{\bibfnamefont{M.}~\bibnamefont{L{\"u}scher}},
  \bibinfo{journal}{Commun.Math.Phys.} \textbf{\bibinfo{volume}{105}},
  \bibinfo{pages}{153} (\bibinfo{year}{1986}).

\bibitem[{\citenamefont{L{\"u}scher}(1991)}]{Luscher:1991n1}
\bibinfo{author}{\bibfnamefont{M.}~\bibnamefont{L{\"u}scher}},
  \bibinfo{journal}{Nucl.Phys.} \textbf{\bibinfo{volume}{B354}},
  \bibinfo{pages}{531} (\bibinfo{year}{1991}).

\bibitem[{\citenamefont{Rummukainen and Gottlieb}(1995)}]{Rummukainen:1995vs}
\bibinfo{author}{\bibfnamefont{K.}~\bibnamefont{Rummukainen}} \bibnamefont{and}
  \bibinfo{author}{\bibfnamefont{S.~A.} \bibnamefont{Gottlieb}},
  \bibinfo{journal}{Nucl. Phys.} \textbf{\bibinfo{volume}{B450}},
  \bibinfo{pages}{397} (\bibinfo{year}{1995}), \eprint{hep-lat/9503028}.

\bibitem[{\citenamefont{Kim et~al.}(2005)\citenamefont{Kim, Sachrajda, and
  Sharpe}}]{Kim:2005gf}
\bibinfo{author}{\bibfnamefont{C.~h.} \bibnamefont{Kim}},
  \bibinfo{author}{\bibfnamefont{C.~T.} \bibnamefont{Sachrajda}},
  \bibnamefont{and} \bibinfo{author}{\bibfnamefont{S.~R.}
  \bibnamefont{Sharpe}}, \bibinfo{journal}{Nucl. Phys.}
  \textbf{\bibinfo{volume}{B727}}, \bibinfo{pages}{218} (\bibinfo{year}{2005}),
  \eprint{hep-lat/0507006}.

\bibitem[{\citenamefont{He et~al.}(2005)\citenamefont{He, Feng, and
  Liu}}]{He:2005ey}
\bibinfo{author}{\bibfnamefont{S.}~\bibnamefont{He}},
  \bibinfo{author}{\bibfnamefont{X.}~\bibnamefont{Feng}}, \bibnamefont{and}
  \bibinfo{author}{\bibfnamefont{C.}~\bibnamefont{Liu}},
  \bibinfo{journal}{JHEP} \textbf{\bibinfo{volume}{07}}, \bibinfo{pages}{011}
  (\bibinfo{year}{2005}), \eprint{hep-lat/0504019}.

\bibitem[{\citenamefont{Lage et~al.}(2009)\citenamefont{Lage, Mei{$\ss$}ner,
  and Rusetsky}}]{Lage:2009}
\bibinfo{author}{\bibfnamefont{M.}~\bibnamefont{Lage}},
  \bibinfo{author}{\bibfnamefont{U.-G.} \bibnamefont{Mei{$\ss$}ner}},
  \bibnamefont{and} \bibinfo{author}{\bibfnamefont{A.}~\bibnamefont{Rusetsky}},
  \bibinfo{journal}{Phys.Lett.} \textbf{\bibinfo{volume}{B681}},
  \bibinfo{pages}{439} (\bibinfo{year}{2009}), \eprint{0905.0069}.

\bibitem[{\citenamefont{Fu}(2012)}]{Fu:2011}
\bibinfo{author}{\bibfnamefont{Z.}~\bibnamefont{Fu}},
  \bibinfo{journal}{Phys.Rev.} \textbf{\bibinfo{volume}{D85}},
  \bibinfo{pages}{014506} (\bibinfo{year}{2012}), \eprint{1110.0319}.

\bibitem[{\citenamefont{Hansen and Sharpe}(2012)}]{Hansen:2012tf}
\bibinfo{author}{\bibfnamefont{M.~T.} \bibnamefont{Hansen}} \bibnamefont{and}
  \bibinfo{author}{\bibfnamefont{S.~R.} \bibnamefont{Sharpe}},
  \bibinfo{journal}{Phys.Rev.} \textbf{\bibinfo{volume}{D86}},
  \bibinfo{pages}{016007} (\bibinfo{year}{2012}), \eprint{1204.0826}.

\bibitem[{\citenamefont{Brice\~no and
  Davoudi}(2013{\natexlab{a}})}]{Briceno:2012yi}
\bibinfo{author}{\bibfnamefont{R.~A.} \bibnamefont{Brice\~no}}
  \bibnamefont{and} \bibinfo{author}{\bibfnamefont{Z.}~\bibnamefont{Davoudi}},
  \bibinfo{journal}{Phys. Rev.} \textbf{\bibinfo{volume}{D88}},
  \bibinfo{pages}{094507} (\bibinfo{year}{2013}{\natexlab{a}}),
  \eprint{1204.1110}.

\bibitem[{\citenamefont{Gockeler et~al.}(2012)\citenamefont{Gockeler, Horsley,
  Lage, Mei{$\ss$}ner, Rakow, Rusetsky, Schierholz, and
  Zanotti}}]{Gockeler:2012yj}
\bibinfo{author}{\bibfnamefont{M.}~\bibnamefont{Gockeler}},
  \bibinfo{author}{\bibfnamefont{R.}~\bibnamefont{Horsley}},
  \bibinfo{author}{\bibfnamefont{M.}~\bibnamefont{Lage}},
  \bibinfo{author}{\bibfnamefont{U.-G.} \bibnamefont{Mei{$\ss$}ner}},
  \bibinfo{author}{\bibfnamefont{P.}~\bibnamefont{Rakow}},
  \bibinfo{author}{\bibfnamefont{A.}~\bibnamefont{Rusetsky}},
  \bibinfo{author}{\bibfnamefont{G.}~\bibnamefont{Schierholz}},
  \bibnamefont{and} \bibinfo{author}{\bibfnamefont{J.}~\bibnamefont{Zanotti}},
  \bibinfo{journal}{Phys. Rev. D} \textbf{\bibinfo{volume}{86}},
  \bibinfo{pages}{094513} (\bibinfo{year}{2012}), \eprint{1206.4141}.

\bibitem[{\citenamefont{Brice\~no}(2014)}]{Briceno:2014oea}
\bibinfo{author}{\bibfnamefont{R.~A.} \bibnamefont{Brice\~no}},
  \bibinfo{journal}{Phys. Rev.} \textbf{\bibinfo{volume}{D89}},
  \bibinfo{pages}{074507} (\bibinfo{year}{2014}), \eprint{1401.3312}.

\bibitem[{\citenamefont{Brice\~no et~al.}(2018)\citenamefont{Brice\~no, Dudek,
  and Young}}]{Briceno:2017max}
\bibinfo{author}{\bibfnamefont{R.~A.} \bibnamefont{Brice\~no}},
  \bibinfo{author}{\bibfnamefont{J.~J.} \bibnamefont{Dudek}}, \bibnamefont{and}
  \bibinfo{author}{\bibfnamefont{R.~D.} \bibnamefont{Young}},
  \bibinfo{journal}{Rev. Mod. Phys.} \textbf{\bibinfo{volume}{90}},
  \bibinfo{pages}{025001} (\bibinfo{year}{2018}), \eprint{1706.06223}.

\bibitem[{\citenamefont{Polejaeva and Rusetsky}(2012)}]{Polejaeva:2012ut}
\bibinfo{author}{\bibfnamefont{K.}~\bibnamefont{Polejaeva}} \bibnamefont{and}
  \bibinfo{author}{\bibfnamefont{A.}~\bibnamefont{Rusetsky}},
  \bibinfo{journal}{Eur.\ Phys.\ J.\ A} \textbf{\bibinfo{volume}{48}},
  \bibinfo{pages}{67} (\bibinfo{year}{2012}), \eprint{1203.1241}.

\bibitem[{\citenamefont{Tan}(2008)}]{Tan:2007bg}
\bibinfo{author}{\bibfnamefont{S.}~\bibnamefont{Tan}}, \bibinfo{journal}{Phys.
  Rev. A} \textbf{\bibinfo{volume}{78}}, \bibinfo{pages}{013636}
  (\bibinfo{year}{2008}), \eprint{0709.2530}.

\bibitem[{\citenamefont{Beane et~al.}(2007)\citenamefont{Beane, Detmold, and
  Savage}}]{Beane:2007qr}
\bibinfo{author}{\bibfnamefont{S.~R.} \bibnamefont{Beane}},
  \bibinfo{author}{\bibfnamefont{W.}~\bibnamefont{Detmold}}, \bibnamefont{and}
  \bibinfo{author}{\bibfnamefont{M.~J.} \bibnamefont{Savage}},
  \bibinfo{journal}{Phys. Rev.} \textbf{\bibinfo{volume}{D76}},
  \bibinfo{pages}{074507} (\bibinfo{year}{2007}), \eprint{0707.1670}.

\bibitem[{\citenamefont{Detmold and Savage}(2008)}]{Detmold:2008gh}
\bibinfo{author}{\bibfnamefont{W.}~\bibnamefont{Detmold}} \bibnamefont{and}
  \bibinfo{author}{\bibfnamefont{M.~J.} \bibnamefont{Savage}},
  \bibinfo{journal}{Phys. Rev.} \textbf{\bibinfo{volume}{D77}},
  \bibinfo{pages}{057502} (\bibinfo{year}{2008}), \eprint{0801.0763}.

\bibitem[{\citenamefont{Hansen and
  Sharpe}(2016{\natexlab{a}})}]{Hansen:2016fzj}
\bibinfo{author}{\bibfnamefont{M.~T.} \bibnamefont{Hansen}} \bibnamefont{and}
  \bibinfo{author}{\bibfnamefont{S.~R.} \bibnamefont{Sharpe}},
  \bibinfo{journal}{Phys. Rev.} \textbf{\bibinfo{volume}{D93}},
  \bibinfo{pages}{096006} (\bibinfo{year}{2016}{\natexlab{a}}),
  \bibinfo{note}{[Erratum: Phys. Rev.D96,no.3,039901(2017)]},
  \eprint{1602.00324}.

\bibitem[{\citenamefont{Detmold and Flynn}(2015)}]{Detmold:2014fpa}
\bibinfo{author}{\bibfnamefont{W.}~\bibnamefont{Detmold}} \bibnamefont{and}
  \bibinfo{author}{\bibfnamefont{M.}~\bibnamefont{Flynn}},
  \bibinfo{journal}{Phys. Rev. D} \textbf{\bibinfo{volume}{91}},
  \bibinfo{pages}{074509} (\bibinfo{year}{2015}), \eprint{1412.3895}.

\bibitem[{\citenamefont{Brice\~no and
  Davoudi}(2013{\natexlab{b}})}]{Briceno:2012rv}
\bibinfo{author}{\bibfnamefont{R.~A.} \bibnamefont{Brice\~no}}
  \bibnamefont{and} \bibinfo{author}{\bibfnamefont{Z.}~\bibnamefont{Davoudi}},
  \bibinfo{journal}{Phys. Rev.} \textbf{\bibinfo{volume}{D87}},
  \bibinfo{pages}{094507} (\bibinfo{year}{2013}{\natexlab{b}}),
  \eprint{1212.3398}.

\bibitem[{\citenamefont{Guo and Gasparian}(2017)}]{Guo:2017ism}
\bibinfo{author}{\bibfnamefont{P.}~\bibnamefont{Guo}} \bibnamefont{and}
  \bibinfo{author}{\bibfnamefont{V.}~\bibnamefont{Gasparian}},
  \bibinfo{journal}{Phys. Lett.} \textbf{\bibinfo{volume}{B774}},
  \bibinfo{pages}{441} (\bibinfo{year}{2017}), \eprint{1701.00438}.

\bibitem[{\citenamefont{Klos et~al.}(2018)\citenamefont{Klos, K{\"o}nig, Hammer,
  Lynn, and Schwenk}}]{Klos:2018sen}
\bibinfo{author}{\bibfnamefont{P.}~\bibnamefont{Klos}},
  \bibinfo{author}{\bibfnamefont{S.}~\bibnamefont{K{\"o}nig}},
  \bibinfo{author}{\bibfnamefont{H.~W.} \bibnamefont{Hammer}},
  \bibinfo{author}{\bibfnamefont{J.~E.} \bibnamefont{Lynn}}, \bibnamefont{and}
  \bibinfo{author}{\bibfnamefont{A.}~\bibnamefont{Schwenk}},
  \bibinfo{journal}{Phys. Rev.} \textbf{\bibinfo{volume}{C98}},
  \bibinfo{pages}{034004} (\bibinfo{year}{2018}), \eprint{1805.02029}.

\bibitem[{\citenamefont{Guo et~al.}(2018)\citenamefont{Guo, D{\"o}ring, and
  Szczepaniak}}]{Guo:2018ibd}
\bibinfo{author}{\bibfnamefont{P.}~\bibnamefont{Guo}},
  \bibinfo{author}{\bibfnamefont{M.}~\bibnamefont{D{\"o}ring}}, \bibnamefont{and}
  \bibinfo{author}{\bibfnamefont{A.~P.} \bibnamefont{Szczepaniak}},
  \bibinfo{journal}{Phys. Rev.} \textbf{\bibinfo{volume}{D98}},
  \bibinfo{pages}{094502} (\bibinfo{year}{2018}), \eprint{1810.01261}.

\bibitem[{\citenamefont{Hansen and Sharpe}(2019)}]{Hansen:2019nir}
\bibinfo{author}{\bibfnamefont{M.~T.} \bibnamefont{Hansen}} \bibnamefont{and}
  \bibinfo{author}{\bibfnamefont{S.~R.} \bibnamefont{Sharpe}},
  \bibinfo{journal}{Ann. Rev. Nucl. Part. Sci.} \textbf{\bibinfo{volume}{69}},
  \bibinfo{pages}{65} (\bibinfo{year}{2019}), \eprint{1901.00483}.

\bibitem[{\citenamefont{Rusetsky}(2019)}]{Rusetsky:2019gyk}
\bibinfo{author}{\bibfnamefont{A.}~\bibnamefont{Rusetsky}}, in
  \emph{\bibinfo{booktitle}{{37th International Symposium on Lattice Field
  Theory}}} (\bibinfo{year}{2019}), \eprint{1911.01253}.

\bibitem[{\citenamefont{Hansen and Sharpe}(2014)}]{Hansen:2014eka}
\bibinfo{author}{\bibfnamefont{M.~T.} \bibnamefont{Hansen}} \bibnamefont{and}
  \bibinfo{author}{\bibfnamefont{S.~R.} \bibnamefont{Sharpe}},
  \bibinfo{journal}{Phys. Rev.} \textbf{\bibinfo{volume}{D90}},
  \bibinfo{pages}{116003} (\bibinfo{year}{2014}), \eprint{1408.5933}.

\bibitem[{\citenamefont{Hansen and Sharpe}(2015)}]{Hansen:2015zga}
\bibinfo{author}{\bibfnamefont{M.~T.} \bibnamefont{Hansen}} \bibnamefont{and}
  \bibinfo{author}{\bibfnamefont{S.~R.} \bibnamefont{Sharpe}},
  \bibinfo{journal}{Phys. Rev.} \textbf{\bibinfo{volume}{D92}},
  \bibinfo{pages}{114509} (\bibinfo{year}{2015}), \eprint{1504.04248}.

\bibitem[{\citenamefont{Brice\~no et~al.}(2017)\citenamefont{Brice\~no, Hansen,
  and Sharpe}}]{Briceno:2017tce}
\bibinfo{author}{\bibfnamefont{R.~A.} \bibnamefont{Brice\~no}},
  \bibinfo{author}{\bibfnamefont{M.~T.} \bibnamefont{Hansen}},
  \bibnamefont{and} \bibinfo{author}{\bibfnamefont{S.~R.}
  \bibnamefont{Sharpe}}, \bibinfo{journal}{Phys. Rev.}
  \textbf{\bibinfo{volume}{D95}}, \bibinfo{pages}{074510}
  (\bibinfo{year}{2017}), \eprint{1701.07465}.

\bibitem[{\citenamefont{Brice\~no et~al.}(2019)\citenamefont{Brice\~no, Hansen,
  and Sharpe}}]{Briceno:2018aml}
\bibinfo{author}{\bibfnamefont{R.~A.} \bibnamefont{Brice\~no}},
  \bibinfo{author}{\bibfnamefont{M.~T.} \bibnamefont{Hansen}},
  \bibnamefont{and} \bibinfo{author}{\bibfnamefont{S.~R.}
  \bibnamefont{Sharpe}}, \bibinfo{journal}{Phys. Rev.}
  \textbf{\bibinfo{volume}{D99}}, \bibinfo{pages}{014516}
  (\bibinfo{year}{2019}), \eprint{1810.01429}.

\bibitem[{\citenamefont{Romero-L\'opez et~al.}(2019)\citenamefont{Romero-L\'opez,
  Sharpe, Blanton, Brice\~no, and Hansen}}]{Romero-Lopez:2019qrt}
\bibinfo{author}{\bibfnamefont{F.}~\bibnamefont{Romero-L\'opez}},
  \bibinfo{author}{\bibfnamefont{S.~R.} \bibnamefont{Sharpe}},
  \bibinfo{author}{\bibfnamefont{T.~D.} \bibnamefont{Blanton}},
  \bibinfo{author}{\bibfnamefont{R.~A.} \bibnamefont{Brice\~no}},
  \bibnamefont{and} \bibinfo{author}{\bibfnamefont{M.~T.}
  \bibnamefont{Hansen}}, \bibinfo{journal}{JHEP} \textbf{\bibinfo{volume}{10}},
  \bibinfo{pages}{007} (\bibinfo{year}{2019}), \eprint{1908.02411}.

\bibitem[{\citenamefont{Hansen et~al.}(2020)\citenamefont{Hansen,
  Romero-L\'opez, and Sharpe}}]{Hansen:2020zhy}
\bibinfo{author}{\bibfnamefont{M.~T.} \bibnamefont{Hansen}},
  \bibinfo{author}{\bibfnamefont{F.}~\bibnamefont{Romero-L\'opez}},
  \bibnamefont{and} \bibinfo{author}{\bibfnamefont{S.~R.} \bibnamefont{Sharpe}}
  (\bibinfo{year}{2020}), \eprint{2003.10974}.

\bibitem[{\citenamefont{Brice\~no et~al.}(2018)\citenamefont{Brice\~no, Hansen,
  and Sharpe}}]{Briceno:2018mlh}
\bibinfo{author}{\bibfnamefont{R.~A.} \bibnamefont{Brice\~no}},
  \bibinfo{author}{\bibfnamefont{M.~T.} \bibnamefont{Hansen}},
  \bibnamefont{and} \bibinfo{author}{\bibfnamefont{S.~R.}
  \bibnamefont{Sharpe}}, \bibinfo{journal}{Phys. Rev.}
  \textbf{\bibinfo{volume}{D98}}, \bibinfo{pages}{014506}
  (\bibinfo{year}{2018}), \eprint{1803.04169}.

\bibitem[{\citenamefont{Blanton et~al.}(2019)\citenamefont{Blanton,
  Romero-L\'opez, and Sharpe}}]{Blanton:2019igq}
\bibinfo{author}{\bibfnamefont{T.~D.} \bibnamefont{Blanton}},
  \bibinfo{author}{\bibfnamefont{F.}~\bibnamefont{Romero-L\'opez}},
  \bibnamefont{and} \bibinfo{author}{\bibfnamefont{S.~R.}
  \bibnamefont{Sharpe}}, \bibinfo{journal}{JHEP} \textbf{\bibinfo{volume}{03}},
  \bibinfo{pages}{106} (\bibinfo{year}{2019}), \eprint{1901.07095}.

\bibitem[{\citenamefont{Hammer et~al.}(2017{\natexlab{a}})\citenamefont{Hammer,
  Pang, and Rusetsky}}]{Hammer:2017uqm}
\bibinfo{author}{\bibfnamefont{H.-W.} \bibnamefont{Hammer}},
  \bibinfo{author}{\bibfnamefont{J.-Y.} \bibnamefont{Pang}}, \bibnamefont{and}
  \bibinfo{author}{\bibfnamefont{A.}~\bibnamefont{Rusetsky}},
  \bibinfo{journal}{JHEP} \textbf{\bibinfo{volume}{09}}, \bibinfo{pages}{109}
  (\bibinfo{year}{2017}{\natexlab{a}}), \eprint{1706.07700}.

\bibitem[{\citenamefont{Hammer et~al.}(2017{\natexlab{b}})\citenamefont{Hammer,
  Pang, and Rusetsky}}]{Hammer:2017kms}
\bibinfo{author}{\bibfnamefont{H.~W.} \bibnamefont{Hammer}},
  \bibinfo{author}{\bibfnamefont{J.~Y.} \bibnamefont{Pang}}, \bibnamefont{and}
  \bibinfo{author}{\bibfnamefont{A.}~\bibnamefont{Rusetsky}},
  \bibinfo{journal}{JHEP} \textbf{\bibinfo{volume}{10}}, \bibinfo{pages}{115}
  (\bibinfo{year}{2017}{\natexlab{b}}), \eprint{1707.02176}.

\bibitem[{\citenamefont{{D\"oring} et~al.}(2018)\citenamefont{{D\"oring},
  Hammer, Mai, Pang, Rusetsky, and Wu}}]{Doring:2018xxx}
\bibinfo{author}{\bibfnamefont{M.}~\bibnamefont{{D\"oring}}},
  \bibinfo{author}{\bibfnamefont{H.~W.} \bibnamefont{Hammer}},
  \bibinfo{author}{\bibfnamefont{M.}~\bibnamefont{Mai}},
  \bibinfo{author}{\bibfnamefont{J.~Y.} \bibnamefont{Pang}},
  \bibinfo{author}{\bibfnamefont{A.}~\bibnamefont{Rusetsky}}, \bibnamefont{and}
  \bibinfo{author}{\bibfnamefont{J.}~\bibnamefont{Wu}}, \bibinfo{journal}{Phys.
  Rev.} \textbf{\bibinfo{volume}{D97}}, \bibinfo{pages}{114508}
  (\bibinfo{year}{2018}), \eprint{1802.03362}.

\bibitem[{\citenamefont{Pang et~al.}(2019)\citenamefont{Pang, Wu, Hammer,
  Mei{$\ss$}ner, and Rusetsky}}]{Pang:2019dfe}
\bibinfo{author}{\bibfnamefont{J.-Y.} \bibnamefont{Pang}},
  \bibinfo{author}{\bibfnamefont{J.-J.} \bibnamefont{Wu}},
  \bibinfo{author}{\bibfnamefont{H.~W.} \bibnamefont{Hammer}},
  \bibinfo{author}{\bibfnamefont{U.-G.} \bibnamefont{Mei{$\ss$}ner}},
  \bibnamefont{and} \bibinfo{author}{\bibfnamefont{A.}~\bibnamefont{Rusetsky}},
  \bibinfo{journal}{Phys. Rev.} \textbf{\bibinfo{volume}{D99}},
  \bibinfo{pages}{074513} (\bibinfo{year}{2019}), \eprint{1902.01111}.

\bibitem[{\citenamefont{Mai and {D\"oring}}(2017)}]{Mai:2017bge}
\bibinfo{author}{\bibfnamefont{M.}~\bibnamefont{Mai}} \bibnamefont{and}
  \bibinfo{author}{\bibfnamefont{M.}~\bibnamefont{{D\"oring}}},
  \bibinfo{journal}{Eur. Phys. J.} \textbf{\bibinfo{volume}{A53}},
  \bibinfo{pages}{240} (\bibinfo{year}{2017}), \eprint{1709.08222}.

\bibitem[{\citenamefont{Mai et~al.}(2017)\citenamefont{Mai, Hu, {D\"oring},
  Pilloni, and Szczepaniak}}]{Mai:2017vot}
\bibinfo{author}{\bibfnamefont{M.}~\bibnamefont{Mai}},
  \bibinfo{author}{\bibfnamefont{B.}~\bibnamefont{Hu}},
  \bibinfo{author}{\bibfnamefont{M.}~\bibnamefont{{D\"oring}}},
  \bibinfo{author}{\bibfnamefont{A.}~\bibnamefont{Pilloni}}, \bibnamefont{and}
  \bibinfo{author}{\bibfnamefont{A.}~\bibnamefont{Szczepaniak}},
  \bibinfo{journal}{Eur. Phys. J.} \textbf{\bibinfo{volume}{A53}},
  \bibinfo{pages}{177} (\bibinfo{year}{2017}), \eprint{1706.06118}.

\bibitem[{\citenamefont{Jackura et~al.}(2019)\citenamefont{Jackura,
  Fern\'andez-Ram\'irez, Mathieu, Mikhasenko, Nys, Pilloni, Salda\~na,
  Sherrill, and Szczepaniak}}]{Jackura:2018xnx}
\bibinfo{author}{\bibfnamefont{A.}~\bibnamefont{Jackura}},
  \bibinfo{author}{\bibfnamefont{C.}~\bibnamefont{Fern\'andez-Ram\'irez}},
  \bibinfo{author}{\bibfnamefont{V.}~\bibnamefont{Mathieu}},
  \bibinfo{author}{\bibfnamefont{M.}~\bibnamefont{Mikhasenko}},
  \bibinfo{author}{\bibfnamefont{J.}~\bibnamefont{Nys}},
  \bibinfo{author}{\bibfnamefont{A.}~\bibnamefont{Pilloni}},
  \bibinfo{author}{\bibfnamefont{K.}~\bibnamefont{Salda\~na}},
  \bibinfo{author}{\bibfnamefont{N.}~\bibnamefont{Sherrill}}, \bibnamefont{and}
  \bibinfo{author}{\bibfnamefont{A.~P.} \bibnamefont{Szczepaniak}}
  (\bibinfo{collaboration}{JPAC}), \bibinfo{journal}{Eur. Phys. J.}
  \textbf{\bibinfo{volume}{C79}}, \bibinfo{pages}{56} (\bibinfo{year}{2019}),
  \eprint{1809.10523}.

\bibitem[{\citenamefont{Hansen and Sharpe}(2017)}]{Hansen:2016ync}
\bibinfo{author}{\bibfnamefont{M.~T.} \bibnamefont{Hansen}} \bibnamefont{and}
  \bibinfo{author}{\bibfnamefont{S.~R.} \bibnamefont{Sharpe}},
  \bibinfo{journal}{Phys. Rev.} \textbf{\bibinfo{volume}{D95}},
  \bibinfo{pages}{034501} (\bibinfo{year}{2017}), \eprint{1609.04317}.

\bibitem[{\citenamefont{Hansen and
  Sharpe}(2016{\natexlab{b}})}]{Hansen:2015zta}
\bibinfo{author}{\bibfnamefont{M.~T.} \bibnamefont{Hansen}} \bibnamefont{and}
  \bibinfo{author}{\bibfnamefont{S.~R.} \bibnamefont{Sharpe}},
  \bibinfo{journal}{Phys. Rev.} \textbf{\bibinfo{volume}{D93}},
  \bibinfo{pages}{014506} (\bibinfo{year}{2016}{\natexlab{b}}),
  \eprint{1509.07929}.

\bibitem[{\citenamefont{Sharpe}(2017)}]{Sharpe:2017jej}
\bibinfo{author}{\bibfnamefont{S.~R.} \bibnamefont{Sharpe}},
  \bibinfo{journal}{Phys. Rev.} \textbf{\bibinfo{volume}{D96}},
  \bibinfo{pages}{054515} (\bibinfo{year}{2017}), \eprint{1707.04279}.

\bibitem[{\citenamefont{Blanton and Sharpe}(2020)}]{BS2}
\bibinfo{author}{\bibfnamefont{T.~D.} \bibnamefont{Blanton}} \bibnamefont{and}
  \bibinfo{author}{\bibfnamefont{S.~R.} \bibnamefont{Sharpe}}
  (\bibinfo{year}{2020}), \eprint{2007.16190}.

\bibitem[{\citenamefont{Sterman}(1993)}]{Sterman:1994ce}
\bibinfo{author}{\bibfnamefont{G.~F.} \bibnamefont{Sterman}},
  \emph{\bibinfo{title}{{An Introduction to quantum field theory}}}
  (\bibinfo{publisher}{Cambridge University Press}, \bibinfo{year}{1993}), ISBN
  \bibinfo{isbn}{978-0-521-31132-8}.

\bibitem[{\citenamefont{Li et~al.}(2020)\citenamefont{Li, Wu, Young, and
  Lee}}]{Wuboost20}
\bibinfo{author}{\bibfnamefont{Y.}~\bibnamefont{Li}},
  \bibinfo{author}{\bibfnamefont{J.-J.} \bibnamefont{Wu}},
  \bibinfo{author}{\bibfnamefont{R.~D.} \bibnamefont{Young}}, \bibnamefont{and}
  \bibinfo{author}{\bibfnamefont{T.-S.~H.} \bibnamefont{Lee}},
  \bibinfo{journal}{in preparation}  (\bibinfo{year}{2020}).

\end{thebibliography}

\end{document}